\documentclass[12pt]{article}%
\usepackage[nosort]{cite}
\usepackage{graphicx}
\usepackage{multicol}
\usepackage{amsfonts}
\usepackage{amssymb}
\usepackage{amsmath}
\usepackage{heck}
\usepackage{afterpage}
\usepackage{setspace}
\usepackage{verbatim}
\usepackage{color}
\usepackage{longtable}
\usepackage{float}
\usepackage{subcaption}
\usepackage{epsfig}
\usepackage{epstopdf}
\usepackage{adjustbox}
\usepackage{tikz}
\usepackage[margin=1in]{geometry}
\usepackage{titletoc}
\usepackage{hyperref}%
\usepackage{mathrsfs}
\usepackage{dsfont}
\usepackage{algorithm}
\usepackage{algpseudocode}
\usepackage{upgreek}
\usepackage{mathtools}
\usepackage{stmaryrd}
\usepackage{pdflscape}

\usepackage{float}
\pdfoutput=1

\usepackage{caption}
\providecommand{\U}[1]{\protect\rule{.1in}{.1in}}

\newsavebox{\mysavebox}

\hypersetup{colorlinks,citecolor=black,filecolor=black,linkcolor=black,urlcolor=black}
\usetikzlibrary{decorations.markings}
\numberwithin{equation}{section}

\usepackage{tikz}
\usetikzlibrary{arrows}
\usetikzlibrary{arrows.meta}
\usetikzlibrary{arrows,decorations.pathmorphing}
\usetikzlibrary{shapes.geometric,calc,arrows, positioning,shapes.misc,decorations.markings}
\tikzset{
  big arrow/.style={
    decoration={markings,mark=at position 1 with {\arrow[scale=2,#1]{>}}},
    postaction={decorate},
    shorten >=0.4pt},
  big arrow/.default=black}

\usetikzlibrary{hobby}
\usetikzlibrary{decorations.markings}

\pgfdeclarelayer{edgelayer}
\pgfdeclarelayer{nodelayer}
\pgfsetlayers{edgelayer,nodelayer,main}
\tikzstyle{none}=[inner sep=0pt]

\tikzstyle{NodeCross}=[draw, shape=circle, cross out, inner sep=0pt, minimum size=10pt,line width=0.25mm]
\tikzstyle{SmallCircle}=[draw, shape=circle, black, fill=black, inner sep=0pt, minimum size=6pt]
\tikzstyle{BigCircle}=[draw, shape=circle, black, fill=black, inner sep=0pt, minimum size=16pt]
\tikzstyle{SmallCircleRed}=[draw, shape=circle, fill={rgb,255: red,191; green,0; blue,0}, inner sep=0pt, minimum size=6pt]
\tikzstyle{BigCircleRed}=[draw, shape=circle, fill={rgb,255: red,191; green,0; blue,0}, inner sep=0pt, minimum size=10pt]
\tikzstyle{SmallCircleBlue}=[draw, shape=circle, fill=blue, inner sep=0pt, minimum size=6pt]
\tikzstyle{BigCircleBlue}=[draw, shape=circle, fill=blue, inner sep=0pt, minimum size=10pt]
\tikzstyle{SmallCirclePurple}=[draw, shape=circle, fill={rgb,255: red,191; green,0; blue,191}, inner sep=0pt, minimum size=6pt]
\tikzstyle{BigCirclePurple}=[draw, shape=circle, fill={rgb,255: red,191; green,0; blue,191}, inner sep=0pt, minimum size=10pt]
\tikzstyle{SmallCircleGreen}=[draw, shape=circle,  fill={rgb,255: red,80; green,200; blue,120}, inner sep=0pt, minimum size=6pt]
\tikzstyle{BigCircleGreen}=[draw, shape=circle,  fill={rgb,255: red,80; green,200; blue,120}, inner sep=0pt, minimum size=10pt]
\tikzstyle{SmallCircleBrown}=[draw, shape=circle,  fill={rgb,255: red,210; green,105; blue,30}, inner sep=0pt, minimum size=6pt]
\tikzstyle{BigCircleBrown}=[draw, shape=circle,  fill={rgb,255: red,210; green,105; blue,30}, inner sep=0pt, minimum size=10pt]
\tikzstyle{Star}=[draw, shape=star, fill=black, star points=8, inner sep=0pt, minimum size=10pt]

\tikzstyle{DashedLine}=[-, densely dashed, line width=0.25mm]
\tikzstyle{DottedLine}=[-, dotted, line width=0.25mm]
\tikzstyle{ThickLine}=[-, line width=0.25mm]
\tikzstyle{RedLine}=[-, draw={rgb,255: red,191; green,0; blue,0}, fill=none, line width=0.5mm]
\tikzstyle{DashedRedLine}=[-, densely dashed, draw={rgb,255: red,191; green,0; blue,0}, fill=none, line width=0.5mm]
\tikzstyle{DottedRed}=[-, dotted, draw={rgb,255: red,191; green,0; blue,0}, fill=none, dotted, line width=0.5mm]
\tikzstyle{BlueLine}=[-, draw=blue, fill=none, line width=0.5mm]
\tikzstyle{DashedBlueLine}=[-, densely dashed, draw=blue, fill=none, line width=0.5mm]
\tikzstyle{DottedBlue}=[-, dotted, draw=blue, fill=none, dotted, line width=0.5mm]
\tikzstyle{PurpleLine}=[-, draw={rgb,255: red,191; green,0; blue,191}, fill=none, line width=0.5mm]
\tikzstyle{DashedPurpleLine}=[-, densely dashed, draw={rgb,255: red,191; green,0; blue,191}, fill=none, line width=0.5mm]
\tikzstyle{DottedPurple}=[-, dotted, draw={rgb,255: red,191; green,0; blue,191}, fill=none, dotted, line width=0.5mm]
\tikzstyle{GreenLine}=[-, draw={rgb,255: red,80; green,200; blue,120}, fill=none, line width=0.5mm]
\tikzstyle{DashedGreenLine}=[-, densely dashed, draw={rgb,255: red,80; green,200; blue,120}, fill=none, line width=0.5mm]
\tikzstyle{DottedGreen}=[-, dotted, draw={rgb,255: red,80; green,200; blue,120}, fill=none, dotted, line width=0.5mm]
\tikzstyle{BrownLine}=[-, draw={rgb,255: red,210; green,105; blue,30}, fill=none, line width=0.5mm]
\tikzstyle{DashedBrownLine}=[-, densely dashed, draw={rgb,255: red,210; green,105; blue,30}, fill=none, line width=0.5mm]
\tikzstyle{DottedBrown}=[-, dotted, draw={rgb,255: red,210; green,105; blue,30}, fill=none, dotted, line width=0.5mm]
\tikzstyle{ArrowLineRight}=[-,  -{Stealth[scale=1.75]}, line width=0.25mm, scale=5]
\tikzstyle{ArrowLineRed}=[-, draw={rgb,255: red,191; green,0; blue,0},  -{Stealth[scale=1.75]}, line width=0.25mm, scale=5]
\tikzstyle{ArrowLineBlue}=[-, draw=blue,  -{Stealth[scale=1.75]}, line width=0.25mm, scale=5]
\tikzstyle{ArrowLinePurple}=[-, draw={rgb,255: red,191; green,0; blue,191},  -{Stealth[scale=1.75]}, line width=0.25mm, scale=5]
\tikzstyle{ArrowLineGreen}=[-, draw={rgb,255: red,80; green,200; blue,120},  -{Stealth[scale=1.75]}, line width=0.5mm, scale=5]
\tikzstyle{ArrowLineBrown}=[-, draw={rgb,255: red,210; green,105; blue,30},  -{Stealth[scale=1.75]}, line width=0.25mm, scale=5]

\tikzset{snake it/.style={decorate, decoration=snake}}
\tikzset{
dashstar/.style={
 dash pattern=on 5pt off 5pt,
 postaction={
  decorate,
  decoration={
   markings,
   mark=between positions 9pt and 1 step 10pt with {
     \node[color=red] {*};
   }
  }
 }
},
dashstarstar/.style={ 
 dash pattern=on 5pt off 10pt,
 postaction={
   decorate,
   decoration={
     markings,
     mark=between positions 10pt and 1
          step 15pt
           with {
            \node at (-2pt,0pt) {\pgfuseplotmark{star}};
            \node at (2pt,0pt) {\pgfuseplotmark{star}};
           }
   }
 }
}
}
\usepackage{pgfplots}
\pgfplotsset{compat=1.16}

\newcommand{\bea}{\begin{eqnarray}}
\newcommand{\eea}{\end{eqnarray}}

\def\bp{\begin{pmatrix}}
\def\ep{\end{pmatrix}}

\newcommand{\lb}{\left(}
\newcommand{\rb}{\right)}
\newcommand{\lbb}{\left[}
\newcommand{\rbb}{\right]}
\newcommand{\lbbb}{\left\{}
\newcommand{\rbbb}{\right\}}
\newcommand{\be}{\begin{equation}}
\newcommand{\ee}{\end{equation}}
\newcommand{\ba}{\begin{aligned}}
\newcommand{\ea}{\end{aligned}}
\newcommand{\Z}{{\mathbb Z}}

\newcommand{\C}{{\mathbb C}}
\newcommand{\Q}{{\mathbb Q}}
\renewcommand{\P}{{\mathbb P}}

\newcommand{\etaDirac}[1]{\eta^{\text{\:\!$\slashed{D}$}}_{\raisebox{-0.3ex}{\scriptsize$L_{#1}$}}}
\newcommand{\etaDelBar}[1]{\eta^{\text{\:\!$\bar{\partial}$}}_{\raisebox{-0.1ex}{\scriptsize$L_{#1}$}}}

\newcommand{\etaDiracEmpty}{\eta^{\text{\:\!$\slashed{D}$}}_{\raisebox{-0.3ex}{\scriptsize$\varnothing$}}}
\newcommand{\etaDelBarEmpty}{\eta^{\text{\:\!$\bar{\partial}$}}_{\raisebox{-0.1ex}{\scriptsize$\varnothing$}}}

\newcommand{\hDirac}[1]{h^{\text{\:\!$\slashed{D}$}}_{\raisebox{-0.3ex}{\scriptsize$L_{#1}$}}}
\newcommand{\hDelBar}[1]{h^{\text{\:\!$\bar{\partial}$}}_{\raisebox{-0.1ex}{\scriptsize$L_{#1}$}}}

\newcommand{\hDelBarEmpty}[1]{h^{\text{\:\!$\bar{\partial}$}}_{\raisebox{-0.1ex}{\scriptsize$\varnothing$}}}

\begin{document}

\begin{flushright}
    UUITP-36/25

    UPR-1335-T
\end{flushright}

\date{December 2025}

\title{Extra-Dimensional $\eta$\:\!-Invariants \\[0mm] and Anomaly Theories}


\institution{PENN}{\centerline{$^{1}$Department of Physics and Astronomy, University of Pennsylvania, Philadelphia, PA 19104, USA}}
\institution{PENNmath}{\centerline{$^{2}$Department of Mathematics, University of Pennsylvania, Philadelphia, PA 19104, USA}}
\institution{Maribor}{\centerline{$^{3}$Center for Applied Mathematics and Theoretical Physics, University of Maribor, Maribor, Slovenia}}

\institution{Uppsala}{\centerline{$^{4}$Department of Physics and Astronomy, Uppsala University, Box 516, SE-75120 Uppsala, Sweden}}
\institution{CMSA}{\centerline{$^{5}$Center of Mathematical Sciences and Applications, Harvard University,
Cambridge, MA 02138, USA}}

\institution{Jefferson}{\centerline{$^{6}$Jeﬀerson Physical Laboratory, Harvard University, Cambridge, MA 02138, USA}}

\authors{
Mirjam Cveti\v{c}\worksat{\PENN, \PENNmath, \Maribor}\footnote{e-mail: \texttt{cvetic@physics.upenn.edu}},
Ron Donagi\worksat{\PENNmath, \PENN}\footnote{e-mail: \texttt{donagi@upenn.edu}},   \\[0.5em]
Jonathan J. Heckman\worksat{\PENN, \PENNmath}\footnote{e-mail: \texttt{jheckman@sas.upenn.edu}},
and Max H\"ubner\worksat{\Uppsala, \CMSA, \Jefferson}\footnote{e-mail: \texttt{max@cmsa.fas.harvard.edu}}
}

\abstract{Anomalies of a quantum field theory (QFT) constitute fundamental non-perturbatively robust data.
In this paper we extract anomalies of 5D superconformal field theories (SCFTs) directly from the underlying extra-dimensional geometry. We show that all of this information can be efficiently extracted from extra-dimensional $\eta$-invariants, bypassing previously established approaches based on computationally cumbersome blowup / resolution techniques. We illustrate these considerations for 5D SCFTs engineered in M-theory by non-compact geometries $X=\mathbb{C}^3/\Gamma$ with finite subgroup $\Gamma\subset SU(3)$, where the anomalies are determined by the $\eta$-invariants of the asymptotic boundary $\partial X=S^5/\Gamma$. Our results apply equally to Abelian and non-Abelian $\Gamma$, as well as isolated and non-isolated singularities.
In the setting of non-isolated singularities we further analyze the interplay of anomaly structures across different strata of the singular locus. Our considerations extend readily to backgrounds  which are not global orbifolds, as well as those which do not preserve supersymmetry.}

\maketitle

\enlargethispage{\baselineskip}

\setcounter{tocdepth}{3}

\tableofcontents

\newpage

\section{Introduction}

In recent years generalized symmetries of quantum field theories (QFTs) have seen increased study via their symmetry theories.\footnote{See, e.g., \cite{Reshetikhin:1991tc, TURAEV1992865,
Barrett:1993ab, Witten:1998wy, Fuchs:2002cm, Kirillov2010TuraevViroIA,
Kapustin:2010if, Kitaev2011ModelsFG, Fuchs:2012dt, Freed:2012bs, Heckman:2017uxe, Freed:2018cec,
Gaiotto:2020iye, Apruzzi:2021nmk, Freed:2022qnc, Kaidi:2022cpf,
Brennan:2024fgj,  Argurio:2024oym, DelZotto:2024tae, Apruzzi:2024htg} for a partial list of references to foundational early work and recent generalizations.} Symmetry theories, including, e.g., symmetry topological field theories (SymTFTs), are auxiliary extra-dimensional theories isolating and organizing structures governing a QFTs symmetries across its global forms.

Such auxiliary theories are strikingly physical in contexts where the QFT is itself realized via an extra-dimensional construction, as is the case for string constructed QFTs, supersymmetric or not, and QFTs with gravitational duals.\footnote{See, e.g., \cite{Antinucci:2022vyk, vanBeest:2022fss, Heckman:2022xgu, Bah:2023ymy,Apruzzi:2023uma, Cvetic:2023pgm, Baume:2023kkf, Yu:2023nyn, Heckman:2024oot, Bonetti:2024cjk, Braeger:2024jcj, Franco:2024mxa, Heckman:2024zdo, Cvetic:2024dzu, GarciaEtxebarria:2024jfv, Tian:2024dgl,  Cvetic:2024mtt, Bonetti:2024etn, Najjar:2024vmm, Heckman:2025lmw} for a partial list of references on extra-dimensional approaches to symmetry theories.} The main idea is that the QFT of interest sits in a local patch of an extra-dimensional geometry which can then be studied in isolation after decoupling all bulk degrees of freedom.

To illustrate, consider string / M-theory backgrounds of the form $M_{d} \times X$ with $M$ flat Minkowski space and non-compact $X$ a topological cone $\mathrm{Cone}(\partial X)$ with a singularity at the tip. This engineers a theory $\mathcal{T}_X$ where the QFT degrees of freedom remain localized at the tip of the cone, and other non-dynamical structures can be associated with singularities and branes which extend from the tip of the cone to asymptotic infinity, i.e., $\partial X$. An important feature of this approach is that it provides a general way to engineer theories which are intrinsically strongly coupled, including, for example 6D and 5D SCFTs.\footnote{See e.g., \cite{Heckman:2018jxk, Argyres:2022mnu} for reviews and references therein.} In addition to purely geometric backgrounds, one can also include additional gravitational sources to realize many examples of the AdS / CFT correspondence, where the spacetime geometry is then of the form $\mathrm{AdS}_{d+1} \times \partial X$.

The SymTFT framework naturally arise in this setting since the cone automatically comes with two ``boundaries.''
One is the locus of the interacting $\mathrm{QFT}_{d}$ itself, while the other is the boundary $\partial X$ at asymptotic infinity.
Schematically, one can realize a $(d+1)$-dimensional bulk which encodes the symmetry data of the $\mathrm{QFT}_{d}$
by reducing along the link geometry $\partial X$, resulting in the schematic relation:
\begin{equation}\label{eq:Mapping}
``\partial X \mapsto \mathrm{SymTFT}(\mathcal{T}_X)."
\end{equation}
This class of TFTs includes quadratic order BF-terms, as follows directly from a computation of the associated higher-form symmetries of the QFT$_d$, which can in turn be extracted from the defect group \cite{DelZotto:2015isa} as in \cite{Albertini:2020mdx, Morrison:2020ool}.\footnote{See also \cite{Baume:2023kkf} for how to extract these terms by dimensional reducing the kinetic terms of the higher-dimensional supergravity theory around the torsion classes of $X$. For a careful treatment involving the formulation of gauge potentials in differential cohomology, see reference \cite{GarciaEtxebarria:2024fuk}.} In \cite{Apruzzi:2021nmk} it was also proposed that (at least in principle) one can use the reduction of the M-theory topological terms $C_3 \wedge G_4 \wedge G_4$ and $C_3 \wedge X_8(R)$ to extract cubic couplings of the SymTFT.

Easier said than done.

In practice, there are many limitations to carrying through the mapping of line \eqref{eq:Mapping}. Singularities of $X$ complicate the reduction and uniform approaches have been developed for settings in which $X$ admits a desingularization. However, favorable desingularizations may not exist or geometric moduli spaces, such as in settings of exceptional holonomy, may lack sufficient control. There is also the practical consideration that the resolution of the singularities of $X$ is often rather challenging and introduces a significant amount of resolution-dependent structure into the problem.\footnote{For a recent tour de force for orbifold singularities, see reference \cite{Tian:2021cif}.} Even assuming this has been accomplished, one is then faced with the question of how to compute relevant intersection / cohomology rings. Another conceptual issue is that if the linking geometry $\partial X$ is also singular, then one has additional physical degrees of freedom on stratified loci. In this case, smoothing such singularities can obscure some of the desired topological couplings of the bulk system. On the other hand, if the singularities are retained the supergravity approximation is inappropriate, and must be supplemented by additional physical inputs.

Faced with these considerations, our aim in this work will be to develop a practical computational scheme for extracting such topological data directly from the extra-dimensional geometry, irrespective of whether $X$ carries stratified singularity structures. The main theme will be to eschew the use of any blowups and instead work directly with the singular geometry $X$ and its (possibly also singular) boundary geometry $\partial X$.

The main idea we develop is that the relevant anomaly data is all captured by a suitable class of $\eta$-invariants defined on the (possibly singular) boundary geometry $\partial X$. On general grounds, the reason one ought to expect such a correspondence is that relevant bulk intersection-theoretic quantities can often be recast in terms of invariants constructed from integration of characteristic classes. For example, the triple intersection numbers $D_{i} \cap D_{j} \cap D_{k}$ of non-compact divisors in a non-compact threefold $X$ can be written as:
\begin{equation}
D_{i} \cap D_{j} \cap D_{k} \sim \underset{X}{\int}  F_{i}F_{j} F_{k}\,,
\end{equation}
where we have used the well-known correspondence between line bundles, their first Chern classes and divisors. On the other hand, it is well-known that index theorems on non-compact spaces receive contributions from a suitable $\eta$-invariant on $\partial X$. We leverage this observation to fully dispense with the bulk, focusing instead on the
structures intrinsic to the singular geometry and its avatar $\partial X$. Indeed, general anomaly inflow considerations 
directly imply that all of the relevant data can be extracted directly from $\partial X$.

The main class of theories we will consider are string / M-theory backgrounds where $X = \mathbb{C}^{n} / \Gamma$ for finite $\Gamma$. In the case of ADE singularities $\mathbb{C}^{2} / \Gamma_{\mathrm{ADE}}$, M-theory on this space engineers 7D Super Yang-Mills (SYM) theory, and similar considerations hold for IIA on the same geometry. M-theory on a supersymmetric background $X = \mathbb{C}^{3} / \Gamma$ typically engineers a 5D SCFT. We show that in this class of backgrounds, 1-form symmetry anomalies as well as mixed 1-form 0-form symmetry anomalies are all captured by differences of suitable $\eta$-invariants evaluated modulo $1$ (i.e., in $\mathbb{Q} / \mathbb{Z}$).
For the subclass of isolated singularities this will simply follow from the Atiyah-Singer-Patodi index theorem. For non-isolated singularities, where supergravity approximations away from the singularity break down (on which the former case relies), we will instead demonstrate this through explicit computation.\footnote{See Appendix \ref{sec:Check} for two representative cases of such computations.}

The relation between  $\eta$-invariants and anomalies we point out has a number of advantages over resolution based methods. First, this perspective presents anomalies purely as data of $\partial X=S^5/\Gamma$ and as such, we expect the presented methods to be minimal, at least in the context of computational complexity. Second, for spaces such as $S^5/\Gamma$ expressions for the relevant $\eta$-invariants are well-known, and the anomalies can be presented in closed form, often as some rational function across families of geometries. Further, our approach applies to both Abelian and non-Abelian groups $\Gamma$, groups preserving supersymmetry or not, and does not require considerations of any smoothings. The latter is a drastic simplification: instead of requiring an understanding of the moduli space of the geometry $X$ or equivalently the moduli space of the physical theory $\mathcal{T}_X$ (and the relevant cohomology theories across it), we have that the anomalies are derived directly from the singular geometry. Third, for the case of non-isolated singularities we will be also able to study symmetry structures across the different strata of the singular locus and their interplay \cite{Cvetic:2024dzu}. Fourth, generalizations to other geometric backgrounds are immediate. Fifth, note that there are a number of extra-dimensional approaches to symmetry theories where reduction procedures, similar to \eqref{eq:Mapping}, are applied to (non-closed) subsets of $X$ of codimension larger than one (with boundaries).\footnote{Here, instead of presenting $X$ as fibered over some radial interval, it is viewed with respect to a projection to some manifold with boundaries/corners (which replaces the radial interval $I$). Then, one dimensionally reduces with respect to the preimage of this projection \cite{Cvetic:2024dzu} (this generalizes the SymTFT sandwich to a higher-dimensional auxiliary structure, referred to as a cheesesteak). Already, when $X$ is not a metric cone with multiple isolated singularities, similar considerations may be applied to realize related generalization known as SymTrees \cite{Baume:2023kkf}. } In such settings working directly with these subsets is preferable in the sense of the first point above.

To proceed then, this paper is organized as follows. In section \ref{sec:ExtraDim} we review features of the dimensional reduction scheme \eqref{eq:Mapping} and generalizations thereof. In section \ref{sec:eta} we review some properties of $\eta$-invariants and collect some technical background material. Section \ref{sec:ADE} discusses the prototype example of ADE singularities $\mathbb{C}^2/\Gamma_{\text{ADE}}$ in M-theory focusing on aspects which will play key roles when considering the Calabi-Yau orbifolds $\mathbb{C}^3/\Gamma$ in section \ref{sec:5D}. At the end of section \ref{sec:ADE} we also discuss related constructions in IIA, including non-supersymmetric orbifold backgrounds. In section \ref{sec:5D} we further discuss the interplay of anomaly data across strata of the singular locus, and the structures which may be reliably computed from the effective 11D supergravity approximation. In section \ref{sec:Conc} we present our conclusions and an outlook. Some computational details, explicitly discussed with respect to two representative families of geometries, are presented in Appendix \ref{sec:Check}.

\section{Top Down Approach to Symmetry Theories}
\label{sec:ExtraDim}

We now lay out and review select aspects of extra-dimensional approaches to symmetry theories, such as SymTFTs, in the context of M-theory. Our discussion will center on anomaly terms of symmetry theories and for further well-understood discussion, which focusses more closely on the structure of the underlying defect group \cite{DelZotto:2015isa} and the associated BF-sector, we refer to \cite{Morrison:2020ool, Bhardwaj:2020phs, DelZotto:2022fnw, DelZotto:2022ras, GarciaEtxebarria:2024fuk, Braeger:2025rov, WIP2}.

Let us begin by providing some perspective and motivation. In this work we will primarily consider M-theory on the purely geometric background $M\times X$. Initially we will assume that the extra-dimensional geometry $X=\text{Cone}(\partial X)$ is a cone with an isolated singularity at its tip. Away from the tip of the cone, where $X$ is smooth and weakly curved, 11D supergravity is a good approximation to the effective physics. The relative theory $\mathcal{T}_X$ can then be viewed as a defect theory with respect to 11D supergravity. In order for this system to be well-defined one needs to impose some consistency conditions (see, e.g., \cite{Witten:2001uq}). The symmetry theory framework then aims to present the subset of consistency conditions, formulated manifestly with regards to $\mathcal{T}_X$, deriving from compatibility with topological 11D bulk structures.

This basic setup and method of studying localized degrees of freedom on extra-dimensional patches, where certain low-energy effective descriptions break down, has many generalizations. We will be mostly concerned with generalizations to non-isolated curvature singularities. Of course, extensions to other singular supergravity backgrounds, by including for example branes, also pose interesting research directions. To proceed then, in this section, we will first lay out standard approaches applicable to isolated curvature singularities, which results in SymTFTs, and then discuss different perspectives for approaching non-isolated singularities.

\subsection{Extra-Dimensional Starting Point}

We begin with the observation that 11D supergravity has a Lagrangian formulation.
So, to first approximation, the reduction procedure in \eqref{eq:Mapping} then produces, whenever $\partial X$ is smooth, a Lagrangian topological field theory with action $\mathcal{S}_{\;\!\text{SymTFT\:\!}}$ simply by expanding the 11D supergravity fields and contracting their internal supports. In our discussion of anomalies, we will focus primarily on contributions to $\mathcal{S}_{\;\!\text{SymTFT}}$ which derive from the Chern-Simons terms of 11D supergravity.

For our purposes, the Chern-Simons terms of 11D supergravity can be taken to be\footnote{In the second line we separated the contributions from the L\:\!-genus and $\hat{\text{A}}$-genus, associated with the M5-brane chiral 2-form and worldvolume fermions, respectively \cite{Alvarez-Gaume:1983ihn, Witten:1996hc, Harvey:1998bx, Diaconescu:2000wy}.}
\be\ba\label{eq:CS}
\mathcal{S}_{\text{CS}}^{\:\!\circ}&=\frac{2\pi }{6}\int_{Y_{12}^\circ} \frac{G_4}{2\pi}  \lbb \lb \frac{G_4}{2\pi} \rb^{\!2}+\frac{p_1^2-4p_2}{32} \rbb\\
&=\frac{2\pi }{6}\int_{Y_{12}^\circ} \frac{G_4}{2\pi}  \lbb \lb \frac{G_4}{2\pi} \rb^{\!2}+\lb \frac{p_1^2}{60}-\frac{7p_2}{60}\rb +\lb \frac{7p_1^2}{480}-\frac{4p_2}{480}\rb\rbb\,.
\ea\ee
Here the integral is over an auxiliary 12-dimensional Spin manifold $Y_{12}^\circ\subset Y_{12}$ with boundary, where $ Y_{12}$ is such that the boundary $(\partial Y)_{11}$ is 11D supergravity spacetime. Further, we denote by $G_4$ the 4-form field strength of the 3-form gauge potential $C_3$ (i.e., locally $G_4 = dC_3$) and Pontryagin classes by $p_k\equiv p_k(Y_{12})$.\footnote{We will repeatedly encounter expressions featuring both characteristic classes $p_k,c_k,w_k,\dots$ and other fields $B_2,C_3,G_4,\dots$. In the latter, subscripts will make reference to the differential degree, for the former this is generically not the case. This clash of notation is unfortunate, but standard. } The space $Y_{12}^\circ$ results from excising a ball centered on the singular locus of $Y_{12}$ and is such that  11D supergravity theory realizes an accurate effective approximation precisely on $(\partial Y)_{11}^\circ\cap (\partial Y)_{11} $.

Note that $G_4/2\pi$ is not necessarily integrally quantized and one can introduce the variables
\be \label{eq:Shift}
\lambda_1=\frac{p_1}{2}\,, \qquad \frac{H_4}{2\pi}=\frac{G_4}{2\pi}-\frac{\lambda_1}{2}\,,
\ee
such that $H_4/2\pi$ is always integrally quantized \cite{Witten:1996md}. Overall, $\mathcal{S}_{\text{CS}}^\circ$ may then be rewritten as
\be \label{eq:ShiftedCS}\ba
\mathcal{S}_{\:\!\text{CS}}^{\:\!\circ}&=\frac{2\pi }{6}\int_{Y_{12}^\circ}   \frac{H_4}{2\pi}  \lbb \lb \frac{H_4}{2\pi} \rb^2+   \frac{7 \lambda_1^2 -p_2}{8}\rbb+\frac{2\pi}{4}\int_{Y_{12}^\circ}  \lb \frac{H_4}{2\pi}\rb^2  \lambda_1 +\frac{2\pi}{32}\int_{Y^\circ_{12}}  \lb  \lambda_1^3-\frac{\lambda_1 p_2}{3} \rb.
\ea \ee
This action is considered mod 1 and Pontryagin classes therefore enter modulo congruences. Such congruences allow for these to be expressed with respect to other characteristic classes. We will be interested in local models for which the 11D spacetime takes the form
\be
(\partial Y)_{11}=\mathbb{R}^{1,d-1}\times  X_{11-d}\,, \qquad X_{11-d}=\text{Cone}((\partial X)_{10-d})\,,
\ee
where $X_{11-d}$ is a metric cone with link $\partial (X_{11-d})\equiv(\partial X)_{10-d}$. Consequently, $Y_{12}^\circ$ is topologically modeled as
\be\label{eq:model}
Y_{12}^\circ\,\sim\, \mathbb{R}^{1,d-1}\times I\times Z_{11-d}\,,
\ee
where $I$ is an open interval and $Z_{11-d}$ is a smooth space with boundary $\partial(Z_{11-d})= (\partial X)_{10-d}$.\footnote{The interval $I$ is parametrized by the radius $r$ of the cone $X_{11-d}=\text{Cone}(\partial X_{10-d})$ and explicitly given by $(r_0,\infty)$ for some $r_0>0$. The low-energy 11D supergravity approximation comes with a cutoff $\Lambda$ which determines how much curvature may be tolerated by this effective theory and therefore $r_0\equiv r_0(\Lambda)$. In the context of symmetry theories, which are topological, all that matters is $r_0>0$ which holds for all $\Lambda$. In this sense the picture is similar to the SymTFT sliver advocated for in holographic settings in \cite{Heckman:2024oot}. In particular, topology changing deformations to $X_{11-d}$ localized to its tip leave the SymTFT (and its physical boundary) invariant until, e.g., cycles introduced by deformation exceed $r_0$ in size, as for example studied in \cite{Baume:2023kkf}. }

\begin{figure}
\centering
\scalebox{0.8}{\begin{tikzpicture}
	\begin{pgfonlayer}{nodelayer}
		\node [style=none] (0) at (0, 2) {};
		\node [style=none] (1) at (0, -2) {};
		\node [style=none] (2) at (4, 2) {};
		\node [style=none] (3) at (0, 1) {};
		\node [style=none] (4) at (3, 1) {};
		\node [style=SmallCircle] (5) at (0, 1) {};
		\node [style=none] (6) at (0.375, 1.375) {};
		\node [style=none] (7) at (0.5, 3.25) {};
		\node [style=none] (8) at (0, 3.75) {$(\partial X)_{10-d}^{\:\!r=r_*}$};
		\node [style=none] (11) at (0, -2) {};
		\node [style=none] (13) at (-0.75, 0) {$X_{11-d}$};
		\node [style=none] (14) at (1.25, 0.5) {$Z_{11-d}^{\:\!r=r_*}$};
		\node [style=none] (15) at (0, -2.75) {};
		\node [style=Star] (16) at (0, -2) {};
		\node [style=none] (17) at (-1, -2) {$r=0$};
		\node [style=none] (18) at (0, -1) {};
		\node [style=none] (19) at (1, -1) {};
		\node [style=none] (20) at  (2, 2.5) {$Y_{12}^\circ$};
        \node [style=none] (21) at (6, 0.5) {};
		\node [style=none] (22) at (6, -1) {};
		\node [style=none] (23) at (7.5, -1) {};
		\node [style=none] (24) at (5.5, -0.25) {$r$};
		\node [style=none] (25) at (6.75, -1.5) {$r_\bot $};
	\end{pgfonlayer}
	\begin{pgfonlayer}{edgelayer}
		\filldraw[fill=red!20, draw=red!20]  (0, -1) -- (1, -1) -- (4, 2) -- (0,2) -- cycle;
		\draw [style=ThickLine] (0.center) to (1.center);
		\draw [style=ThickLine] (1.center) to (2.center);
		\draw [style=ThickLine] (2.center) to (0.center);
		\draw [style=BlueLine] (3.center) to (4.center);
		\draw [style=ArrowLineRight, bend left=45] (7.center) to (6.center);
		\draw [style=RedLine] (18.center) to (0.center);
		\draw [style=DashedLine] (18.center) to (19.center);
        \draw [style=ArrowLineRight] (22.center) to (23.center);
		\draw [style=ArrowLineRight] (22.center) to (21.center);
	\end{pgfonlayer}
\end{tikzpicture}}
\caption{We sketch the internal dimensions of $Y_{12}$ projected onto the radial coordinates of $X_{11-d}$ and $Z_{11-d}$ denoted $r,r_\bot$ respectively. 11D supergravity is a good approximation on the red line which carries at every point, at radius $r$  in  $X_{11-d}$,  the fiber $(\partial X)_{10-d}$. This fiber collapses at $r=0$ forming the singularity. At fixed radius $r=r_*$ the radial slice of $X_{11-d}$ is $(\partial X)_{10-d}^{\:\!r=r_*}$. The space $Z_{11-d}^{\:\!r=r_*}$ is smooth and has a boundary given by this copy of the link and fills it in along a transverse auxiliary dimension (blue). The spaces $(\partial X)_{10-d}^{\:\!r=r_*}$ as well as $Z_{11-d}^{\:\!r=r_*}$ are topologically equivalent for all $r_*> 0$. The dashed line denotes the cut used to construct $Y_{12}^\circ$ via excision of a small ball centered on the singularity. The space $Y_{12}^\circ$ projects onto the shaded region.    }
\label{Fig:Triangle}
\end{figure}
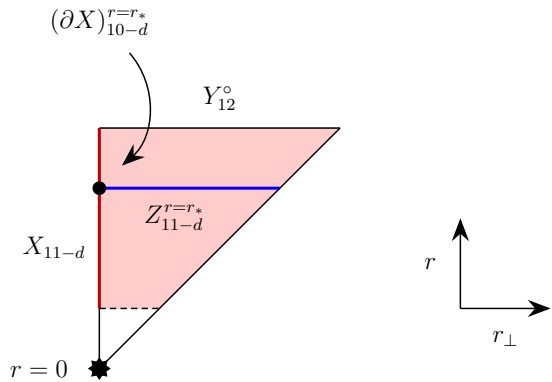

The auxiliary space $Z_{11-d}$ could be realized by a copy of some desingularization of $X_{11-d}$. However, in the settings we will discuss, one can also take $Z_{11-d}$ to be a copy of the singular space $X_{11-d}$ itself and, allowing $Y_{12}^\circ$ to be singular in this sense, $Y_{12}^\circ$ always exists. Our final computations will not depend on any particular choice of $Z_{11-d}$. Further, intermediate steps will be index theoretic and readily extend to the category of orbifolds. Overall, $Z_{11-d}$ always exists and using it, the Chern-Simons terms \eqref{eq:ShiftedCS} may be formulated using singular cohomology. In other words, the supergravity approximation suggests, focusing on anomaly terms, that the reduction procedure \eqref{eq:Mapping} may be parsed using singular cohomology, whenever $\partial X_{10-d}$ is smooth, and that the corresponding SymTFT sector on $\mathbb{R}^{1,d-1}\times I$ is obtained by reduction on $Z_{11-d}$ starting from \eqref{eq:model}.\footnote{Uplifts to differential cohomology, as widely considered in the literature, would allow for an intrinsically 11D formulation, but we will not require such methods in this work.} See figure \ref{Fig:Triangle} for a sketch of the setup.

Above we focused on contributions to the SymTFT action from the 11D Chern-Simons terms. Often, BF-type terms are also considered and utilized to manifestly record quantization conditions on background fields and which background fields are quantum mechanically conjugate variables in the SymTFT. Parsing these structures using such BF-type terms is natural in extra-dimensional constructions of symmetry theories due to the original supergravity fields initially being $U(1)$-valued and arising outside of the context of topological field theory. In the simplest cases, relevant BF-type terms derive from the kinetic term of 11D supergravity proportional to $\frac{1}{2} G_4 G_7$, see for example Appendix B of \cite{Baume:2023kkf}. The setups considered throughout this work may be treated similarly, in particular, these considerations will allow us to simply read off the BF-type terms given the defect group \cite{DelZotto:2015isa, Albertini:2020mdx, Morrison:2020ool} (which is straightforwardly determined from the geometry as briefly reviewed in subsection \ref{ssec:DGS}).

\subsection{Stratified Systems}
\label{sec:stratified}

In many situations of interest the conical geometry $X$ includes singularities which extend from the tip of the cone to asymptotic infinity. Moreover, these singularities can in principle produce further intersections, leading to even further singular behavior (see, e.g., \cite{Acharya:2023bth}). This sort of stratification of singular structures complicates the identification of a bulk SymTFT / SymTh with an extra-dimensional geometry since the supergravity approximation is not really valid near these additional singular loci.

In more detail, consider the asymptotic boundary $(\partial X)_{10-d}$ is singular and the singularity at the tip of $X_{11-d}=\text{Cone}((\partial X)_{10-d})$ is non-isolated and is embedded into higher codimension singular strata which extend radially out to the boundary. In the presence of these additional singularities, 11D supergravity is only a good approximation on
\be
(\partial X)_{10-d}^\circ\equiv (\partial X)_{10-d} \setminus T((\partial \mathscr{S})_{D-d-1} )\,,
\ee
which is a manifold with boundary. Here $\mathscr{S}_{D-d}$ denotes the full singular locus within $X_{11-d}$, which we will assume consists of a singularity enhancement at the tip of $X_{11-d}$ and a generic singular stratum of dimension $D-d>0$ (intersecting the boundary $(\partial X)_{10-d} $ in $(\partial \mathscr{S})_{D-d-1}$), and $T(\cdot)$ denotes an open tubular neighborhood in $ (\partial X)_{10-d} $.

The non-compact strata of $\mathscr{S}_{D-d}$ define additional QFT sectors in higher dimensions. From this broader perspective,
one can view the QFT$_d$ as an edge mode theory in a higher-dimensional system (see, e.g., \cite{Acharya:2023bth}).
In the cases encountered in this work, we can often interpret these strata as flavor branes, i.e., they admit an interpretation as a higher-dimensional gauge theory, and so by abuse of terminology we shall always refer to such sectors as ``flavor branes'' in this work.
The inclusion of these flavor brane sectors in the reduction on a linking geometry can schematically be viewed as contributing a BF-like theory for a continuous non-abelian gauge theory \cite{Bonetti:2024cjk}.

As such, in cases with singular $(\partial X)_{10-d}$, there are essentially two ways to set up topological systems related to symmetries of the localized degrees of freedom. The first treats the degrees of freedom localized to the various singular strata of $\mathscr{S}_{D-d}$ uniformly, see \cite{Cvetic:2024dzu}. This procedure iterates the original SymTFT sandwich construction of \cite{Apruzzi:2021nmk}, as previously discussed for smooth links. One introduces a second radial coordinate in $(\partial X)_{10-d}$ centered on $(\partial \mathscr{S})_{D-d-1}$ and repeats the discussion for smooth links now focusing on the space $(\partial X)_{10-d}$ and smooth links of the singularities modeled on
\be
\partial((\partial X)^\circ)_{9-d}\equiv \partial \:\! T((\partial \mathscr{S})_{D-d-1})\,.
\ee
The construction therefore now results in two topological field theories in adjacent dimensions. One derives from $\partial((\partial X)^\circ)_{9-d}$ and another, which is an edge mode to it, derives from $(\partial X)^\circ_{10-d}$. The former is associated with $(\partial \mathscr{S})_{D-d-1}$ and captures symmetries of degrees of freedom localized to the generic singular stratum, the latter refines these to symmetry structures of the $d$-dimensional degrees of freedom localized to the singularity enhancement.

In contrast, the second topological system characterizing symmetries may only be realized when the $d$-dimensional system can be isolated, i.e., when there exist limits decoupling the non-topological excitations of the $D$-dimensional system. In such cases the physical boundary condition associated in the previous construction to $\mathscr{S}_{D-d}$ away from the tip of the cone becomes non-dynamical and the second SymTFT sandwich, which opens up normal to it, can now be contracted. The result is a standard $(d+1)$-dimensional SymTFT, as with smooth links, but now with contributions from topological structures localized to $\mathscr{S}_{D-d}$.

The first approach is set up precisely such that symmetry theories are associated with smooth subsets of $(\partial X)_{10-d}$. As such, the 11D Chern-Simons term \eqref{eq:ShiftedCS} is a good higher-dimensional starting point for a reduction procedure formulated using singular cohomology. Our work will be primarily motivated by this iterated SymTFT construction, which records and relates symmetry structures across singular strata. Such book-keeping is of particular interest when symmetry structures associated with different strata interact, as, e.g., in the setting of 2-group symmetries \cite{Cordova:2018cvg, Benini:2018reh, Cvetic:2022imb, DelZotto:2022joo}, and we will aim to compute anomalies of such combined systems, sharpening observations of \cite{Cvetic:2024dzu}.

In contrast, working directly with the singular link, as in the second approach is less clear, since neither the supergravity approximation may be trusted, nor the use of singular cohomology. These challenges are quantified precisely by the first approach, which identifies the symmetry data which can be reliably derived from the ambient supergravity theory using singular cohomology. Of course, this still fails to access the symmetry data localized to the singularities $(\partial \mathscr{S})_{D-d-1}$ which are truly beyond 11D supergravity and singular cohomology.\footnote{For example, in type II such data may be accessed, in part, with more stringy cohomology theories, such as Chen-Ruan orbifold cohomology \cite{Chen:2000cy, Braeger:2025rov}, which is sensitive to twisted sectors and correctly identifies Cartan subalgebras of flavor symmetries localized at singularities.}

In summary, by iterating the standard SymTFT construction across the singular strata of the space $X_{11-d}$ we isolate topological structures which are reliably computable from the bulk supergravity approximation to M-theory, utilizing a reduction scheme based on singular cohomology.
\footnote{Going forward we will omit indices indicating dimensions to not overcrowd notation. Similarly we leave wedge and cup products implicit.}


\subsection{Defect Groups and Generalized Symmetries}
\label{ssec:DGS}

Our primary focus throughout this work will center on anomalies of global symmetries and their computations from an extra-dimensional perspective. To frame this discussion we collect some background material and set notation regarding the geometric and field-theoretic characterizations of these symmetries and the defect operators which naturally carry their representations.

In relation to the discussion so far, these should be viewed as symmetry theory substructures, e.g., counts, braidings and extension structures underlying their extended operators. The material collected here is essentially an abridged account of developments covered in more detail in
\cite{DelZotto:2015isa, Morrison:2020ool, Albertini:2020mdx, Bhardwaj:2020phs, Tian:2021cif, DelZotto:2022fnw, Heckman:2022muc, Antinucci:2022vyk, vanBeest:2022fss, Heckman:2022xgu, Bah:2023ymy,Apruzzi:2023uma, Cvetic:2023pgm, Baume:2023kkf, Yu:2023nyn, Heckman:2024oot, Bonetti:2024cjk, Braeger:2024jcj, Franco:2024mxa, Heckman:2024zdo, Cvetic:2024dzu, GarciaEtxebarria:2024jfv, Tian:2024dgl,  Cvetic:2024mtt, Bonetti:2024etn, Heckman:2025lmw, Bertolini:2025wyj}.

Consider the theory $\mathcal{T}_X$ engineered by $X=\text{Cone}(\partial X)$ in IIA/IIB string theory or M-theory. We begin by introducing the defect group $\mathbb{D}$ (see \cite{DelZotto:2015isa}) of extended operators of $\mathcal{T}_X$ constructed from branes wrapping non-compact cycles in $X$. Then, the generalized symmetries of interest will be characterized by their action on $\mathbb{D}$.

Let us begin with the case where $\partial X$ is smooth, i.e., we have an isolated singularity with no deformations.
The defect group of $\mathcal{T}_X$ is defined to be the finite abelian group \cite{DelZotto:2015isa, Morrison:2020ool, Albertini:2020mdx}\footnote{Here we are using the fact that $X$ is rigid, and in particular that $H_{k+1}(X, \partial X) / H_{k+1}(X) \cong H_{k}(\partial X; \Z)$.}
\begin{equation}\label{eq:DefectGroup}
\mathbb{D} \equiv  {\bigoplus_n}\, \mathbb{D}^{(n)}\quad \text{with} \quad \mathbb{D}^{(n)} \cong \underset{p\mathrm{\text{\:\!-}branes}}{\bigoplus}~ \underset{p - k = n}{\bigoplus} \mathrm{Tor}\,H_k(\partial X;\Z)\,,
\end{equation}
where $\mathbb{D}^{(n)}$ is the subgroup of $n$-dimensional defects in $\mathcal{T}_X$ that result from wrapping $p$\:\!-branes on cones over classes in $\mathrm{Tor}\,H_{p-n}(\partial X;\Z)$. The natural electromagnetic pairing between $p$\:\!-branes and the bilinear linking form on the torsional homology groups equip the defect group with a Dirac pairing.

Here we assumed that integral homology accurately characterizes brane charges in the context of defect constructions, which is, for example, accurate for the cases of $X=\mathbb{C}^2/\Gamma_{\text{ADE}}$ with finite $\Gamma_{\text{ADE}}\subset SU(2)$ and $X=\mathbb{C}^3/\Gamma$ with finite $\Gamma\subset SU(3)$ in M-theory.\footnote{The supersymmetric geometries here are simple enough, as checked through brane probe analysis \cite{Closset:2019juk, DelZotto:2022fnw}, that singular homology is a sufficient approximation to the homology theory characterizing configurations of M2- and M5-branes in the context of defect constructions. Then, all further conclusions basically derive from $H_1(S^5/\Gamma;\Z)\cong \text{Ab}(\Gamma/\Gamma_{\text{fix}})$ which is due to Armstrong's theorem \cite{Armstrong_1968}. See also  \cite{Albertini:2020mdx, Morrison:2020ool, DelZotto:2022fnw}.} In these two setups it is well-known that
\be\ba
X&=\mathbb{C}^2/\Gamma_{\text{ADE}}\,:\qquad&& \mathbb{D}=\mathbb{D}^{(1)}_{\text{elec}}\oplus \mathbb{D}^{(4)}_{\text{mag}} \cong \text{Ab}(\Gamma_{\text{ADE}})\oplus  \text{Ab}(\Gamma_{\text{ADE}})^\vee\,,\\[0.5em]
X&=\mathbb{C}^3/\Gamma\,:\qquad &&\mathbb{D}=\mathbb{D}^{(1)}_{\text{elec}}\oplus \mathbb{D}^{(2)}_{\text{mag}}\cong   \text{Ab}(\Gamma)\oplus  \text{Ab}(\Gamma)^\vee \,,
\ea \ee
where electric line operators are constructed by wrapping M2-branes on non-compact 2-cycles constructed as cones over asymptotic 1-cycles, and the electromagnetically dual defect operators are constructed from M5-branes constructed from cones over the asymptotic $k$-cycles which link with the 1-cycles.

When $\partial X$ is singular (and $X$ has non-compact singular loci), as can happen for $\mathbb{C}^3/\Gamma$, then the geometric characterization \eqref{eq:DefectGroup} relates more indirectly to the defect group of $\mathcal{T}_X$. For example, there are often cycles in $H_k(\partial X;\Z)$ for which all representatives stretch between the singular loci in $\partial X$, wrapping branes on such cycles may, by the same logic as in \eqref{eq:DefectGroup}, be interpreted as constructing defects for the degrees of freedom localized to the non-compact singular loci of $X$. Defects of $\mathcal{T}_X$ constructed from cones over these classes are then naturally localized to junctions or end points to these defects of the ambient systems. These structures can be disentangled, see \cite{Cvetic:2024dzu}.
For later use, we will simply record the result for $X=\mathbb{C}^3/\Gamma$ in M-theory \cite{DelZotto:2022fnw}. Denoting the subgroup of $\Gamma$ generated by elements with fixed points on $S^5\subset \mathbb{C}^3$ by $\Gamma_{\text{fix}}$ one finds for non-isolated Calabi-Yau singularities
\be\ba\label{eq:DefectGpNonAb}
X=\mathbb{C}^3/\Gamma\,:\qquad \mathbb{D}&= \mathbb{D}^{(1)}_{\text{elec}}\oplus \mathbb{D}^{(2)}_{\text{mag}}\oplus \dots \cong  \text{Ab}(\Gamma/\Gamma_{\text{fix}})\oplus  \text{Ab}(\Gamma/\Gamma_{\text{fix}})^\vee\oplus \dots\,.
\ea \ee

The generalized symmetries we will be interested in are 1-form symmetries acting on the line defects $\mathbb{D}^{(1)}_{\text{elec}}$. The topological operators realizing these actions may be constructed from M5-branes wrapped over the asymptotic $k$-cycles which link with the 1-cycles used in the construction of the line defects from M2-branes \cite{Heckman:2022muc}. This 1-form symmetry group will be denoted $A$ and is
\be\label{eq:1FS}
A\cong (\mathbb{D}^{(1)}_{\text{elec}})^\vee\cong  \text{Ab}(\Gamma/\Gamma_{\text{fix}})^\vee \,.
\ee

Let us spell out some further structures encountered for 5D SCFTs $\mathcal{T}_X$ engineered by $X=\mathbb{C}^3/\Gamma$ in M-theory. Notably, by equation \eqref{eq:1FS}, the order of the 1-form symmetry group $A$ decreases as $\Gamma_{\text{fix}}$ grows. However, notice that the quotient by $\Gamma_{\text{fix}}$ leads to a collection of non-compact ADE singularities in $\mathbb{C}^3/\Gamma$. Denoting the combined defect group of lines of this higher dimensional system (given by 7D SYM theories) by $ (\mathbb{D}^{(1)}_{\text{elec}})'$ it is a geometric fact \cite{Cvetic:2022imb}, given the geometric characterization of defects in \eqref{eq:DefectGroup}, that we have an extension of groups
\be\label{eq:SES2gp}
 1\rightarrow (\mathbb{D}^{(1)}_{\text{elec}})' \rightarrow \text{Ab}(\Gamma)\rightarrow \mathbb{D}^{(1)}_{\text{elec}}\rightarrow 1\,.
\ee
This realizes the simplest instance in which symmetry data, here in the form of defect groups, interplay across singular strata.\footnote{Many instances of this are of course well-studied in a field-theoretic context where a defect in an ambient QFT, supporting its own quantum system, is characterized by how bulk symmetry structures extend and restrict onto the defect, see, e.g., the recent works \cite{Copetti:2024onh, Cordova:2024iti, Antinucci:2024izg}.  } In limits decoupling the non-topological degrees of freedom of the higher dimensional system these structures assemble to a 2-group symmetry from the perspective of the 5D SCFT. The non-compact singularities contribute a 5D flavor symmetry with Lie algebra
\be
\mathfrak{f}\equiv \bigoplus_i \mathfrak{g}_{\text{ADE}}^{(i)} \,.
\ee
One then has the 2-group sequence \cite{Kapustin:2013uxa, Cordova:2018cvg, Benini:2018reh, Lee:2021crt, Cvetic:2022imb, DelZotto:2022joo}
\be\label{eq:4termseqOG}
1\rightarrow A\rightarrow \widetilde{A} \rightarrow \widetilde{F} \rightarrow F \rightarrow 1\,,
\ee
where $\widetilde{A}\cong \text{Ab}(\Gamma)^\vee$, ${A}\cong \text{Ab}(\Gamma/\Gamma_{\text{fix}})^\vee$ is the 1-form symmetry group, $\widetilde{F}=\times_i\, \widetilde{G}_{\text{ADE}}^{(i)}  $ is the simply connected group with Lie algebra $\mathfrak{f}$, in particular, $ \widetilde{G}_{\text{ADE}}^{(i)} $ is the simply connected ADE group with Lie algebra $\mathfrak{g}_{\text{ADE}}^{(i)}$, and $F$ is the flavor symmetry group of the 5D system. See \cite{Cvetic:2022imb} for further discussion.

\section{Interlude: Anomalies and $\eta\:\!$-Invariants}
\label{sec:eta}

One of the primary goals of the present work is to develop a practical computational scheme for extracting the anomalies of symmetries directly from the boundary geometry $\partial X$ of a singular conical geometry $X = \mathrm{Cone}(\partial X)$. In this section we first motivate this correspondence, and then proceed to review some more technical aspects of how this will work in practice. In subsequent sections we deploy this formalism in a number of settings.

As mentioned in the Introduction, the main idea is that the data of anomalies is expected to be encoded in the boundary $\partial X$ of extra-dimensional geometry. For Calabi-Yau geometries $X$ with an explicit resolution $\widetilde{X}$, one can in principle extract these anomaly coefficients by enumerating the non-compact cycles of the geometry. Wrapping branes on such divisors and quotienting by the states obtained from branes wrapped on compact cycles yields the defect group, and after taking a suitable polarization one can extract the generalized symmetries of the engineered QFT$_d$ directly from geometry. In principle, one can also extract all anomaly data by then considering intersections between these cycles. In addition to the practical and conceptual issues mentioned in the Introduction, there are also subtleties with reading off the corresponding anomalies directly from intersection-theoretic data due to screening effects. In addition, it is well-known that intersection numbers can be resolution dependent.

Faced with this situation, we shall seek a method for extracting anomalies which does not require us to blowup the singularities at the tip of the cone $X$. Along these lines, we shall now argue that $\eta$-invariants defined on $\partial X$ provide an alternative approach which bypasses these intersection-theoretic ambiguities and complications.

To frame the discussion to follow, recall that for a given resolution $\widetilde{X}$, there is a correspondence between line bundles and divisors. In particular, the first Chern class specifies an element of $H^{2}(\widetilde{X}; \mathbb{Z})$ which we can equally well view as specifying a cycle in $H_{n-2}(\widetilde{X},\partial \widetilde{X} ;\mathbb{Z})$, with $n = \mathrm{dim} \,X$. Indeed, intersection numbers can instead be tracked in terms of top forms constructed from products of Chern classes, integrated over $X$. For example, if $X$ is a threefold, we have a triple intersection of divisors which can instead be computed with an integral over quantized fluxes:
\begin{equation}
\label{eq:tripint}
D_i \cap D_j \cap D_k \sim \underset{X}{\int} F_i F_j F_k\,,
\end{equation}
in the obvious notation. 

Now, since $X$ is non-compact, the divisors $D_i$ may be non-compact and the above expressions are not necessarily integral. In addition, any two non-compact divisors may differ by compact classes, associated with dynamical excitations, such that they are, in an appropriate sense, grouped into screening equivalence classes by the QFT. The invariant data is
captured by $\eta$-invariants evaluated $\mathrm{mod} \, 1$, as will follow from the APS index theorem. 
One of our tasks will be to show how this correspondence works in detail, and in particular, to explain how we can wholly dispense with the bulk geometry $X$, focusing solely on $\partial X$ instead. 

Let us emphasize once more that although equation \eqref{eq:tripint} motivates the discussion to follow, the intrinsic data entering our computations deals directly with a singular $X$ and follows purely from $\partial X$. In the computations to follow this is reflected in the fact that we only need make reference to $Z$, namely the auxiliary space introduced so that the 11D Chern Simons terms remain well-defined, and of which only the boundary $\partial Z=\partial X$ is a subset of the 11D supergravity spacetime.   

We will be interested in $\eta$\:\!-invariants associated to $\partial X$ of the Dirac operator $\slashed{D}$ and the $\bar\partial$-operator, twisted by vector bundles. These operators are considered on $Z$ and induce tangential operators to the sphere quotient $\partial Z$. The former couples to vector bundles ${V}'$ over $ Z$, the latter couples only to the restriction of ${V}'$ to $\partial X$, denoted ${V}$. We denote the $\eta$\:\!-invariants by $\eta^{\text{\:\!$\slashed{D}$}}_{\raisebox{-0.3ex}{\scriptsize${V}$}},\eta^{\text{\:\!$\bar\partial$}}_{\raisebox{-0.3ex}{\scriptsize${V}$}}$ accordingly.

These $\eta$\:\!-invariants originate from index theoretic considerations. The two instances of the Atiyah-Patodi-Singer index theorem we will be interested in, when $Z$ is smooth, are \cite{Atiyah:1975jf, Atiyah:1976jg, Atiyah:1976qjr}
\be\ba\label{eq:IndexTheorems}
\text{Index}(\slashed{D},V',Z)&=\int_{Z} \text{ch}(V')\hat{\text{A}}_{Z} -\frac{1}{2}\Big(\eta^{\text{\:\!$\slashed{D}$}}_{\raisebox{-0.3ex}{\scriptsize${V}$}}(\partial Z)+h^{\text{\:\!$\slashed{D}$}}_{\raisebox{-0.3ex}{\scriptsize${V}$}}(\partial Z)\Big) \,,\\[0.5em]
\text{Index}(\bar\partial,V',Z)&=\int_{Z} \text{ch}(V'){\text{Td}}_{Z} -\frac{1}{2}\Big(\eta^{\text{\:\!$\bar\partial$}}_{\raisebox{-0.3ex}{\scriptsize${V}$}}(\partial Z)+h^{\text{\:\!$\bar\partial$}}_{\raisebox{-0.3ex}{\scriptsize${V}$}}(\partial Z)\Big)\,.
\ea \ee
Here the bulk contributions are given by the $\hat{\text{A}}$-roof genus $\hat{\text{A}}_{Z}$ and the Todd class ${\text{Td}}_{Z}$ of the bulk manifold $Z$ contracted with the Chern character $ \text{ch}(V')$ of the bundle $V'$. The boundary contributions are
\be\label{eq:etaANDh}\ba
\eta^{\text{\:\!$\slashed{D}$}}_{\raisebox{-0.3ex}{\scriptsize${V}$}}(\partial Z)&=\lim_{\epsilon\rightarrow 0^+}\sum_k \exp(-\epsilon |\lambda_k|)\text{\;\!sign}(\lambda_k)\,, \\[0.25em]
h^{\text{\:\!$\slashed{D}$}}_{\raisebox{-0.3ex}{\scriptsize${V}$}}(\partial Z)&=\text{dim\,Ker}(\slashed{D}_\partial,L)\,,
\ea \ee
where $\lambda_k$ are the eigenvalues of the induced tangential Dirac operator $\slashed{D}_\partial$ acting on Dirac fermions twisted by $V$. Similar expressions hold for the $\bar\partial$\:\!-operator. Further, these considerations generalize to the category of orbifolds \cite{Atiyah1966ALF, ojm/1200771835} for which the bulk integrals in \eqref{eq:IndexTheorems} receive additional contributions from the twisted sectors of the inertia stack of the orbifold.

To proceed then, and to set notation, we review the computation of $\eta$-invariants associated with twisted Dirac operators for arbitrary $S^{2n-1}/\Gamma$ with finite $\Gamma\subset U(n)$. Our discussion here follows Degeratu's work \cite{10.1093/qmath/han016}, which extends work of B\"ar \cite{Bar1996}, and is situated in the wider setting of orbifold index theory. In \cite{10.1093/qmath/han016} the $\eta$\:\!-invariants are computed directly from the eigenvalues and eigenspaces of the Dirac operator. There, the discussion begins with noting that $S^{2n-1}$ can be realized as the quotient $\widetilde{U}(n)/\widetilde{U}(n-1)$ where $\widetilde{U}(n)$ is a double cover of $U(n)$ with respect to the embedding $U(n)\rightarrow SO(2n)$ and projection $\text{Spin}(2n)\rightarrow SO(2n)$. The spin structures and therefore the Dirac operator on $S^{2n-1}$ are invariant under $\widetilde{U}(n)$ and therefore much of the problem can be reduced to representation theory of $\widetilde{U}(n)$. Key in these considerations is the half-determinant representation $D_{1/2}$ of $\widetilde{U}(n)$ which is the square root, with respect to the spin structure, of the determinant representation $\Lambda^n(\mathbb{C}^n)=D_{1/2}\otimes D_{1/2}$. Ultimately, following result is proved:\smallskip

\noindent {\bf Theorem (Degeratu):} Let $\Gamma$ be a finite subgroup of $U(n)$ admitting a lift $\tilde{\epsilon}:\Gamma \rightarrow \widetilde{U}(n)$ and $\chi$ be a representation of $\Gamma$. Then the $\eta$\:\!-invariant of the orbifold $S^{2n-1}\!/\Gamma$ twisted by the orbi-bundle $\mathscr{V}_\chi$  corresponding to $\chi$ is
\be\label{eq:Degeratu}
\frac{1}{2}\eta^{\text{\:\!$\slashed{D}$}}_{\raisebox{-0.3ex}{\scriptsize$\mathscr{V}_\chi$}}(S^{2n-1}/\Gamma)=\frac{(-1)^n}{|\Gamma|}\sum_{\gamma\;\! \in\;\! {\text{FF}}(\Gamma) } \frac{\text{Tr}_{D_{1/2}}( \tilde \epsilon(\gamma) )}{\text{det}(1-\gamma)} \chi(\gamma)\,,
\ee
which depends on the choice of $\tilde{\epsilon}$. Here FF$(\Gamma)$ denotes the subset of $\Gamma$ containing group elements $\gamma$ which act fixed point free on $S^{2n-1}$.\smallskip

Denoting the vector space underlying the representation by $R_\chi$ one has
\be
\mathscr{V}_\chi=(R_\chi \times S^{2n-1})/\Gamma\,.
\ee
Whenever the subgroup $\Gamma_{\text{fix}}\subset\Gamma$, generated by elements in the complement of FF$(\Gamma)$, act trivially on $R_\chi $ then the orbi-bundle $\mathscr{V}_\chi$ is a vector bundle, which we will denote as ${V}_\chi$, i.e., vector bundles are associated with a subring in $\text{Rep}(\Gamma)$.

With this, it only remains to evaluate the summands. For this, consider a $\gamma$ represented as a matrix in $U(n)$ which we may diagonalize as
\be
\gamma \sim \text{diag}(\exp(2\pi i w_1^{\gamma}/\text{ord}(\gamma)),\dots, \exp(2\pi i w_n^{\gamma}/\text{ord}(\gamma))\,,
\ee
where $\text{ord}(\gamma)$ is the order of $\gamma\in \Gamma$ and $0<w_k^\gamma<\text{ord}(\gamma)$ are non-zero positive integer weights. The trace then evaluates to
\be
\text{Tr}_{D_{1/2}}( \tilde \epsilon(\gamma) )=\epsilon(\gamma) \prod_{k=1}^n \exp\!\lb\frac{2\pi i w_k^{\gamma}}{2\text{\:\!ord}(\gamma)}\rb\,,
\ee
with the sign $\epsilon(\gamma)=\pm (-1)^{2s(\gamma)}$ where $s(\gamma)=\sum_{k=1}^n({w_k^{\gamma}}/{2\text{\:\!ord}(\gamma)})$. Then, overall we have
\be\label{eq:EtaInvNonAb}
\frac{\text{Tr}_{D_{1/2}}( \tilde \epsilon(\gamma) )}{\text{det}(1-\gamma)} =\pm(-1)^{2s(\gamma)}\prod_{k=1}^n \lbb \exp\!\lb-\frac{2\pi i w_{k_{}}^{\gamma}}{2\text{\:\!ord}(\gamma)^{}}\rb^{\!\,}-\exp\!\lb\frac{2\pi i w_{k_{}}^{\gamma}}{2\text{\:\!ord}(\gamma)^{}}\rb\rbb^{-1}\,.
\ee
The natural choice of sign $\epsilon$ or lift $\tilde{\epsilon}$ will be ``$+$", which is induced by $S^{2n-1}/\Gamma$ being the link of a singularity $\mathbb{C}^{2n}/\Gamma$ (or, said more physically, the spin structure with respect to which the fermionic degrees of freedom at the tip of the cone $\mathbb{C}^{2n}/\Gamma$ will be formulated).

Similar considerations for the $\bar\partial$-operator result, with analogous notation, in
\be\label{eq:Degeratu2}
\frac{1}{2}\eta^{\text{\:\!$\bar\partial$}}_{\raisebox{-0.3ex}{\scriptsize$\mathscr{V}_\chi$}}(S^{2n-1}/\Gamma)=\frac{(-1)^n}{|\Gamma|}\sum_{\gamma\;\! \in\;\! {\text{FF}}(\Gamma) } \frac{ \chi(\gamma)}{\text{det}(1-\gamma)}\,.
\ee

Having laid out the relevant machinery, we now put it to use.

\section{Illustrative Examples in 7D and 6D }
\label{sec:ADE}

In this section we present some illustrative examples which show how to read off anomaly data directly from $\eta$-invariants.
The examples we consider are based on supersymmetric string / M-theory backgrounds involving the ADE singularities $\mathbb{C}^{2} / \Gamma_{\mathrm{ADE}}$ for $\Gamma_{\mathrm{ADE}}$ a finite subgroup of $SU(2)$. M-theory on this background realizes 7D $\mathcal{N} = 1$ Super Yang-Mills theory with corresponding ADE gauge symmetry, and IIA on the same background realizes 6D $\mathcal{N} = (1,1)$ Super Yang-Mills theory with corresponding ADE gauge symmetry.\footnote{The case of type IIB on an ADE singularity instead results in a 6D $\mathcal{N} = (2,0)$ superconformal field theory, where the structure of higher-form symmetries is qualitatively different.} We also show how the same methods also apply in non-supersymmetric IIA backgrounds with orbifold singularities of the form $\mathbb{R}^{4} / \Gamma$ for $\Gamma$ a finite subgroup of $U(2)\subset \mathrm{Spin}(4)$ such that the closed string twisted sector contains a tachyon.

To present a uniform discussion, we focus on isolated singularities.\footnote{In the supersymmetric case there is no loss of generality, but in the non-supersymmetric setting one can in principle entertain lower codimension singularities. For a discussion of higher-form symmetries in this and related situations involving non-supersymmetric six-manifolds, see references \cite{Braeger:2024jcj, Braeger:2025rov}.} Even in this ``simplified'' setting we show that $\eta$-invariant techniques detects quadratic refinements of anomalies based on purely intersection-theoretic methods.

\subsection{Mixed Anomalies of 7D Super-Yang Mills Theory}
\label{sec:7D}

First, let us lay out the extra-dimensional construction of our system and how certain mixed anomalies of interest are computed from bottom up field-theoretic techniques. The latter serves both as a check when recomputing anomalies from the extra-dimensional geometry in the spirit of \eqref{eq:Mapping} and will clarify the physics of this computation.

To proceed then, let us consider M-theory on $X=\mathbb{C}^2/\Gamma_{\text{ADE}}$ with finite subgroup $ \Gamma_{\text{ADE}}\subset SU(2)$. The cone $X=\text{Cone}(S^{3}/\Gamma_{\text{ADE}})$ exhibits an isolated singularity at its tip and has a smooth link $\partial X=S^{3}/\Gamma_{\text{ADE}}$. To simplify the discussion further let us specialize to the A-type singularity with $\Gamma_{\text{ADE}}\cong \Z_N$. The theory $\mathcal{T}_X$ under consideration is then maximally supersymmetric 7D Yang-Mills theory with $\mathfrak{su}_N$ gauge algebra. This is seen, for example, in a IIA dual frame which consists of a stack of $N$ D6-branes with frozen center of mass mode.

This theory has a conserved current $*(j_3)\equiv (*j)_4$ proportional to the instanton density $\text{Tr}(F_2 F_2)$ which couples to a 3-form background field $C_3$ as
\be\label{eq:Coupling}
S_{\text{7D}}^{\:\!\text{inst}}= \frac{2\pi}{16\pi^3}  \int_{\text{7D}}C_{3\,} \text{Tr}( F_2 F_2)\,.
\ee
Following standard techniques \cite{Kapustin:2014gua, Gukov:2020btk}  we can now compute a mixed anomaly between the 2-form symmetry with background $C_3$ and  the 1-form center symmetry with background $\hat B_2$, valued in $U(1)$, which arises for the choice of gauge group $SU(N)$. For this one promotes the $SU(N)$ gauge theory to a $U(N)$ gauge theory and turning on a background of the center symmetry one has:\:\!\footnote{Note that here we have (following \cite{Kapustin:2014gua}) embedded the $\mathfrak{su}(N)$-valued gauge field $A_1$ into a $\mathfrak{u}(N)$-valued gauge field $\hat{A}_1$ as
\be
\hat A_1=A_1+\frac{1}{N}\text{Id}^N A_1'\,,
\ee
where $A'$ is a background $\mathfrak{u}_1$ gauge field that satisfies $B_2=dA_1'/N$. The 1-form symmetry
is then $(A_1,B_2)\rightarrow (A_1+\lambda_1,B_2+d\lambda_1)$. The second Chern class $c_2$ of the gauge bundle then evaluates according to
\be
c_2=\frac{1}{8\pi^2}\lbb \lb  \text{Tr}\, \hat F\rb^2-\text{Tr}\lb \hat F\wedge \hat F\rb \rbb=\frac{1}{8\pi^2}\lbb N^2B_2\wedge B_2-\text{Tr}\lb \hat F\wedge \hat F\rb \rbb\,.
\ee
}
\be\ba\label{eq:Coupling3}
S_{\text{7D}}^{\:\!\text{inst}}&\rightarrow \frac{2\pi}{16\pi^3}  \int_{\text{7D}}C_{3\,} \text{Tr}\!\lbb ( \hat F_2-\hat B_2\text{Id}^N)  (\hat F_2-\hat B_2\text{Id}^N)\rbb\\[0.25em] &=-\int_{\text{7D}} C_{3} c_2 + \alpha \int_{\text{7D}}   C_3  B_2 B_2  \,,
\ea\ee
where $\hat F_2$ is a $\mathfrak{u}_N$-valued field strength, $\text{Id}^N$ is an identity matrix, $(2\pi/N)B_2= \hat B_2$ has been renormalized such that periods are quantized in $\Z_N$, $c_2$ is the second Chern class of the gauge bundle with gauge algebra $\mathfrak{u}_N$, and the coefficient $\alpha$ is computed to be:
\be\label{eq:fieldtheory}
\alpha= -\frac{N-1}{2N} \,.
\ee

Our focus will be on the second term in the second line of \eqref{eq:Coupling3} with coefficient $\alpha$.\footnote{The first term in the second line is trivial mod 1, i.e., there is, as expected, no gauge anomaly. See \cite{Gukov:2020btk} for further discussion.} The corresponding SymTFT term in 8D is
\be\ba\label{eq:Coupling2}
\mathcal{S}_{\;\!\text{SymTFT}}^{(\text{anom})}&\supset \alpha \int_{\text{8D}}   G_4 B_2 B_2  \,,
\ea\ee
with $G_4=dC_3$ locally. There is actually also another relevant term proportional to $p_1B_2 B_2$ which comes from a coupling very similar to \eqref{eq:Coupling}. This term is made explicit for example through the Wess-Zumino terms of the D6-brane stack (which involves the $\hat{\text{A}}$-roof genus) in the IIA duality frame. Including this term \eqref{eq:Coupling2} becomes
\be\ba\label{eq:Coupling4}
\mathcal{S}_{\;\!\text{SymTFT}}^{(\text{anom})}&\supset \alpha \int_{\text{8D}}   H_4 B_2 B_2  \,,
\ea\ee
with $H_4$ as give in \eqref{eq:Shift}. For completeness, we also give the BF-sector accompanying \eqref{eq:Coupling4}. Given that $B_2$ describes a $\Z_N$ symmetry we have canonically
\be\label{eq:Coupling12}
\mathcal{S}_{\;\!\text{SymTFT}}^{(\text{BF,1})}= \frac{N}{2\pi} \int_{\text{8D}} \hat B_5 d\hat B_2= \frac{2\pi}{N} \int_{\text{8D}}  B_5 d B_2\,,
\ee
with $(2\pi/N) B_5=\hat B_5$ and as before $(2\pi/N) B_2=\hat B_2$. This sector is associated to the $\Z_N$ line and 4-surface defects given in \eqref{eq:DefectGpNonAb}. For the continuous symmetry associated to the $U(1)$-valued field strength $H_4$ we have canonically \cite{Brennan:2024fgj}
\be\label{eq:Coupling13}
\mathcal{S}_{\;\!\text{SymTFT}}^{(\text{BF,\:\!2})}= \frac{1}{2\pi} \int_{\text{8D}}  {h}_4 H_4\,,
\ee
with $\mathbb{R}$-valued gauge field ${h}_4$. Overall, the SymTFT subsector under consideration is then the sum of the terms displayed in \eqref{eq:Coupling4}, \eqref{eq:Coupling12} and \eqref{eq:Coupling13}, explicitly
\be
\mathcal{S}_{\;\!\text{SymTFT}}\supset \frac{N}{2\pi}\int_{\text{8D}} \hat B_5 d\hat B_2 +\frac{1}{2\pi} \int_{\text{8D}}  {h}_4 H_4+ \alpha \frac{N^2}{(2\pi)^2} \int_{\text{8D}}   H_4 \hat B_2 \hat B_2 \,,
\ee
where we have presented terms in a uniform normalization taking values in $U(1)$.

With this result and context established field-theoretically, our goal now is to derive $\alpha$ following the extra-dimensional approach outlined in \eqref{eq:Mapping}.

Indeed, straightforwardly following the reduction procedure \eqref{eq:model}, i.e., reducing \eqref{eq:ShiftedCS} on some smooth auxiliary space $Z$ with boundary $\partial Z=\partial X$ by expanding $H_4=B_2T_2+\dots$, one produces the term \eqref{eq:Coupling4} with following extra-dimensional expression for the anomaly
\be\label{eq:congruence}\ba
\alpha&=\frac{1}{2} \int_{Z}T_2 \lb T_2+w_2\rb\quad \text{mod}\,1\,.
\ea \ee
Here $T_2$ is a free class in $H^2(Z;\Z)$ which restricts on $\partial Z$ to a generator $t_2$ of $H^2(\partial Z;\Z)\cong \Z_N$ and $w_2$ denotes the second Stiefel-Whitney\footnote{Let us further comment on the role of $w_2$ in \eqref{eq:congruence}. Given any integral lattice, a vector $w$ such that
\be\label{eq:charvec}
v^2=w\cdot v\quad \text{mod}\,2\,,
\ee
for all lattice elements $v$ is referred to as a characteristic vector of that lattice.  There are infinite characteristic vectors and the difference of any two such vectors is divisible by 2. It turns out that any integral lift of the second Stiefel-Whitney class is a characteristic vector on $H^2(Z';\Z)$ where $Z'$ is closed 4-manifold. In this context, \eqref{eq:congruence} should be understood as the action of 3D Spin-Chern-Simons theory (recall we assume $\partial Z$ to be Spin in \eqref{eq:CS}) and the independence on the choice of bulk follows standard arguments by considering the difference of two bulk choices and employing \eqref{eq:charvec}. For further details see \cite{MooreTasi}. Dually, in homology, the second Stiefel-Whitney class is key for \eqref{eq:congruence} to define a quadratic refinement of the bilinear linking form on 1-cycles, which we will call a self-linking. The set of quadratic refinements of this fixed bilinear form is an affine torsor over $H^1(\partial Z;\Z_2)$, i.e., any two choices differ by an element of $H^1(\partial Z;\Z_2)$, exactly as the set of Spin structures which they are in correspondence with here. More explicitly, self-linkings of closed curves $\gamma_1$ depend on the choice of framing, i.e., a trivialization of the curves normal bundle, which are classified by $\Z\cong \pi_1(SO(2))$, i.e., the winding of the orthonormal frame along the curve. A naive ``division by 2" in the definition of the linking form suffers from an ambiguity, the corresponding self-linking computed with respect to two framings differing by $1\in \Z$ will differ by 1/2. The spin-structure on $\partial Z$ fixes this ambiguity \cite{CSKnotsFraming}, as $w_2=0$ in cohomology there is a closed 1-cochain $\xi_1$ with values in $\Z_2$ such that $d\xi_1=w_2$ as cochains. By adding an integral $\int_\gamma \xi_1$, which suffers from an identical ambiguity, to the naive definition, all changes in framining now lead to integer shifts. The quadratic refinements are now well-defined, but which quadratic refinement is realized still depends on the choice of spin structure on $\partial Z$. The spin structure chosen follows from the original 11D setup. Recall that $\partial X=\partial Z$ so the spin structure on $\partial Z$ arises as a restriction of that chosen to formulate the 11D fermion content.} class of $Z$.\footnote{The computation is standard and straightforward, following essentially steps outlined in \cite{Apruzzi:2021nmk}, and we note here the main trick, which is a formula by Wu \cite{Pont} relating characteristic classes modulo 4 as
\be\ba\label{eq:relation}
p_1&=\mathfrak{P}(w_2)-\mathfrak{B}(w_1w_2)-2(w_1\textnormal{Sq}^1(w_2)+w_0w_4) \\
&=\mathfrak{P}(w_2)+2w_4\,,
\ea\ee
where in the second line, given the setting of this work, we have specialized with $w_0=1$, $w_1=0$ to oriented manifolds. Notation is such that $w_i$ is the $i$-th Stiefel-Whitney class, $\mathfrak{P}$ is the Pontryagin square operation, $\mathfrak{B}$ the Bockstein homomorphism for the sequence $\Z\rightarrow\Z\rightarrow \Z_2$ and $\textnormal{Sq}^k$ is the $k$-th Steenrod square. The terms of \eqref{eq:congruence} quadratic and linear in $T_2$ originate from the terms cubic and quadratic in $H_4$ in \eqref{eq:ShiftedCS} respectively, with the latter computed using \eqref{eq:relation}. } Next, whenever $Z$ is complex, we can equivalently write
\be\label{eq:congruence2}\ba
\alpha&=\frac{1}{2} \int_{Z}T_2 \lb T_2+c_1\rb\quad \text{mod}\,1\,,
\ea \ee
where $c_1\equiv c_1(TZ)$ is the first Chern class of the tangent bundle of $Z$ which realizes an integral lift of the second Stiefel-Whitney class, i.e., $c_1=w_2$ mod 2. This expression should be recognized in the context of the Todd class
\be
\text{Td}_{Z} = 1+\frac{1}{2}c_1+\frac{1}{12}(c_1^2+c_2)\,,
\ee
which relates $\alpha$ to the index theory of $\bar{\partial}_{L'}$ which is the $\bar\partial$-operator twisted by some line bundle $L'$ on $Z$ with first Chern class $c_1(L')=T_2$. We will denote by $L=L'|_{\partial Z}$ the restriction of the bulk line bundle $L'$ to the boundary $\partial Z=\partial X$. The first Chern class of this line bundle is $c_1(L)=t_2$. Then, from the Atiyah-Patodi-Singer index theorem \eqref{eq:IndexTheorems} one has
\be\label{eq:IndexTheorem0}
\text{Index}(\bar{\partial},L',Z)=\int_{Z} \text{ch}(L')\text{Td}_{Z}-\frac{1}{2}(\etaDelBar{}(\partial Z)+\hDelBar{}(\partial Z)) \,,
\ee
where $\hDelBar{}$ is the dimension of the kernel the induced tangential $\bar{\partial}$-operator twisted by $L$ and $\etaDelBar{}$ is the Atiyah-Patodi-Singer $\eta$\:\!-invariant. For line bundles $L'$ one has
\be
\text{ch}(L')=\exp\!\lb {c_1} \rb=\exp\!\lb {T_2} \rb=1+T_2+T_2^2/2\,,
\ee
and therefore
\be\label{eq:IndexTheorem3}
\text{Index}(\bar{\partial},L',Z)=\frac{1}{2}\int_Z T_2(T_2+c_1)-\frac{1}{24}\int_Z  \lb \text{Tr}(\Omega_2)^2+\text{Tr}(\Omega_2\Omega_2)  \rb -\frac{1}{2}(\etaDelBar{}(\partial Z)+\hDelBar{}(\partial Z)) \,,
\ee
where $2\pi\Omega_2$ is the curvature 2-form. The first term is an integral lift of $\alpha$, the second term is independent of the line bundle $L'$ and that the last term depends only on boundary data. We can therefore remove the second term by subtracting the index of the untwisted $\bar\partial$-operator
\be\label{eq:IndexTheorem2}
\text{Index}(\bar{\partial},\varnothing,Z)=-\frac{1}{24}\int_Z  \lb \text{Tr}(\Omega_2)^2+\text{Tr}(\Omega_2\Omega_2)  \rb  -\frac{1}{2}(\etaDelBarEmpty(\partial Z)+\hDelBarEmpty{}(\partial Z)) \,,
\ee
and project back down onto $\alpha$ by considering expressions modulo 1. Here $\varnothing$ denotes that no further bundle enters via twisting. This results in\footnote{Here we have added an index `\:\!$L$\:\!' to $\alpha$ indicating its dependence on the line bundle $L$ which was implicit in \eqref{eq:congruence2} through the dependence on $T_2$. }
\be
\alpha_L=\frac{1}{2}(\etaDelBar{}(\partial Z)+\hDelBar{}(\partial Z)) -\frac{1}{2}(\etaDelBarEmpty{}(\partial Z)+\hDelBarEmpty{}(\partial Z)) \qquad \text{mod 1}\,.
\ee
Further, to reiterate the connection to \eqref{eq:Mapping}, note that $\partial X=\partial Z$ is physical, by considering equations modulo 1, as relevant to SymTFT discussions. That is, $\alpha_L$ is manifestly independent of the auxiliary dimension introduced back in line \eqref{eq:CS} used to make the Chern-Simons like coupling of M-theory well-defined. Now, the boundary $\partial Z=\partial X$ is positively curved and by Lichnerowicz-type vanishing theorems we have $\hDelBarEmpty{}, \hDelBar{}=0$.

Overall, our final result for expressing the mixed anomaly in \eqref{eq:Coupling4} via extra-dimensional structure depending purely on the boundary $\partial X$ is
\be\label{eq:neat}
\alpha_L=\frac{1}{2}\etaDelBar{}(\partial X) -\frac{1}{2}\etaDelBarEmpty{}(\partial X)\qquad \text{mod 1}\,.
\ee

This extra-dimensional approach should of course match the field-theoretic computation \eqref{eq:fieldtheory}. To proceed then, we recall that we explicitly have $\partial X=S^3/\Z_N$ and now turn to evaluate our extra-dimensional expression for this case. Given the lens spaces $S^3/\Z_N(r,s)$ equation \eqref{eq:Degeratu2} evaluates as\footnote{Here the notation $\partial X=S^3/\Z_N(r,s)$ denotes a quotient where $S^3$ is realized as the unit sphere in $\mathbb{C}^2$ with complex coordinates $z_1,z_2$ and then acted on by the $\Z_N$ action $(z_1,z_2)\mapsto (\omega^r z_1,\omega^s z_2)$ with $\omega=2\pi i/N$ and $\text{gcd}(N,r)=\text{gcd}(N,s)=1$.}
\be\label{eq:EtaLens}
\etaDelBar{}(S^3/\Z_N(r,s)) =\frac{2}{N}\sum_{k=1}^{N-1}\frac{1}{1-\omega^{k r }}\frac{1}{1-\omega^{k s }}\,\omega^{Q_Lk}\,, \qquad \omega=\exp(2\pi i/N)\,,
\ee
with $Q_L$ the charge of the line bundle $L$ and $r,s$ the weights of the $\Z_N$ action. To define the charge $Q_L$ of the complex line bundle $L$, recall that every complex line bundle $L$ on the lens space $\partial X=S^3/\Z_N$ is topologically equivalent to some
\be
L_Q=(R_Q\times S^3)/\Z_N\,,
\ee
where $R_Q$ is an irreducible complex representation of $\Z_N$ of charge $Q\in \Z_N$. Then, $Q_L\in \Z_N$ is simply defined as the charge of $R_Q \in \text{Rep}(\Z_N)\cong \Z_N$ of the equivalent $L_Q$. For example, the trivial line bundle has $Q=0$. More generally, the charge of a line bundle $L$ with first Chern class $t_2$ is determined via the canonical isomorphism
\be\label{eq:holos}
H^2(S^3/\Z_N;\Z )\cong \text{Hom}(H_1(S^3/\Z_N;\Z),U(1))\,,
\ee
which recasts the torsional first Chern class as a holonomy mapping and, by $H_1(S^3/\Z_N;\Z)\cong \Z_N$, as some representation $R_Q$ of $\Z_N$. With this, for A-type quotients with $(r,s)=(1,N-1)$ relevant to the initial 7D supersymmetric Yang-Mills theory one now readily computes
\be\label{eq:AnomalyViaEta}
\alpha_L=\frac{1}{N}\sum_{k=1}^{N-1}\frac{\omega^{k}-1}{(1-\omega^{-k})(1-\omega^{k}) }= -\frac{1}{N}\sum_{k=1}^{N-1}\frac{1}{1-\omega^{-k}} =-\frac{N-1}{2N}\,,
\ee
for line bundles $L$ of unit charge $Q_L=1$, matching the field theory result \eqref{eq:fieldtheory}. The choice of line bundle simply amounts to a choice of generator of $\Z_N$.

Let us make some further comments. First, the anomaly computation \eqref{eq:neat} should be understood in the context of the bilinear linking form $\ell:H^2(\partial X;\Z)\times H^2(\partial X;\Z)\rightarrow \Q/\Z$ on the boundary $\partial X$. The mapping
\be\label{eq:quadraticrefinement}
\mathscr{Q}\,:~H^2(\partial X;\Z)~\rightarrow~ \Q/\Z\,, \quad t_2=c_1(L)~\mapsto~ \frac{1}{2}\etaDelBar{}(\partial X) -\frac{1}{2}\etaDelBarEmpty{}(\partial X)\quad \text{mod}\,1\,,
\ee
is a quadratic refinement of the bilinear linking form, i.e.,
\be
\ell(t_2,t_2')\equiv \mathscr{Q}(t_2+t_2')-\mathscr{Q}(t_2)-\mathscr{Q}(t_2')\,.
\ee
See \cite{Hopkins:2002rd} for further discussion.

Second, the explicit value of the anomaly $\alpha_L\in \Q/\Z$ depends on extra-dimensional choices, i.e., the line bundle $L$. That is, we can consider alternative choices for $T_2$ entering \eqref{eq:congruence} and thereby different line bundles $L'$ with $c_1(L')=T_2$, e.g., by redefining $T_2\rightarrow mT_2$ for any $m$ with gcd$(m,N)=1$. Such a redefinition results in $\alpha\rightarrow m^2\alpha$.\footnote{This follows from the explicit formula \eqref{eq:EtaLens} (essentially, because $\partial Z$ is spin, the $w_2$ term in \eqref{eq:congruence} does not spoil quadratic scalings).} Equivalently, in line \eqref{eq:Coupling4} one can change the $\Z_N$ generator $B_2$ makes reference to via the redefinition $B_2\rightarrow m B_2$. Therefore, more invariantly, the anomaly should be viewed as given by the entire quadratic form $\mathscr{Q}$. From this perspective one can ask in what sense the anomaly is ``non-vanishing." One way to quantify this is by the image of $\mathscr{Q}$. Denote the size of the image by $|\text{Im}(\mathscr{Q})|$. A direct computation gives $|\text{Im}(\mathscr{Q})|/N\leq 1/2$ for any $ N$, i.e., reparametrizations fail to sweep out all possible values of the anomaly which is valued in $\Z_{N},\Z_{2N}$ for odd, even $N$ respectively. For example, when $N=5$ only $\alpha=2/5,3/5$ occur, i.e., $|\text{Im}(\mathscr{Q})|=2$. The anomaly is non-trivially non-zero. Further note, that when gcd$(m,N)\neq 1$ scalings map us into a subgroup. Then, whenever $2^3|N$ or $p^2|N$ with odd prime $p$, there exist subgroups for which the mixed anomaly $\alpha$ vanishes.

Third, other ADE singularities may of course be analyzed analogously. Let us sketch this for quotients by the D-type dicyclic group $D_{2n}$ with $8n$ elements and $2n+3$ conjugacy classes. Using the conjugacy classes we can count the Dynkin nodes and, with this parametrization, identify the engineered gauge algebra as $\mathfrak{so}(4n+4)$. For such singularities, instead of \eqref{eq:EtaLens}, one now starts from \eqref{eq:Degeratu2} which for the current setup takes the form
\be\label{eq:EtaLens2}
\etaDelBar{\rho}(\partial X) =\frac{2}{8n}\sum_{g\in D_{2n}}\frac{\chi_\rho(g)}{\text{det}(1-g)}\,, \qquad \rho=t,s,c,v\,.
\ee
The denominator is with respect to the 2D represention realized in the geometric engineering construction, i.e., the D-type singularity $\mathbb{C}^2/D_{2n}$ itself. The numerator features the character of the four 1D representations of $D_{2n}$ which are the trivial, spinor, co-spinor, vector representation abbreviated $t,s,c,v$ respectively. As before, the fields of the SymTFT follow from reduction of $H_4$ along the classes
\be
H^2(S^3/D_{2n};\Z)\cong \Z_2^{(s)}\oplus \Z_2^{(c)}\,, \qquad \Z_2^{(v)} \subset \Z_2^{(s)}\oplus \Z_2^{(c)}\,,
\ee
resulting in background fields $B_2^{(s,c,v)}$. Two of these fields are linearly independent. We now aim to compute the anomaly terms\footnote{Of course, instead of picking out a $\Z_2$ subgroup of the center symmetry, one could have treated $\Z_2^2$ center in one shot by starting with an irreducible representation of dimension 2 in \eqref{eq:EtaLens2}.}
\be\ba
 \alpha_{L_\sigma} \int_{\text{8D}}   H_{4\:\!}^{\,\!}B_2^{(\sigma)} B_2^{(\sigma)} \,, \qquad \sigma=s,c,v\,,
\ea\ee
which fix the mixed anomalies quadratic in the $B_2$'s uniquely. With
\be\ba
\frac{1}{2}\etaDelBarEmpty{}(\partial X)&=\frac{1}{8n}\lbb n+ \sum_{\ell=1}^{4n-1} \frac{1}{(1-e^{\pi i\ell/2n})(1-e^{-\pi i\ell/2n})}\rbb\,,
\ea \ee
and
\be\ba
\frac{1}{2}\etaDelBar{v}(\partial X) &=\frac{1}{8n}\lbb -n+ \sum_{\ell=1}^{4n-1}\frac{1}{(1-e^{\pi i\ell/2n})(1-e^{-\pi i\ell/2n})}\rbb\,,  \\[0.3em]
\frac{1}{2}\etaDelBar{s}(\partial X) &=\frac{1}{8n}\lbb \sum_{\ell=0}^{4n-1} \frac{i\exp(i\pi\ell )}{(1-e^{\pi i\ell/2n})(1-e^{-\pi i\ell/2n})} + \sum_{\ell=1}^{4n-1} \frac{1}{(1-e^{\pi i\ell/2n})(1-e^{-\pi i\ell/2n})}\rbb \,, \\[0.3em]
\frac{1}{2}\etaDelBar{c}(\partial X) &=\frac{1}{8n}\lbb \sum_{\ell=0}^{4n-1} \frac{-i\exp(i\pi\ell )}{(1-e^{\pi i\ell/2n})(1-e^{-\pi i\ell/2n})}  + \sum_{\ell=1}^{4n-1} \frac{1}{(1-e^{\pi i\ell/2n})(1-e^{-\pi i\ell/2n})}\rbb\,,
\ea \ee
we compute
\be
 \alpha_{L_\sigma} =\frac{1}{2}\etaDelBar{\sigma}(\partial X) -\frac{1}{2}\etaDelBarEmpty{}(\partial X)=\begin{cases} ~\;\!\frac{1}{2}\, \quad\: \qquad \qquad \sigma=v\\\frac{3-n}{4} ~~\text{mod}\,1\,, \quad  \sigma=s,c \end{cases}\,,
\ee
which agrees with intersection theoretic methods (which we mention momentarily in \eqref{eq:ncdivisor}).

Fourth, let us highlight some technical features we will encounter in other settings. Note that there are three distinct ways the geometric data is parametrized above. For the first two, which are with respect to the boundary $\partial X=\partial Z$, consider, e.g., \eqref{eq:EtaLens} and \eqref{eq:holos}. There we have either a cohomological or a representation theoretic presentation of the line bundle $L$. The former is preferred by integrated index densities while the latter is preferred by explicit expressions of $\eta$\:\!-invariants, which are representation theoretic. The third parametrization that has been in use in the literature is in contrast homological. Although not explicitly discussed by us here one can also compute the bulk integral \eqref{eq:congruence2} via the intersection ring of $Z$. For example, when $Z$ is explicitly chosen to be a crepant resolution of the singularity $\mathbb{C}^2/\Z_N$, with $N-1$ exceptional curves $E_k$, then $T_2$ can be dualized via the intersection pairing to
\be\label{eq:ncdivisor}
T^{\:\!2} \equiv \frac{1}{N}\sum_{k=1}^{N-1}k E_k\,,
\ee
see \cite{Hubner:2022kxr} for further details. With respect to this homological parametrization and the anomaly is computed by $T^{\:\!2} \cdot T^{\:\!2} /2$ modulo 1. This should be thought of as specifying the line bundle $L'$ via a divisor $T^2$ of $Z$. When matching and checking different approaches, in the general case, we will have to carefully translate between these different presentations of data.

Finally, note that there is yet another way to arrive at \eqref{eq:AnomalyViaEta}. As beautifully explained in \cite{EGUCHI1980213} the terms entering index theorems, such as in \eqref{eq:IndexTheorem0}, are not invariant upon deforming $Z$, individual contributions to the index can be traded off between terms. In particular, one could have picked $Z=X=\mathbb{C}^2/\Gamma_{\text{ADE}}$, i.e., a singular $Z$. Away from the tip of the cone both the metric and bundle curvature are flat, the integrated index densities have now been localized to the tip of the cone and can be evaluated using fixed point methods \cite{Atiyah1966ALF}. This fits into a wider programme of orbifold index theory developed for example, in part, in \cite{ojm/1200771835, nmj/1118786571}.

\subsection{Mixed Anomalies of Non-SUSY 6D String Backgrounds}

Our discussion so far has been topological. In section \ref{sec:7D} we considered supersymmetric ADE singularities in order to readily check our extra-dimensional approach against field-theoretic approaches, but the holomorphicity underlying these examples was never utilized. Overall, the general expectation is that extra-dimensional approaches to symmetry theories apply to non-supersymmetric and supersymmetric backgrounds alike, as for example demonstrated in \cite{Braeger:2024jcj, Braeger:2025rov} for a variety of settings.

Here, we will briefly consider IIA string theory on $\mathbb{R}^4/\Gamma$ with $\Gamma$ some finite subgroup of $U(2)\subset \text{Spin}(4)$.\footnote{Due to fermionic degrees of freedom in the system the background is specified by a subgroup $\Gamma\subset  \text{Spin}(4)$. We shall assume that the group action has a fixed point at the tip of the cone, but that the system contains no additional singularities.\footnote{See \cite{Braeger:2024jcj, Braeger:2025rov} for discussion on these more general cases.} The metric is only acted on by the pushforward subgroup in $SO(4)$ which is the isometry group of $S^3$. It is the latter which is relevant for the anomaly considerations presented here.} The closed string spectrum of the 6D system engineered by such a background has twisted sector tachyons \cite{Adams:2001sv, Harvey:2001wm, Vafa:2001ra, Morrison:2004fr} and is unstable against their condensation. However, as these localize to fixed loci, this condensation turns out to be neatly characterized by desingularizations of  $\mathbb{R}^4/\Gamma$, i.e., one has a rolling solution with $X\big|_{t=t_0}=\mathbb{R}^4/\Gamma$ as initial condition at some initial time $t_0$. Even with such an instability it is a well-posed problem to discuss defect groups and generalized symmetries of such systems, and their anomalies \cite{Braeger:2024jcj}. Further, the geometry remains reliable in parametrizing these structures, even when microscopically they are now tied to fermionic data \cite{Braeger:2024jcj, Braeger:2025rov}. Then, the discussion of the previous section straightforwardly carries over, the 6D system has a mixed anomaly between a discrete and continuous 1-form symmetry
\be\ba\label{eq:CouplingNonSUSY}
\mathcal{S}_{\;\!\text{SymTFT}}^{(\text{anom})}&\supset \alpha_L \int_{\text{7D}}   G_3 B_2 B_2  \,, \qquad \alpha_L=\frac{1}{2}\etaDelBar{}(\partial X) -\frac{1}{2}\etaDelBarEmpty{}(\partial X)\quad \text{mod 1}\,,
\ea\ee
with field strength $G_3=dC_2$ where $C_2$ is the background field for the continuous 1-form symmetry and $B_2$ is the background field for the discrete 1-form symmetry. For example, the space $\mathbb{C}^2/\Z_N(1,1)$ has $\alpha_L=(1-N)/2N$ and $\mathbb{C}^2/\Z_{3N+1}(1,3)$ has $\alpha_L=-3N/(6N+2)$ which we both evaluated via \eqref{eq:EtaLens} with respect to a line bundle $L$ of charge $Q_L=1$.

\section{Anomalies of 5D SCFTs}
\label{sec:5D}

We now turn to study the anomalies of 1-form symmetries of 5D SCFTs engineered by Calabi-Yau orbifolds $X=\mathbb{C}^3/\Gamma$ in M-theory. We separate these geometries into three classes:

\begin{itemize}
\item Case 1: Isolated singularities $\mathbb{C}^3/\Z_N$.

\item Case 2: Non-isolated singularities $\mathbb{C}^3/\Z_N$.

\item Case 3: All other cases $\mathbb{C}^3/\Gamma$, i.e., non-cyclic $\Gamma$.

\end{itemize}

The cases are such that in case 1 we can focus on generalizing our discussion in section \ref{sec:ADE} with minimally added technical complications. From there, we then focus in case 2 on the discussion of symmetry data in stratified systems, as sketched in section \ref{sec:stratified}. Case 3 then serves to demonstrate that methods straightforwardly extend to non-cyclic quotients and non-toric geometries.

\subsection{Isolated Orbifold Singularities ($\Gamma\cong \Z_N$)}
\label{sec:IsolatedCY3}

We now consider 5D SCFTs engineered by the Calabi-Yau orbifold singularities $X=\mathbb{C}^3/\Gamma$ with finite $\Gamma\subset SU(3)$ in M-theory. Demanding an isolated singularity already determines $\Gamma\cong \Z_N$ with odd $N$.\footnote{For an isolated singularity $\Gamma$ must be cyclic. Consider $\Gamma\cong \Z_N$ and assume that $N$ were even, then all weights of the group action would have to be odd for an isolated singularity, however in this case the sum of the three weights would be odd, in particular, the sum does not vanish modulo $N$ and $X$ would not be Calabi-Yau, violating the assumption $\Gamma\subset SU(3)$.} The defect group \eqref{eq:DefectGpNonAb} is then simply
\be
\mathbb{D}=\mathbb{D}^{(1)}_{\text{elec}}\oplus \mathbb{D}^{(2)}_{\text{mag}}\cong \Z_N\oplus \Z_N\,,
\ee
and the line operators $\mathbb{D}^{(1)}_{\text{elec}}$ are acted on by a 1-form symmetry $A\cong \Z_N$ with background field $B_2$. Focusing purely on this 1-form symmetry we observe that there may be two types of anomalies as recorded by the SymTFT couplings
 \be\label{eq:Anom}\ba
\mathcal{S}_{\;\!\text{SymTFT}}^{(\text{anom})}&= 2\pi  \int_{\text{6D}}  \lb \beta B_2^3+ \gamma B_{2\:\!}  (p_1/4)  \rb+\dots \qquad \text{mod 1}\,,
\ea\ee
where $p_1$ is the first Pontryagin class of the SymTFT spacetime which we will assume to be divisible by 4. With this the couplings $\beta,\gamma\in \Q/\Z$ quantify a 1-form self-anomaly and a mixed 1-form-gravitational anomaly.

The action \eqref{eq:Anom} is considered mod 1 and consequently $\beta, \gamma$ are physical only up to certain redundancies. These redundancies are quantified by the index theory of the Dirac operator defined on the 6D SymTFT spacetime $M$. Whenever $B_2$ is compactly supported (allowing us to neglect boundary contributions), and whenever we can view $B_2$ as a 2-cocycle, we have by the Atiyah-Singer index theorem of the Dirac operator, twisted by a line bundle with first Chern class $B_2$, the identity
\be\label{eq:Dirac}
0=\frac{1}{6} \int_{M} \lb B_2^3-\frac{1}{4} B_2p_1\rb \qquad \text{mod}\,1\,.
\ee
With this $\beta,\gamma$ are not unique in \eqref{eq:Anom}. The pair $(\beta,\gamma)$ suffers from a $\Z_6$ ambiguity.\footnote{Let us contrast this to two related $\Z_6$ ambiguities. Consider for example the $C_3G_4G_4$ term. Reducing it naively in singular cohomology on the link $\partial Z$ one finds a linking number ``multiplied by 1/6." Division by $6$ is ill-defined over $\Q/\Z$ and thus a $\Z_6$ ambiguity arises. Identical considerations apply to $\beta$. The point is that \eqref{eq:Dirac} renders a combination of these ambiguities non-physical. The remaining of these naive ambiguities is removed by having included an auxiliary bulk in \eqref{eq:CS} for well-definition of the 11D Chern-Simons term.} This ambiguity is such that the combination
\be
\alpha=\beta+\gamma\,,
\ee
remains fixed. Importantly, note that $\alpha$ does not set any physical anomaly itself as $\beta,\gamma$ transform differently under redefinitions of $B_2$, and consequently $\beta+\gamma$ is not appropriately covariant.\footnote{For example, for the 5D SCFT engineered by the orbifold $\mathbb{C}^3/\Z_5(1,1,3)$ the combination $\alpha$ takes, for the four different choices of generator of $\Z_5$, the values
\be
\alpha=0,-\frac{1}{5},+\frac{1}{5},0\,.
\ee In contrast, anomalies being non-vanishing should not depend on generator choices.}

Nonetheless $\alpha$ has a favorable extra-dimensional presentation with respect to $\partial X$, much as in \eqref{eq:neat},
allowing for closed form evaluation, which also results in closed form expressions
for $\beta$ and $\gamma$ up to their common $\Z_6$ ambiguity.

We now discuss this boundary presentation following similar steps as in section \ref{sec:ADE}. To begin, let $Z$ be some smooth Spin manifold with boundary $\partial Z=S^5/\Z_N$ and $T_2\in H^2(Z;\Z)$ a free class that restricts on the boundary $S^5/\Z_N$ to a generator $t_2\in H^2(S^5/\Z_N;\Z)\cong \Z_N$. Then, essentially as in reference \cite{Apruzzi:2021nmk}, expanding \eqref{eq:CS} with $G_4=B_2T_2+\dots$ and matching to \eqref{eq:Anom} one finds 
\be\label{eq:bulkint}
\alpha=\int_{Z}\lb \frac{1}{6}T_2^3-\frac{1}{24}T_2 p_1\!\rb \quad \text{mod}\,1\,.
\ee
This expression \eqref{eq:bulkint} for $\alpha$ should be recognized in the context of the $\hat{\text{A}}$-genus
\be
\hat{\text{\:\!A}}_{Z} = 1-\frac{1}{24}p_1\,,
\ee
which relates $\alpha$ to the index of the Dirac operator $\slashed{D}$ twisted by some line bundle $L'$ over $Z$ with first Chern class $c_1(L')=T_2$. This is made precise by the Atiyah-Patodi-Singer index theorem
\be\label{eq:IndexTheoremZ}
\text{Index}(\slashed{D},L',Z)=\int_{Z} \text{ch}(L')\hat{\text{A}}_{Z} -\frac{1}{2}\Big(\etaDirac{}(\partial Z)+\hDirac{}(\partial Z)\Big) \,.
\ee
We have the Chern character ch$(L')=\exp( {T_2} )$ and that $L\equiv L'|_{\partial Z}$. The boundary terms are set by, respectively, the Atiyah-Patodi-Singer $\eta$-invariant and the dimension of the kernel of the induced twisted tangential Dirac operator given in \eqref{eq:etaANDh}.

One now obtains a closed form expression for $\alpha$ as a function of $\partial Z=S^5/\Z_N(m_1,m_2,m_3)$.\footnote{Here $S^5/\Z_N(m_1,m_2,m_3)$ denotes a $\Z_N$ action with positive integer weights $m_i$ presented as $0\leq m_i <N$. The action is $(z_1,z_2,z_3)\mapsto (\omega^{m_1}z_1,\omega^{m_2} z_2, \omega^{m_3}z_3)$ on the three complex coordinates of a $\mathbb{C}^3$ with respect to which $S^5$ is thought of as the unit sphere. For isolated CY singularities we have odd $N$ and $\text{gcd}(m_k,N)=1$.} Note that $\partial Z$ has positive sectional curvature implying $\text{dim\,Ker}(\slashed{D}_\partial,L)=0$ by Lichnerowicz-type vanishing theorems. Therefore, considering equations modulo 1 and evaluating \eqref{eq:Degeratu} we have
\be\label{eq:alphaDirac}
\alpha_L=\frac{1}{2}\eta^{\text{\:\!$\slashed{D}$}}_{\raisebox{-0.3ex}{\scriptsize$L$}}=\frac{1}{N}\sum_{k=1}^{N-1}(-1)^k\omega^{k\:\!Q_L}\prod_{i=1}^3\frac{1}{\omega^{-km_i/2}-\omega^{km_i/2}}\quad \text{mod 1}\,,
\ee
with primitive root of unity $\omega=\exp(2\pi i/N)$.\footnote{This expression is trivially a difference of two $\eta$-invariants as $\etaDiracEmpty=0$ due to the untwisted Dirac operator having non-trivial index theory only in dimension 0 mod 4.} Here, exactly as in \eqref{eq:neat}, we have made the dependence on the line bundle $L$ explicit via adding an index `$L$' to $\alpha$. Again, $Q_L$ denotes the charge of the line bundle: Given some 1D irreducible representation $R_Q$ of $\Z_N$ of charge $Q$ we can construct a complex line bundle $L_Q=(R_Q\times S^5)/\Z_N$ and any complex line bundle, as classified by its first Chern class, has a representative given by some $L_Q$.

When evaluating \eqref{eq:IndexTheoremZ} we must, in order to consistently evaluate the bulk and boundary terms with respect to the same basis, either determine the first Chern class $c_1(L_Q)$ or alternatively determine the charge $Q_L$ of the line bundle $L$ with first Chern class $c_1(L)=t_2$. We achieve this by expressing both of these parametrizations of line bundles with respect to a preferred basis of the cohomology ring of $S^5/\Z_N$, which in particular contains the group $H^2(S^5/\Z_N;\Z)$, as put forward by Kawasaki \cite{Kawasaki1973CohomologyOT}.

To proceed then, we first review Kawasaki's parametrization. The cohomology ring of $S^5/\Z_N(m_1,m_2,m_3)$ may be related to the cohomology ring of $\mathbb{CP}^2$ following \cite{Kawasaki1973CohomologyOT}. This relation establishes a set of preferred generators for the groups $H^{2k}(S^5/\Z_N;\Z)\cong \Z_N$ corresponding to the generators of the free groups $H^{2k}(\mathbb{CP}^2;\Z)\cong \Z$ for $k=1,2$. Let us denote the former by $v_2,w_4$ and the latter by $u_2,u_2^2$ respectively. Then
\be\label{eq:Kawasaki0}
v_2= q u_2\,, \qquad w_4 = p u_2^2\,,
\ee
together with $Nv_2=Nw_4=0$ and integers $q,p\in\Z$ defined by
\be\label{eq:Kawasaki1}\ba
q&=\text{lcm}\lbbb \frac{m_{i_1}m_{i_2}}{\text{gcd}(m_{i_1},m_{i_2})}\,\Big|\, 1 \leq i_1<i_2 \leq 3\rbbb=\text{lcm}\lbbb  \frac{m_{1}m_{2}}{\text{gcd}(m_{1},m_{2})},  \frac{m_{2}m_{3}}{\text{gcd}(m_{2},m_{3})},  \frac{m_{1}m_{3}}{\text{gcd}(m_{1},m_{3})} \rbbb \\[0.5em]
p&=\text{lcm}\lbbb \frac{m_{i_1}m_{i_2}m_{i_3}}{\text{gcd}(m_{i_1},m_{i_2},m_{i_3})}\,\Big|\, 1 \leq i_1<i_2<i_3 \leq 3\rbbb=  \frac{m_{1}m_{2}m_3}{\text{gcd}(m_{1},m_{2},m_3)}=m_1m_2m_3 ~\,.
\ea \ee
This determines the ring structure of cohomology ring. However, this parametrization is also favorable in expressing the link pairing $\ell: H^2(S^5/\Z_N;\Z)\times H^4(S^5/\Z_N;\Z)\rightarrow \Q/\Z$ with respect to the cohomology ring of $\mathbb{CP}^2$. For this one defines the integer
\be\label{eq:Kawasaki2}
r=\text{lcm}\lbbb \frac{m_{i_1}m_{i_2}m_{i_3}m_{i_4}}{\text{gcd}(m_{i_1},m_{i_2},m_{i_3})}\,\Big|\, 1 \leq i_1<i_2<i_3<i_4 \leq 4\rbbb= m_1m_2m_3N\,,
\ee
with $m_4= N$ and introduces $x_6=ru_2^3$. Then, one has
\be\label{eq:LinkingCharge}
\ell(v_2,w_4)=\frac{qp}{r}=\frac{q}{N}\qquad \text{mod}\,1\,,
\ee
which should be viewed as formally cupping $v_2$ with $w_4$ mod $x_6$. Identifying the link pairing with the holonomy mapping (viewing $v_2$ as a field strength and $w_4$ as dual to a curve) the isomorphism
\be\label{eq:HoloMap}
H^2(S^5/\Z_N)\cong \text{Hom}(H_1(S^5/\Z_N),U(1))\cong  \text{Hom}(\Z_N,U(1))\,,
\ee
is seen to assign to a line bundle $L$ on $S^5/\Z_N$ with first Chern class $c_1(L)=v_2$ the representation of charge $Q(v_2)=Q_L=q\in \Z_N$. We will denote this preferred line bundle therefore as $L_{q}$ and fix it as a generator of the Picard group of $S^5/\Z_N$.

Finally, there is again another way to parametrize line bundles $L$ on $S^5/\Z_N$ by appealing to the isomorphism $H^2(S^5/\Z_N;\Z)\cong H_3(S^5/\Z_N;\Z)$ and by thinking of $S^5/\Z_N$ as the boundary to $\mathbb{C}^3/\Z_N$. The line bundle $L$ can be thought of as a restriction of a line bundle $L'$ on $\mathbb{C}^3/\Z_N$ (or its toric resolution if one wants to work in the smooth category) to the boundary  $S^5/\Z_N$. The bulk complex line bundle may then be characterized by a bulk divisor. From this perspective, line bundles which are non-trivial on the boundary $S^5/\Z_N$ are characterized by some non-compact divisor of $\mathbb{C}^3/\Z_N$ modulo compact divisors. There are three natural non-compact divisors obtained by setting one of the three complex coordinates to zero. Let us consider $D_{z_3}\equiv\{ z_3=0\}\subset \mathbb{C}^3/\Z_N$.\footnote{The choices $D_{z_1},D_{z_2}$ are equally good, however, we have parametrized the examples written out explicitly in this note such that formulae are especially simple for the choice $D_{z_3}$.} This divisor meets the boundary in a 3-cycle. From this perspective there exists yet another integer $s_3$ such that $s_3D_{z_3}$, restricted to the boundary, is dual to the Kawasaki generator $v_2\in H^2(S^5/\Z_N;\Z)\cong \Z_N$. For example, this integer is such that $s_3D_{z_3}$ has a triple intersection mod 1 given by $q^3/r$ which is $v_2^3$ mod $x_6$.

\begin{table}[]
{\renewcommand*{\arraystretch}{1.35}

\hspace{-13pt}\begin{tabular}{||c ||   c| c || c| c || c| c ||}
 \hline
 $\Z_N(m_1,m_2,m_3)$ & $\raisebox{0.05ex}{$\frac{1}{2}\eta^{\text{\:\!$\slashed{D}$}}_{\raisebox{-0.3ex}{\scriptsize$L_q$}}\!(S^5\!/\:\!\Z_N)$}$ & $q$ &  $\beta_{L_q}$ & $\gamma_{L_q}$ & $s_3$ & $\text{Ind}(\slashed{D})_{s_3D_{z_3}}$  \\ [0.5ex]
 \hline\hline
 $\frac{1}{2n+3}(1,1,2n+1)$ & $ \frac{(n+1)(n-1)}{3(2n+3)}$  & $-2$  & $-\frac{2n-1}{6(2n+3)}$ & $\frac{2n^2+2n-3}{6(2n+3)}$  & 2 & 0 \\[0.5ex]
 \hline

  $\frac{1}{12n+5}(1,3,12n+1)$ & $ \frac{(2n-1)(6n-10)}{12n+5}$  & $-12$  & $\frac{24}{12n+5}$ & $\frac{12n^2-26n-14}{12n+5}$  & 4 & $-20$  \\[0.5ex]
 \hline
 $\frac{1}{30n+7}(1,5,30n+1)$ & $ \frac{(5n-9)(15n-11)}{30n+7}$  & $-30$  & $-\frac{5(30n-53)}{2(30n+7)}$ & $\frac{150n^2-230n-67}{2(30n+7)}$ & 6 & $-140$  \\[0.5ex]
       \hline
        $\frac{1}{56n+9}(1,7,56n+1)$ & $ \frac{4(7n-13)(28n-23)}{3(56n+9)} $  & $ -56$ & $-\frac{14(56 n-103)}{3(56 n+9)}$  & $\frac{2 (392 n^2-658 n-123)}{3 (56 n+9)}$ & 8 & $-504$  \\[0.5ex]
       \hline
\end{tabular}
}
\caption{List of $\eta$-invariants of $S^5/\Z_N$ associated to the Dirac operator $\protect\slashed{D}_{L_q'}$ on $\mathbb{C}^3/\Z_N(m_1,m_2,m_3)$ twisted by a line bundle $L_q'$. The line bundle $L_q'$ is not unique, however it uniquely restricts to the boundary $S^5/\Z_N(m_1,m_2,m_3)$ realizing a line bundle $L_q$ with charge $Q_{L_q}=q\in \Z_N$ with reference to Kawasaki's preferred generator, i.e., simultaneously $c_1(L_q)=v_2\in H^2(S^5/\Z_N)$ and choice of generator $v_2\equiv 1\in \Z_N$. Further, classes in $H^2(S^5/\Z_N)$ are Poincar\'e dual to 3-cycles $H_3(S^5/\Z_N)\cong \Z_N$. The equation $z_3=0$ cuts out a non-compact divisor of $\mathbb{C}^3/\Z_N(m_1,m_2,m_3)$ denoted $D_{z_3}$ which intersects the boundary $S^5/\Z_N$ in a 3-cycle. The column labelled $s_3$ specifies this 3-cycle, as an element $s_3^{-1} \in \Z_N$, with respect to Kawasaki's generator. The divisor $s_3D_{z_3}$ is one good choice for $L_q'$. For example, in the toric resolution of $\mathbb{C}^3/\Z_N$ the triple intersection of $s_3D_{z_3}$ modulo 1 and the linking of $v_2^2$ with $v_2$ in $S^5/\Z_N$ agree. We also give the Dirac index which, on the other hand, depends on bulk data of $L_q'$ here fixed by $s_3D_{z_3}$ (and computed with respect to a particular toric resolution as specified by the Delaunay Triangulation of the toric diagram, see figure \ref{fig:toricDs} for the diagrams of the first family). The anomalies $\beta,\gamma$ are the pure 1-form anomaly and the mixed 1-form-gravitational anomaly. }
\label{tab:TableForIsolated}
\end{table}

In table \ref{tab:TableForIsolated} we list some 1-parameter families of toric isolated Calabi-Yau 3-fold singularities and compute their combination $\alpha$ with respect to the Kawasaki choice of generator and further give the integers $q,s_3$ (mod $N$) relating these to other parameterizations picking out their own preferred generators of $\Z_N$.\footnote{These 1-parameter families of examples combine into one 2-parameter family of examples
\be
X=\mathbb{C}^3/\Z_{k(k+1)n+k+2}(1,k,k(k+1)n+1)\,,
\ee
with $k$ odd (table \ref{tab:TableForIsolated} gives $k=1,3,5,7$) and $q=-k(k+1)$ and $s_3=k+1$ and
\be\label{eq:unneccessary}
\frac{1}{2}\eta^{\text{\:\!$\slashed{D}$}}_{\raisebox{-0.4ex}{\scriptsize$L_q$}}\!(S^5\!/\:\!\Z_{k(k+1)n+k+2})=\frac{\mu n^2-\nu n+\rho}{k(k+1)n+k+2}\,,
\ee
where $\mu,\nu,\rho$ are order 4 polynomials in $k$ given by
\be\ba
\mu&=  \frac{1}{3} (1-2 k)^2 k^2\,,\\[0.25em]
\nu&= \frac{1}{3} (k-1) k (2 k-1) (6 k+1)\,, \\[0.25em]
\rho&=\frac{1}{3} k \left(8 k^3-14 k^2+2 k+3\right)\,. \\[0.25em]
\ea\ee.}

Let us comment on our computations. First, we checked these against the bulk intersection computations directly evaluating \eqref{eq:bulkint} for the choice $Z$ being the toric resolution of $\mathbb{C}^3/\Z_N$. Indeed, this is how we determined the integer $s_3$. See appendix \ref{sec:Check1} for an example of such a computation. Second, note that the $\eta$\:\!-invariant, for the families of geometries parametrized by $n$, computes for any of our examples to the form
\be
\frac{1}{2}\eta^{\text{\:\!$\slashed{D}$}}_{\raisebox{-0.3ex}{\scriptsize$L_q$}}(\partial X_{(n)})=\frac{P^{(q)}_2(n)}{N(n)}\,,
\ee
with a degree two polynomial $P_2^{(q)}$ whose 3 coefficients are functions of the 3 weights $m_k$. This structure makes it straightforward to extrapolate and determine closed form expressions for multi-family classes of examples, using code, such as in \eqref{eq:unneccessary}. Third, except for the twisted Dirac index, which depends on explicit choices for the bulk $Z$ and extension data of the boundary line bundle, the results in table \ref{tab:TableForIsolated} only depend on the boundary and bundle $L$.

Finally, let us give closed form boundary expressions for the 1-form self-anomaly and mixed 1-form gravitational anomaly. From $\frac{1}{2}\eta^{\text{\:\!$\slashed{D}$}}_{\raisebox{-0.3ex}{\scriptsize$L_Q$}},\frac{1}{2}\eta^{\text{\:\!$\slashed{D}$}}_{\raisebox{-0.3ex}{\scriptsize$L_{2Q}$}}$ we obtain a linear system for $\beta_{L_Q},\gamma_{L_Q}$. It can be solved up to the unphysical $\Z_6$ ambiguity giving the closed from expressions\;\!\footnote{See appendix \ref{sec:Check1} for further details of this unphysical ambiguity and how it enters expected homogeneity properties of the anomalies as a function of $Q$.}
\be\label{eq:betagamma}\ba
\beta_{L_Q}&=+\frac{1}{12}\eta^{\text{\:\!$\slashed{D}$}}_{\raisebox{-0.3ex}{\scriptsize$L_{2Q}$}}(\partial X) -\frac{1}{6} \eta^{\text{\:\!$\slashed{D}$}}_{\raisebox{-0.3ex}{\scriptsize$L_{Q}$}}(\partial X)   \qquad \,\;\!\text{mod}\,1 \,,\\[0.5em]
\gamma_{L_Q}&= -\frac{1}{12}\eta^{\text{\:\!$\slashed{D}$}}_{\raisebox{-0.3ex}{\scriptsize$L_{2Q}$}}(\partial X)  + \frac{2}{3} \eta^{\text{\:\!$\slashed{D}$}}_{\raisebox{-0.3ex}{\scriptsize$L_{Q}$}}(\partial X)  \qquad\; \text{mod}\,1 \,.
\ea \ee

\subsection{Non-Isolated Orbifold Singularities ($\Gamma\cong \Z_N$)}
\label{sec:NonIsolatedCyclic}

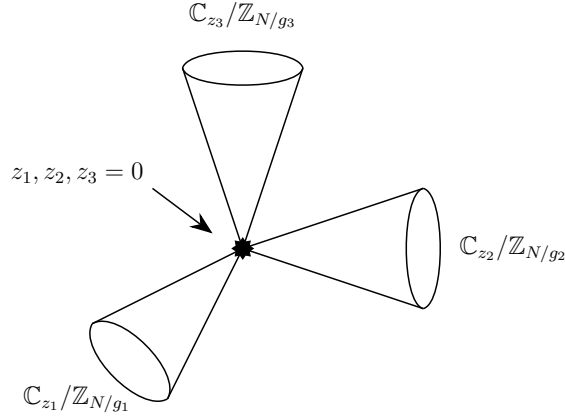
\begin{figure}
\centering
\scalebox{0.8}{
\begin{tikzpicture}
	\begin{pgfonlayer}{nodelayer}
		\node [style=none] (0) at (-1, 3) {};
		\node [style=none] (1) at (1, 3) {};
		\node [style=none] (2) at (3, 1) {};
		\node [style=none] (3) at (3, -1) {};
		\node [style=none] (4) at (0, 0) {};
		\node [style=none] (5) at (-2.5, -1.25) {};
		\node [style=none] (6) at (-1.25, -2.5) {};
		\node [style=Star] (7) at (0, 0) {};
		\node [style=none] (8) at (-2.75, -2.5) {$\mathbb{C}_{z_1}/\Z_{N/g_1}$};
		\node [style=none] (9) at (4.5, 0) {$\mathbb{C}_{z_2}/\Z_{N/g_2}$};
		\node [style=none] (10) at (0, 3.875) {$\mathbb{C}_{z_3}/\Z_{N/g_3}$};
		\node [style=none] (11) at (-0.5, 0.25) {};
		\node [style=none] (12) at (-1.5, 1) {};
		\node [style=none] (13) at (-2.75, 1.25) {$z_1,z_2,z_3=0$};
		\node [style=none] (14) at (0, -3) {};
	\end{pgfonlayer}
	\begin{pgfonlayer}{edgelayer}
		\draw [style=ThickLine] (0.center) to (4.center);
		\draw [style=ThickLine] (4.center) to (1.center);
		\draw [style=ThickLine] (2.center) to (4.center);
		\draw [style=ThickLine] (4.center) to (3.center);
		\draw [style=ThickLine] (4.center) to (5.center);
		\draw [style=ThickLine] (4.center) to (6.center);
		\draw [style=ThickLine, bend left=75, looseness=0.50] (1.center) to (0.center);
		\draw [style=ThickLine, bend left=75, looseness=0.75] (5.center) to (6.center);
		\draw [style=ThickLine, bend right=75, looseness=0.50] (2.center) to (3.center);
		\draw [style=ThickLine, bend left=75, looseness=0.50] (0.center) to (1.center);
		\draw [style=ThickLine, bend left=75, looseness=0.50] (2.center) to (3.center);
		\draw [style=ThickLine, bend right=285, looseness=0.75] (6.center) to (5.center);
		\draw [style=ArrowLineRight] (12.center) to (11.center);
	\end{pgfonlayer}
\end{tikzpicture}}
\caption{We sketch the geometry $X=\mathbb{C}^3/\Z_N(m_1,m_2,m_3)$. The picture depicts the singularities $\mathscr{S}$ which consist of up to three 2D cones cut out by coordinate planes. Each cone supports an A-type ADE singularity in codimension-4. This singularity enhances at the tip of the cone at $z_1,z_2,z_3=0$ where the singularity enhances to codimension-6 (where the 5D SCFT is supported). }
\label{fig:5DSCFTwFlavor}
\end{figure}

Next, consider 5D SCFTs engineered by non-ioslated singularities $X=\mathbb{C}^3/\Z_N(m_1,m_2,m_3)$ with $g_k\equiv \text{gcd}(N,m_k)>1$ for at least one weight $m_k$. For each such $g_k$ the sphere $S^5\subset \mathbb{C}^3$ has the circle cut out by $z_i,z_j=0$ with $\{i,j,k\}=\{1,2,3\}$ as a fixed locus. In $S^5/\Z_N$ this fixed locus projects to a circle worths of $A_{g_k-1}$ A-type ADE singularity. These circles fill into (up to three) cones, also cut out by $z_i,z_j=0$ with $\{i,j,k\}=\{1,2,3\}$, of $A_{g_k-1}$ singularities in $X$ which are modeled on $\mathbb{C}/\Z_{N/g_k}$. Away from the tip of the cone $X$ these are disjoint. See figure \ref{fig:5DSCFTwFlavor}. Next, introduce the product $g=g_1g_2g_3$. Then, the defect group \eqref{eq:DefectGpNonAb} is 
\be\label{eq:DGwFixed}
\mathbb{D}=\mathbb{D}^{(1)}_{\text{elec}}\oplus \mathbb{D}^{(2)}_{\text{mag}}\cong \Z_{N/g}\oplus \Z_{N/g}\,,
\ee
and the line operators $\mathbb{D}^{(1)}_{\text{elec}}$ are acted on by a 1-form symmetry $A\cong \Z_{N/g}$ with background field $B_2^{({N/g})}$. Focusing purely on this 1-form symmetry we observe that there may be anomalies as recorded by the SymTFT couplings
 \be\label{eq:Anom2}\ba
 2\pi  \int_{\text{6D}}  \lbb \beta \lb B_2^{(N/g)}\rb^3+ \gamma B_{2\:\!}^{(N/g)}  (p_1/4)  \rbb+\dots \qquad \text{mod 1}\,,
\ea\ee
much as in \eqref{eq:Anom}, and again $\beta,\gamma \in \Q/\Z$ set a 1-form self-anomaly, mixed 1-form-gravitational anomaly respectively. However, the system now also has a flavor symmetry with Lie algebra (see subsection \ref{ssec:DGS}):
\be
\mathfrak{f}\equiv\mathfrak{su}({g_1})\oplus \mathfrak{su}({g_2})\oplus \mathfrak{su}({g_3})\,,
\ee
which has two consequences of interest for us here. First, denoting the field strength of the associated background fields by $F_2^{(k)}$, which are valued in $\mathfrak{su}({g_k})$, one possibly has further mixed anomalies $\delta_k,\epsilon_k $ which are recorded by the SymTFT couplings
\be\ba
\frac{2\pi}{2(2\pi)^2}\sum_{k=1}^3 \delta_k \int_{\text{6D}}  B_2^{(N/g)\:\!} \text{Tr}(F_2^{(k)}F_2^{(k)}) \quad \text{and}\quad \frac{2\pi}{6(2\pi)^3}\sum_{k=1}^3 \epsilon_k \int_{\text{6D}} \text{Tr}(F_2^{(k)}F_2^{(k)}F_2^{(k)})\,.
\ea \ee
These contributions to the SymTFT are strictly beyond singular cohomology (the Cartan contributions are expected to derive from Chen-Ruan orbifold cohomology, but even then seeing the non-Abelian structure would require further work). Overall, the SymTFT anomaly sector we are therefore interested in has action
\be
\mathcal{S}_{\;\!\text{SymTFT}}=\mathcal{S}_{\;\!\text{SymTFT}}^{(\text{BF})}+\mathcal{S}_{\;\!\text{SymTFT}}^{(\text{anom})}\,,
\ee
considered modulo 1, where the anomaly terms are given by\footnote{Here our normalization convention is such that $F_2^{(k)}/2\pi$ is quantized to have integer periods, while $B_2^{(N/g)}$ takes values in $\Z_{N/g}$. These mixed conventions can of course be streamlined, e.g., by defining \be
B_2^{(N/g)}=  (N/g)\hat{B}_2^{(N/g)}\!/2\pi\,,
\ee
which takes values in $U(1)$.}
\be\label{eq:SymTFTwFlavor}\ba
\mathcal{S}_{\;\!\text{SymTFT}}^{(\text{anom})}&=   2\pi  \int_{\text{6D}}  \Big[ \beta \lb B_2^{(N/g)}\rb^3\!+ \gamma\:\! B_{2\:\!}^{(N/g)}  (p_1/4)  \\&~~~\, +\frac{1}{2(2\pi)^2}\sum_{k=1}^3 \delta_k  B_2^{(N/g)\:\!} \text{Tr}(F_2^{(k)}F_2^{(k)}) \\&~~~\,+  \frac{1}{6(2\pi)^3}\sum_{k=1}^3 \epsilon_k\text{Tr}(F_2^{(k)}F_2^{(k)}F_2^{(k)}) \Big]+\dots \,,
\ea \ee
considered modulo 1, and with BF-terms, included here for completeness, given by
\be
\mathcal{S}_{\;\!\text{SymTFT}}^{(\text{BF})}=\frac{2\pi}{N/g}\int_{\text{6D}} B_{3\:\!}^{(N/g)}  dB_{2\:\!}^{(N/g)} + \frac{1}{2\pi} \sum_{k} \int_{\text{6D}}  \text{Tr} (h_4^{(k)} F_2^{(k)}) \,,
\ee
where the first term is inferred from the defect group \eqref{eq:DGwFixed}, associated with discrete $\Z_{N/g}$ symmetries, and the second term is determined following, e.g., the discussion in \cite{Bonetti:2024cjk}, with  $h_4^{(k)}$ gauge fields valued in $\mathfrak{su}(g_k)$.

Second, the 0-form flavor symmetry and the 1-form symmetry may combine into a 2-group with 4-term sequence\footnote{The tilde here and in related notation, e.g., $\widetilde{B}_2^{(N)}$, are unrelated to resolutions of the geometry, e.g., $\widetilde{X}$.}
\be\label{eq:4termseq}
1\rightarrow A\rightarrow \widetilde{A} \rightarrow \widetilde{F} \rightarrow F\rightarrow 1\,,
\ee
where $\widetilde{A}\cong \Z_N^\vee$, ${A}\cong \Z_{N/g}^\vee$ is the 1-form symmetry group, $\widetilde{F}=SU(g_1)\times SU(g_2)\times SU(g_3)$ is the simply connected group with Lie algebra $\mathfrak{f}$ and $F$ is the flavor symmetry group of the system, see discussion in the context of \eqref{eq:4termseqOG}. Whenever gcd$(N,g)\neq 1$ this is the sequence of a non-trivial 2-group structure, and, e.g., the short exact sequence \eqref{eq:SES2gp} does not split.

To this 2-group structure an interesting anomaly may be attached. Let us motivate this anomaly by ``naive''
computations in singular cohomology. For this begin with the observation
\be\label{eq:PicSub}
H^2(\partial X;\Z)\cong (\Gamma/\Gamma_{\text{fix}})^\vee \subset \text{Pic}(\partial X)\cong \Gamma^\vee\,,
\ee
 with $\Gamma_{\text{fix}}\cong \Z_g$, the 1-form symmetry group $(\Gamma/\Gamma_{\text{fix}})^\vee\cong \Z_{N/g}^\vee\cong A$ and the Picard group $\text{Pic}(\partial X)$. Here dualization via $\vee$ is crucial for the subgroup relation. The subgroup $H^2(\partial X;\Z)$ of the Picard group is in correspondence with complex line bundles on $\partial X$. Elements outside of this subgroup are invertible sheaves (or orbi-line bundles). These are of course ordinary bundles away from the singularities, i.e., we have
\be
 H^2(\partial X^\circ;\Z)\cong  \text{Pic}(\partial X)\cong \Gamma^\vee\,,
\ee
where $\partial X^\circ$ is a manifold with boundary obtained by excision of a tubular neighborhood $T(\partial \mathscr{S})$ of the asymptotic singular locus. See the discussion in section \ref{sec:stratified}.

With this now consider the stratified construction in \cite{Cvetic:2024dzu} which assigns a symmetry theory to $\partial X^\circ$ characterizing symmetry structures of the 5D SCFT. At a technical level, one requires a reduction scheme for manifolds with boundary, we refer to \cite{Cvetic:2024dzu} for further details and will simply sketch relevant steps here. To proceed then, consider expanding the $C_3G_4G_4$ term with respect to $H^m(\partial X^\circ;\Z)$ and $H^m(\partial X^\circ,\partial^2 X^\circ;\Z)$ modulo identifications induced by the long exact sequence in relative homology of the pair $(\partial X^\circ,\partial^2 X^\circ)$.\footnote{This step generalizes an expansion with respect to $H^m(\partial X;\Z)$ and seprates out the structures which can be reliably computed using singular cohomology.} As in a normal reduction scheme we must contract the internal cocycles and, from these considerations, the following triple product arises
\be\ba
(H^2(\partial X^\circ,\partial^2 X^\circ;\Z)/H^1(\partial^2 X^\circ)) \times (H^2(\partial X^\circ,\partial^2 X^\circ;\Z)/H^1(\partial^2 X^\circ))  \times H^2(\partial X^\circ;\Z) \rightarrow \Q/\Z\,,\\[0.5em]
\ea\ee
where all groups are torsional and which may be rewritten as
\be
\text{Tor}\, H^2(\partial X;\Z) \times \text{Tor}\, H^2(\partial X;\Z) \times \text{Tor}\, H^2(\partial X^\circ;\Z) \rightarrow \Q/\Z\,.
\ee
Geometrically, the triple product is realized as a cup product between the first two factors, resulting in a class in $\text{Tor}\, H^4(\partial X;\Z) $, which is then linked with the final factor. Written with respect to 2-group data this triple product is simply
\be
\mathscr{T}\,:~A\times A\times \widetilde{A}\rightarrow \Q/\Z\,.
\ee
Restricting the last entry to the subgroup $A\subset  \widetilde{A}$ we have $\mathscr{T}|_{A}:A^3=H^2(\partial X;\Z)^3\rightarrow \Q/\Z$ which relates to the 1-form self-anomaly, more precisely evaluating on $(1,1,1)$ gives $6\beta$.

Unfortunately, the naive computation of expanding $C_3G_4G_4$ (and other terms) in singular cohomology classes of course runs into the issue that the link pairing above is multiplied by $1/6$, which is ill-defined over $\Q/\Z$. In the above considerations employing $\partial X^\circ$ and $\partial^2 X^\circ$ adding an auxiliary bulk to settle this division by $6$ is not straightforward. Even so, note that $\mathscr{T}|_{A}$ is precisely the triple product which refined via $1/6$ to the 1-form self-anomaly $\beta$. Indeed, ignoring the singularities we would have written identical expression to \eqref{eq:bulkint} with $T_2$ inducing a generator of $H^2(\partial X;\Z)$.

Overall, we are lead to two natural conjectures. First, the anomalies $\beta,\gamma$ in \eqref{eq:SymTFTwFlavor} remain computable via $\eta$-invariants even in the presence of singularities. Second, we expect to be able to refine anomaly computations
according to the short exact sequences:
{\renewcommand*{\arraystretch}{1.35}
\be\label{eq:Ref}
\begin{array}{c} 1\rightarrow A \rightarrow \widetilde{A} \rightarrow \widetilde{A}/A \rightarrow 1 \\ 1\rightarrow H^2(\partial X;\Z) \rightarrow H^2(\partial X^\circ;\Z)  \rightarrow H^2(\partial X^\circ;\Z)/H^2(\partial X;\Z) \rightarrow 1 \\  1\rightarrow \Z_{N/g}^\vee \rightarrow \Z_N^\vee \rightarrow \Z_g^\vee \rightarrow 1 \\ \end{array}
\ee}

\noindent Here we have given the short exact sequence for symmetries, its incarnation in geometry and its explicit evaluation for the examples under consideration.

Our goal, is now to first compute the anomalies $\beta,\gamma,\delta_k,\epsilon_k$, given in \eqref{eq:SymTFTwFlavor}, and subsequently discuss refinements due to \eqref{eq:Ref}.

\subsubsection{Anomaly Computations for Singular Boundaries via $\eta$-Invariants}

We now compute the anomalies $\beta,\gamma,\delta_k,\epsilon_k$ presented in \eqref{eq:SymTFTwFlavor} beginning with the pair $\beta,\gamma$.

To begin, we deform the 5D SCFT $\mathcal{T}_X$ to its Coulomb branch. Geometrically, this deformation is realized by some toric resolution of $X=\mathbb{C}^3/\Z_N$. Previously massless excitations may pick up masses, however the lattice of charges realized by dynamical excitations remains unchanged. As a consequence the 1-form symmetry is unaltered, including anomalies related thereto \cite{DelZotto:2022ras}. The toric resolution in particular smooths the boundary $\partial X$ and applying our results from section \ref{sec:IsolatedCY3} we immediately conclude
\be
\alpha_{\widetilde{L}}=\beta_{\widetilde{L}}+\gamma_{\widetilde{L}}=\frac{1}{2}\eta^{\text{\:\!$\slashed{D}$}}_{\raisebox{-0.55ex}{\scriptsize$\widetilde{L}$}}\!\!\;\!\lb \partial \widetilde{X} \rb \quad \text{mod 1}\,,
\ee
where $\widetilde{L}$ is a line bundle on $\partial \widetilde{X}= \widetilde{S^5/\Z_N}$. The applicability and interpretation of this result requires some further specifications. First, note that we have
\be
H^2(\partial X;\Z)\cong \Z_{N/g}\,, \quad H^2(\partial \widetilde{X};\Z)\cong \Z^{\:\!\text{rk}(\mathfrak{f})}\oplus \Z_{N/g}\,,
\ee
with $\text{rk}(\mathfrak{f})$ denoting the rank of the flavor symmetry algebra $\mathfrak{f}$.\footnote{The torsion subgroup is unaffected by resolution, which can be explicitly seen from the perspective of vanishing cycles. Dualizing to singular homology $\Z_{N/g}$ is accounted for by $H_1(\partial X;\Z)$ due to the universal coefficient theorem and adding or removing 2-cycles leaves this group of 1-cycles invariant.} Therefore, line bundles on $\partial X$ and line bundles on $\partial \widetilde{X}$ with torsional first Chern class are in 1:1 correspondence. We will denote the latter by $\widetilde{L}_Q$ and importantly, due to the subgroup relation \eqref{eq:PicSub}, the charge $Q$ takes values in $\Z_{N/g}\subset \Z_N$ or equivalently takes values in $\Z_N$ quantized in units of $g$
\be\label{eq:quant}
Q\in \{0,g,2g,3g, \dots\} \cong \Z_{N/g}\subset \Z_N\,.
\ee
Contracting the exceptional locus we obtain a pushforward $\widetilde{L}_Q\rightarrow L_Q$ where $L_Q$ is simply a line bundle on $\partial X$ with charge $Q$ as introduced below \eqref{eq:alphaDirac}. By direct computation we verify across many toric examples that the $\eta$-invariants of interest are invariant with respect to toric resolutions
\be\label{eq:EtaInvInv}
\frac{1}{2}\eta^{\text{\:\!$\slashed{D}$}}_{\raisebox{-0.55ex}{\scriptsize$\widetilde{L}_Q$}}\!\!\;\!( \partial \widetilde{X} )=\frac{1}{2}\eta^{\text{\:\!$\slashed{D}$}}_{\raisebox{-0.55ex}{\scriptsize${L}_Q$}}\!\!\;\!\lb \partial {X} \rb\,.
\ee
Here, the left hand side is computed via the bulk integral given in \eqref{eq:bulkint}. The right hand side is computed as in \eqref{eq:Degeratu} which we now make explicit. See appendix \ref{sec:Check2} for an example of a bulk computation.


To give the $\eta$-invariant of the orbifold $S^5/\Z_N$ introduce the subset of $\Gamma\cong \Z_N$ consisting of group elements which act without fixed point on $S^5$. We write
\be
\text{FF}(\Gamma)=\{\gamma \in \Gamma\,:\, 1\not\in \text{Eigenvalues}(\gamma_{SU(3)})\}\,,
\ee
in the obvious notation. For $\Gamma\cong \Z_N$ we have $\Gamma\setminus \text{FF}(\gamma)\cong \Z_{g}\cong \Gamma_{\text{fix}}$ but whenever $\Gamma$ is non-cyclic we will have $\Gamma\setminus\text{FF}(\gamma)\not\cong \Gamma_{\text{fix}}$ and $|\Gamma\setminus\text{FF}(\gamma)|<|\Gamma_{\text{fix}}|$. In any case, one then has from \eqref{eq:Degeratu} with these definitions
\be\label{eq:EtaFixed}\ba
\frac{1}{2}\eta^{\text{\:\!$\slashed{D}$}}_{\raisebox{-0.3ex}{\scriptsize$L$}}&=\frac{1}{N}\sum_{\gamma \;\!\in\;\! \text{FF}(\Z_N)}(-1)^k\omega^{k\:\!Q_L}\prod_{i=1}^3\frac{1}{\omega^{-km_i/2}-\omega^{km_i/2}}\\[0.5em]
&=\frac{1}{N}\sum_{\ell=0}^{g-1}\lb \sum_{k\;\!=\;\!\ell N/g\:\!+\:\!1}^{(\ell+1) N/g\:\!-\:\!1}(-1)^k\omega^{k\:\!Q_L}\prod_{i=1}^3\frac{1}{\omega^{-km_i/2}-\omega^{km_i/2}}\rb\,,
\ea \ee
and then for the combination $\alpha_L=\beta_L+\gamma_L$ of anomalies simply
\be
\alpha_L=\frac{1}{2}\eta^{\text{\:\!$\slashed{D}$}}_{\raisebox{-0.3ex}{\scriptsize$L$}}\quad \text{mod}\,1\,.
\ee
Here $Q_L$ is quantized in units of $g$, see \eqref{eq:quant}, and $L$ is a line bundle with first Chern class in $H^2(\partial X;\Z)\cong \Z_{N/g}$. Further, by \eqref{eq:EtaInvInv}, with these definitions we have
\be\label{eq:betagamma2}\ba
\beta_{L_Q}&=+\frac{1}{12}\eta^{\text{\:\!$\slashed{D}$}}_{\raisebox{-0.3ex}{\scriptsize$L_{2Q}$}}(\partial X) -\frac{1}{6} \eta^{\text{\:\!$\slashed{D}$}}_{\raisebox{-0.3ex}{\scriptsize$L_{Q}$}}(\partial X)   \qquad \,\;\!\text{mod}\,1 \,,\\[0.5em]
\gamma_{L_Q}&= -\frac{1}{12}\eta^{\text{\:\!$\slashed{D}$}}_{\raisebox{-0.3ex}{\scriptsize$L_{2Q}$}}(\partial X)  + \frac{2}{3} \eta^{\text{\:\!$\slashed{D}$}}_{\raisebox{-0.3ex}{\scriptsize$L_{Q}$}}(\partial X)  \qquad\; \text{mod}\,1 \,,
\ea \ee
exactly as in \eqref{eq:betagamma} which are similarly subject to an unphysical $\Z_6$ ambiguity.

In table \ref{tab:TableForNonIsolated} we list some 1-parameter families of toric Calabi-Yau 3-folds with non-isolated singularities and compute their combination $\alpha$ and the anomalies $\beta,\gamma$ with respect to the Kawasaki choice of generator and further give the integers $q,s_3$ relating these to other parameterizations picking out their own preferred generators of $\Z_{N/g}$.\footnote{Kawasaki's parametrization of the cohomology ring of $S^5/\Z_N$ as given in  \eqref{eq:Kawasaki0}, \eqref{eq:Kawasaki1} and  \eqref{eq:Kawasaki2} holds irrespective whether the action on $S^5$ is fixed point free or not.}

\begin{table}[]
\centering
{\renewcommand*{\arraystretch}{1.35}

\begin{tabular}{||c ||   c| c || c| c || c|| c || c |}
 \hline
 $\Z_N(m_1,m_2,m_3)$ & $\raisebox{0.05ex}{$\frac{1}{2}\eta^{\text{\:\!$\slashed{D}$}}_{\raisebox{-0.3ex}{\scriptsize$L_q$}}\!(S^5\!/\:\!\Z_N)$}$ & $q$ &  $\beta_{L_q}$ & $\gamma_{L_q}$ & $s_3$ & $\raisebox{0.05ex}{$\frac{1}{2}\eta^{\text{\:\!$\slashed{D}$}}_{\raisebox{-0.3ex}{\scriptsize$L_{Q=-1}$}}\!(S^5\!/\:\!\Z_N)$}$ & $\mathfrak{f}$ \\ [0.5ex]
 \hline\hline
 $\frac{1}{2n+2}(1,1,2n)$ & $ \frac{n(n-1)}{6(n+1)}$  & $-2$  & $-\frac{n-1}{6(n+1)}$ & $\frac{n^2-1}{6(n+1)}$  & $-1$ &$\frac{n(n+2)}{12(n+1)}$& $\mathfrak{su}_2$ \\[0.5ex]
 \hline
  $\frac{1}{6n+3}(1,2,6n)$ & $ \frac{n(n-1)}{2(2n+1)}$  & $-3$  & $-\frac{n-1}{6(2n+1)}$ & $\frac{3n^2-2n-1}{6(2n+1)}$  & $-1$ & $ \frac{n(n+1)}{6(2n+1)} $  & $\mathfrak{su}_3$ \\[0.5ex]
 \hline
 $\frac{1}{12n+4}(1,3,12n)$ & $ \frac{n(n-1)}{3n+1}$  & $-4$  & $-\frac{n-1}{6(3n+1)}$ & $\frac{6n^2-5n-1}{6(3n+1)}$ & $-1$ & $\frac{n^2}{4(3n+1)}$ & $\mathfrak{su}_4$  \\[0.5ex]
       \hline
        $\frac{1}{20n+5}(1,4,20n)$ & $ \frac{5n(n-1)}{3(4n+1)} $  & $ -5$ & $-\frac{n-1}{6(4n+1)}$  & $\frac{10n^2-9n-1}{6(4n+1)}$ & $-1$ &$\frac{n(n-1)}{3(4n+1)}$ & $\mathfrak{su}_5$ \\[0.5ex]
       \hline
\end{tabular}
}
\caption{List of $\eta$-invariants of $S^5/\Z_N$ associated to the Dirac operator $\protect\slashed{D}_{L_q'}$ on $\mathbb{C}^3/\Z_N(m_1,m_2,m_3)$ twisted by a line bundle $L_q'$. We consider non-isolated singularities $\mathbb{C}^3/\Z_N(m_1,m_2,m_3)$ which have singular links $S^5/\Z_N$. The line bundle $L_q'$ is not unique, however it uniquely restricts to the boundary $S^5/\Z_N(m_1,m_2,m_3)$ realizing a line bundle $L_q$ with charge $Q_{L_q}=q\in \Z_N$ with reference to Kawasaki's preferred generator, i.e., simultaneously $c_1(L_q)=v_2\in H^2(S^5/\Z_N;\Z)$ and choice of generator $v_2\equiv 1\in \Z_N$. Further, classes in $H^2(S^5/\Z_N;\Z)$ are Poincar\'e dual to 3-cycles $H_3(S^5/\Z_N;\Z)\cong \Z_N$. The equation $z_3=0$ cuts out a non-compact divisor of $\mathbb{C}^3/\Z_N(m_1,m_2,m_3)$ denoted $D_{z_3}$ which intersects the boundary $S^5/\Z_N$ in a 3-cycle. The column labelled $s_3$ specifies this 3-cycle, as an element $s_3^{-1} \in \Z_N$, with respect to the Kawasaki's generator. The divisor $s_3D_{z_3}$ is one good choice for $L_q'$. For example, in the toric resolution of $\mathbb{C}^3/\Z_N$ the triple intersection of $s_3D_{z_3}$ modulo 1 and the linking of $v_2^2$ with $v_2$ in $S^5/\Z_N$ agree.  The anomalies $\beta,\gamma$ are the pure 1-form anomaly and the mixed 1-form-gravitational anomaly. We also give the $\eta$-invariants where we have lifted the quantization condition restricting us to $H^2(S^5/\Z_N;\Z)\subset \text{Pic}(S^5/\Z_N)$.}
\label{tab:TableForNonIsolated}
\end{table}

Next, we turn to discuss the anomalies $\delta_k$. These originate from a Wess-Zumino term
\be\label{eq:WZterm}
S_{\text{WZ}}^{(k)}=\frac{2\pi}{2(2\pi)^2}\int C_3 \text{Tr}(F_2^{(k)}F_2^{(k)})+\dots
\ee
supported on the $k$-th component of the non-compact codimension-4 singularities (of which there are up to three, see figure \ref{fig:5DSCFTwFlavor}).\footnote{In the IIA setting with a stack of D6-branes, this is the familiar WZ term coupling the RR 3-form potential to the instanton density.}

To proceed, note that asymptotically each such stack is wrapped on $S^1/\Z_N=S^1/\Z_{N/g_k}$ where we have quotiented out the subgroup fixing the circle. We have
\be
[S^1/\Z_{N/g_k}]= g_1g_2g_3/g_k=g_ig_j \in \Z_{N/g}\cong H_1(S^5/\Z_N;\Z)\,, \qquad \{i,j,k\}=\{1,2,3\}\,,
\ee
which is Poincar\'e-dual to $g_ig_j \in \Z_{N/g}\subset  H^4(S^5/\Z_N;\Z)\cong \Z_N$ with respect to Kawasaki's generator. Let us denote the corresponding 4-cocycle by $t_4^{(ij)}$. Then, via the standard trick of completing the integral in \eqref{eq:WZterm} to the full space by extending the integrand by the cocycle dual to the support of the integral, the above term evaluates via linking pairing as
\be\label{eq:OtherTerm}\ba
S_{\text{WZ}}^{(k)}&=2\pi \ell(t_4^{(ij)}, t_2) \int_{\text{6D}} B_2^{(N/g)} \frac{\text{Tr}(F_2^{(k)}F_2^{(k)})}{2(2\pi)^2}+\dots \\[0.5em]
&=\frac{2\pi qg_ig_j}{N} \int_{\text{6D}} B_2^{(N/g)}  \frac{\text{Tr}(F_2^{(k)}F_2^{(k)})}{2(2\pi)^2}+\dots\,,
\ea \ee
where the anomaly $\delta_k$ is determined by comparing to \eqref{eq:SymTFTwFlavor} and evaluated by \eqref{eq:LinkingCharge} to:
\be
\delta_k=\ell(t_4^{(ij)}, t_2)=qg_ig_j/N\,,  \qquad \{i,j,k\}=\{1,2,3\}\,.
\ee

Next, we turn to discuss the anomalies $\epsilon_k$. 
As we now explain, this descend from the localized 7D Super Yang-Mills theory sector.
Observe that we can introduce a chemical potential for the third Chern class of the gauge bundle, i.e., a Wess-Zumino term:
\be\label{eq:WZterm2}
S_{\text{WZ}}^{(k)}=\frac{2\pi}{2(2\pi)^2}\int C_1 \text{Tr}(F_2^{(k)}F_2^{(k)}F_2^{(k)})+\dots
\ee
which is supported on the $k$-th component of the non-compact codimension-4 singularities (of which there are up to three, see figure \ref{fig:5DSCFTwFlavor}).\footnote{In the IIA setting this term would arise from the WZ term involving the coupling of the RR 1-form potential
to the third Chern character. In the lift to M-theory the RR 1-form potential becomes the component of the spin connection along the circle used for the reduction.}
We focus now on the local model $\text{Loc}_k$ of the $k$-th flavor brane which fits into the fibration
\be
\mathbb{C}^2/\Z_{g_k}\hookrightarrow \text{Loc}_k \rightarrow \lb (\mathbb{C}/\Z_{N/g_k})\setminus \{0\}\rb .
\ee
In the base we have excised the origin $z_1,z_2,z_3=0$ where the different flavor branes meet (and the singularity enhances and where the 5D SCFT is localized). Importantly, $\text{Loc}_k$ is not a direct product. The base has a non-trivial circle and traversing it the overall quotient is such that the normal geometry is acted on by a $\Z_{N/g_k}(m_i,m_j)$ with $\{i,j,k\}=\{1,2,3\}$. Focusing on the Hopf circle in $S^3/\Z_{g_k}\subset \mathbb{C}^2/\Z_{g_k}$ this gives a circle bundle with flat connection $C_1$. From here we compute
\be
\epsilon_k=\frac{m_k}{N}\,,
\ee
see \cite{Cvetic:2024dzu} for further discussion.

\subsubsection{Anomaly Refinements via $\eta$\:\!-Invariants}
\label{ssec:RefinedAnomaly}

One conclusion of our discussion on the pure 1-form anomaly and mixed 1-form-gravitational anomaly for 5D SCFTs was that due to the subgroup relation
\be
\Z_{N/g}\cong H^2(\partial X;\Z)\subset H^2(\partial X^\circ)\cong \text{Pic}(\partial X)\cong \Z_N\,,
\ee
the charge $Q$ relevant in computing the sum
\be
\alpha_{L_Q}=\beta_{L_Q}+\gamma_{L_Q}=\frac{1}{2}\eta^{\text{\:\!$\slashed{D}$}}_{\raisebox{-0.3ex}{\scriptsize$L_Q$}}(S^5/\Z_N)\quad \text{mod}\,1\,,
\ee
was quantized in units of $g$, see \eqref{eq:quant}, or equivalently that $Q$ must be associated with a line bundle $L_Q$ (rather than an orbi-bundle). Said differently, we have the function
\be\label{eq:Unrestricted}
\eta^{\text{\:\!$\slashed{D}$}}(S^5/\Z_N)\,:\quad \text{Rep}(\Z_N)\rightarrow \Q/\Z\,, \qquad  Q \mapsto \frac{1}{2}\eta^{\text{\:\!$\slashed{D}$}}_{\raisebox{-0.3ex}{\scriptsize$L_Q$}}(S^5/\Z_N)\quad \text{mod}\,1\,,
\ee
and studied its restriction to the 1-form symmetry subgroup $A\cong \Z_{N/g}$ with the result
\be\label{eq:CleanCase}
\text{Rep}(\Z_{N/g})\rightarrow \Q/\Z\,, \qquad  Q \mapsto \frac{1}{2}\eta^{\text{\:\!$\slashed{D}$}}_{\raisebox{-0.3ex}{\scriptsize$L_Q$}}= \beta_1 Q^3 + \gamma_1\:\! Q\quad \text{mod}\,1\,,
\ee
whose homogenous pieces we related to anomalies, i.e., the $\eta$\:\!-invariant plays the role of a generating function for the SCFT 1-form anomalies.\footnote{Here, anomalies are expressed  respect to some choice of generator $1\in \Z_{N/g}$, this makes automatic that and $\beta_Q=\beta_1Q^3$ and $\gamma_Q=\gamma_1Q$ have the correct transformation properties to multiply a term cubic and linear in the 1-form background as in \eqref{eq:SymTFTwFlavor}. Note that the $\eta$\:\!-invariant takes this functional form only mod 1 (over $\Q$ quadratic pieces appear). }

Naturally, the question arises which anomaly data is recorded by the unrestricted function \eqref{eq:Unrestricted} associated with $\widetilde{A}\cong \Z_N$ in \eqref{eq:Ref}. With this, the obvious way in proceeding to refine the 1-form anomaly discussion is then to simply drop this quantization condition. For any element $\mathscr{L}\in \text{Pic}(\partial X)$ of the Picard group we define mod 1
\be\label{eq:Refined}\ba
\widetilde{\alpha}_{\mathscr{L}}=\frac{1}{2}\eta^{\text{\:\!$\slashed{D}$}}_{\raisebox{-0.3ex}{\scriptsize$\mathscr{L}$}}&=\frac{1}{N}\sum_{\gamma \;\!\in\;\! \text{FF}(\Z_N)}(-1)^k\omega^{k\:\!Q_{\mathscr{L}}}\prod_{i=1}^3\frac{1}{\omega^{-km_i/2}-\omega^{km_i/2}}\\[0.5em]
&=\frac{1}{N}\sum_{\ell=0}^{g-1}\lb \sum_{k\;\!=\;\!\ell N/g\:\!+\:\!1}^{(\ell+1) N/g\:\!-\:\!1}(-1)^k\omega^{k\:\!Q_{\mathscr{L}}}\prod_{i=1}^3\frac{1}{\omega^{-km_i/2}-\omega^{km_i/2}}\rb\,.
\ea \ee
The elements $\mathscr{L}$ are invertible sheaves or orbifold line bundles which, unlike the line bundles so far, carry additional structure localized to the non-compact singularities / flavor branes.

We pause to demonstrate via an example that there is indeed strictly more information in $\widetilde{\alpha}_{\mathscr{L}}$ than in $\alpha_L$. Explicitly, consider the family of geometries $X_n=\mathbb{C}^3/\Z_{2n}(1,1,2n-2)$. For these we have
\be\ba\label{eq:alphaDiracAp4}
\frac{1}{2}\eta^{\text{\:\!$\slashed{D}$}}_{\raisebox{-0.3ex}{\scriptsize$L_{-Q}$}}&=\frac{1}{2n}\lb \sum_{k=1}^{n-1}+\sum_{k=n+1}^{2n-1}\rb (-1)^k\omega^{-Qk}\frac{1}{\omega^{-k/2}-\omega^{k/2}}\frac{1}{\omega^{-k/2}-\omega^{k/2}}\frac{1}{\omega^{-k(2n-2)/2}-\omega^{k(2n-2)/2}}\\[0.5em]
&~~~\,=\frac{(Q-n) \left(-4 n Q+2 Q^2+3 (-1)^Q-3\right)}{48 n} \\[0.5em]
&~~~\,=\frac{Q^3}{24 n}+\frac{n Q}{12}+\frac{(-1)^Q Q}{16 n}-\frac{Q}{16 n}-\frac{Q^2}{8}-\frac{(-1)^Q}{16}+\frac{1}{16}\,,
\ea\ee
in $\Q$. Fixing $n$, there are up to 8 types of terms given by monomials $Q^k$ with $k=0,\dots, 3$ multiplied by $(-1)^Q$ or not. The function $(-1)^Q$ is related to $\Z_g\cong \Z_2$ in this case and, more generally, $g$-th roots of unity appear. In contrast to \eqref{eq:CleanCase}, even when \eqref{eq:alphaDiracAp4} is considered mod 1 terms other than those proportional to $Q^3$ and $Q$ need to be specified.  See tables \ref{tab:TableForNonIsolated} and \ref{tab:TableForNonIsolated2} for some further evaluated examples.

We now conjecture which anomaly information is encoded in \eqref{eq:Refined}. For this we include background fields for the flavor brane 1-form symmetries into our symmetry theory. At a computational level this is straightforward, we simply apply the methods which produced \eqref{eq:Coupling3} starting from  \eqref{eq:Coupling} to the SymTFT action \eqref{eq:SymTFTwFlavor} and obtain
\be\label{eq:CDHHT}\ba
\mathcal{S}_{\;\!\text{CDHHT}}^{(\text{anom})}&=   2\pi  \int_{\text{6D}}  \Big[ \beta \lb B_2^{(N/g)}\rb^3\!+ \gamma\:\! B_{2}^{(N/g)}  (p_1/4)   +\sum_{k=1}^3 \widetilde{\delta}_k  B_2^{(N/g)}  B_2^{(g_k)}  B_2^{(g_k)}  +  \sum_{k=1}^3 \widetilde{\epsilon}_k \lb B_2^{(g_k)}\rb^3 \Big] \,,
\ea \ee
where $B_2^{(g_k)}$ is the background of the 1-form symmetry of the 7D SYM theory with gauge algebra $\mathfrak{su}(g_k)$ which constitute flavor branes to the 5D SCFT. As such \eqref{eq:CDHHT} should be viewed as governing anomaly structures in the combined 5D/7D system. The coefficients $\widetilde{\delta}_k,  \widetilde{\epsilon}_k \in \Q/\Z$ are computed to satisfy
\be
-\lbb \frac{g_k-1}{2g_k}\rbb {\delta}_k=\widetilde\delta_k\,, \qquad \lbb \frac{(g_k-1)(g_k-2)}{6g_k^2}\rbb {\epsilon}_k=\widetilde \epsilon_k\,.
\ee
Note, in \eqref{eq:CDHHT} we have dropped the labelling ``SymTFT." Nonetheless, the above results can be given a symmetry theory interpretation, following \cite{Cvetic:2024dzu}, which is that $\mathcal{S}_{\;\!\text{CDHHT}}^{(\text{anom})}$ is associated with a 6D symmetry theory, derived from $\partial X^\circ$ which is relative to the 7D symmetry theory associated to $\partial^2 X^\circ$ which is associated to the flavor branes. See \cite{Cvetic:2024dzu} for the discussion of the 7D theory and the 6D/7D boundary conditions.

Returning to our original observation centered on \eqref{eq:Unrestricted} we conjecture that the anomalies $\beta,\gamma, \widetilde{\delta}_k,  \widetilde{\epsilon}_k$ are all encoded in the twisted $\eta$\:\!-invariants \eqref{eq:Refined} of the boundary $S^5/\Z_N$. 

The reason for this conjecture is \eqref{eq:Ref} suggesting that $B_2^{(N/g)} ,B_2^{(g)}$ combines into a $B_2^{(N)}$ associated internally with a generator of $H^2(\partial X^\circ;\Z)\cong \text{Pic}(\partial X)\cong \Z_N$ and the fact that \eqref{eq:CDHHT} only contains the fields $B_2^{(N/g)},B_2^{(g)}$. 

The relevance of this to our discussion is as follows. The space $\partial X^\circ$ is also a topological model for the boundary of
\be
X\setminus T( \mathscr{S})\,,
\ee
i.e., the full internal geometry with a tubular neighborhood of the full singular locus excised. As such $X\setminus T( \mathscr{S})$ is the maximal region on which the 11D supergravity approximation is accurate and the action \eqref{eq:CDHHT}, computed with respect to the smooth manifold with boundary $\partial X^\circ$, records the strongest topological constraints on the physics of $T( \mathscr{S})$ which may be derived from 11D supergravity.

Given this maximality, a good question is: Does the symmetry data encoded in $\mathcal{S}_{\;\!\text{CDHHT}}^{(\text{anom})}$ uniquely determine 
a 5D SCFT among the set of 5D SCFTs? Among the set of 5D SCFTs with extra-dimensional construction in M-theory on a Calabi-Yau three-folds $X=\mathbb{C}^3/\Z_N$, does the function \eqref{eq:Unrestricted} uniquely determine a 5D SCFT?

As a related comment, observe that the restriction \eqref{eq:CleanCase} cannot distinguish between 5D SCFTs with trivial $\mathbb{D}^{(1)}$.

\subsection{Non-Isolated Orbifold Singularities (Non-Cyclic $\Gamma$)}

We now discuss anomalies of 5D SCFTs associated with quotients by non-cyclic $\Gamma$, starting with non-Abelian $\Gamma$ and then, for completeness, also discuss the case $\Gamma\cong \Z_N\times \Z_M$. Given the evidence in previous section we assume that the anomalies $\beta,\gamma$ of 5D SCFTs engineered by $\mathbb{C}^3/\Gamma$, as written in \eqref{eq:Anom}, are computed for arbitrary quotients by $\eta$\:\!-invariants, i.e., \eqref{eq:EtaInvInv} is assumed to hold beyond toric examples.

To proceed, we focus now on the subset of M-theory backgrounds $\mathbb{C}^3/\Gamma$ with non-Abelian $\Gamma\subset SU(3)$ which engineer 5D SCFTs with non-trivial defect group.\footnote{See \cite{Tian:2021cif} for explicit computations of the toric resolutions.} Such groups $\Gamma$ and their presentations were given in \cite{DelZotto:2022fnw}, and were found to all be small subgroups of $SU(3)$ \cite{yaugorenstein}.\footnote{A group $G \subset \text{GL}(n, \C)$ is referred to as small if it contains no elements $g\in G$ with exactly $n-1$ many eigenvalues equal to 1 (i.e., $G$ contains no reflections).} Small subgroups derive from finite subgroups of $U(2)$ and as such are related to the binary dihedral ($D$), binary tetrahedral ($T$), binary octahedral ($O$), binary icosahedral ($I$) groups.

\begin{table}[]
\centering
{\renewcommand*{\arraystretch}{1.35}
\begin{tabular}{||c ||   c| c | c| c || }
 \hline
 $\Gamma$ & $|\Gamma|$ & $ |\text{FF}(\Gamma)|$ & Ab$(\Gamma/\Gamma_{\text{fix}})$ & Flavor $\mathfrak{f}$   \\ [0.5ex]
 \hline\hline
 $D_{n,q}^{(m\,\text{odd})}$ & $ 4qm$  & $4q(m-1)$  & $ \Z_{m}$ & $\mathfrak{so}_{2q+4}$ \\[0.5ex]
 \hline

  $D_{n,q}^{(m\,\text{even})}$ & $4qm$  & $4q(m-1)$  & $\Z_{2m}$ & $\mathfrak{so}_{2q+4}$   \\[0.5ex]
 \hline
$T_m^{\:\!(1,5)}$ & $24m$  & $24(m-1)$  & $\Z_m$ & $\mathfrak{e}_6$   \\[0.5ex]
       \hline
       $T_m^{\:\!(3)}$ & $ 24m $  & $ 24(m-1)$ & $\Z_{3m}$  & $\mathfrak{e}_6$   \\[0.5ex]
       \hline
        $O_m$ & $ 48m $  & $ 48(m-1)$ & $\Z_m$  & $\mathfrak{e}_7$  \\[0.5ex]
       \hline
       $I_m$ & $120m$  & $ 120(m-1)$ & $\Z_m$  & $\mathfrak{e}_8$   \\[0.5ex]
       \hline
\end{tabular}
}
\caption{List of small subgroups $\Gamma\subset SU(3)$ and some of their properties. }
\label{tab:NonAbData}
\end{table}

We now present a self-contained parametrization of these small groups, taken from \cite{DelZotto:2022fnw}, and discuss their properties. Generators of all small groups are given by the four matrices
\be
\ba
A= \left(
\begin{array}{ccc}
 0 & i & 0 \\
 i & 0 & 0 \\
 0 & 0 & 1 \\
\end{array}
\right)\,, \quad B&=
\left(
\begin{array}{ccc}
 0 & -1 & 0 \\
 1 & 0 & 0 \\
 0 & 0 & 1 \\
\end{array}
\right)\,, \quad  C=\frac{1}{2}\left(
\begin{array}{ccc}
1+i & -1+i & 0 \\
1+i & 1-i & 0 \\
 0 & 0 & 2 \\
\end{array}
\right)\,, \\[0.75em] &\hspace{-30pt}D=\frac{1}{\sqrt{5}}\left(
\begin{array}{ccc}
 {e^{-\frac{2  \pi i }{5}}-e^{\frac{2  \pi i}{5}}} & {e^{\frac{4  \pi i }{5}}-e^{-\frac{4  \pi i }{5}}} & 0 \\
 {e^{\frac{4  \pi i }{5}}-e^{-\frac{4  \pi i }{5}}} & {e^{\frac{2  \pi i }{5}}-e^{-\frac{2 \pi i }{5}}} & 0 \\
 0 & 0 & \sqrt{5} \\
\end{array}
\right)\,,
\ea
\ee
and the two matrix families
\be
\ba
X_n=\left(
\begin{array}{ccc}
 \omega_{n} & 0 & 0 \\
 0 &  \omega_{n}^{-1} & 0 \\
 0 & 0 & 1 \\
\end{array}
\right)\,, \quad  Y_n=\left(
\begin{array}{ccc}
 \omega_{n} & 0 & 0 \\
 0 &  \omega_{n} & 0 \\
 0 & 0 & \omega_n^{-2}  \\
\end{array}
\right)\,,
\ea
\ee
with primitive $n$-th root of unity $\omega_n=\exp(2\pi i/n)$. With respect to these, the small finite subgroups $\Gamma \subset SU(3)$ are generated as:
\be\ba
D_{n,q}^{(m\,\text{odd})}&=\langle A, X_{2q}, Y_{2m}\rangle \qquad &&\text{with $m=n-q$ odd, $1<q<n$, and gcd$(n,q)=1$}\,, \\
D_{n,q}^{(m\,\text{even})}&=\langle  X_{2q}, Y_{4m}A\rangle&&\text{with $m=n-q$ even, $1<q<n$, and gcd$(n,q)=1$}\,, \\
T_m^{\:\!(1,5)}&=\langle A,C,X_4,Y_{2m} \rangle && \text{with $m=1$ or $m=5$ mod 6, and gcd$(m,2)=1$} \,,\\
T_m^{\:\!(3)}&=\langle A, X_4, Y_{6m}C\rangle && \text{with $m=3$ mod 6, and gcd$(m,2)=1$}\,, \\
O_m&= \langle A,C, X_8, Y_{2m} \rangle && \text{with gcd$(m,6)=1$}\,, \\
I_m&=\langle B,D,  X_5^3, Y_{2m} \rangle && \text{with gcd$(m,30)=1$}\,,
\ea \ee
where notation $\langle A,B,\dots \rangle$ denotes the group generated by $A,B,\dots$. See table \ref{tab:NonAbData} for some of their properties. The non-trivial defect group computes from $ \text{Ab}(\Gamma/\Gamma_{\text{fix}})$ following \eqref{eq:DefectGpNonAb}. The 1-form anomalies compute from a sum over elements in $\text{FF}(\Gamma)$ following \eqref{eq:EtaInvNonAb}.

\begin{table}[]
\hspace{-45pt}
{\renewcommand*{\arraystretch}{1.35}
\begin{tabular}{||c| c  ||   c| c | c || }
 \hline
 $\Gamma$ &  $\text{Ab}(\Gamma/\Gamma_{\text{fix}})$  & $\raisebox{0.05ex}{$\frac{1}{2}\eta^{\text{\:\!$\slashed{D}$}}_{\raisebox{-0.3ex}{\scriptsize$L_\rho$}}\!(S^5\!/\:\!\Z_N)$}$  &  $\beta_\rho$ & $\gamma_\rho$  \\ [0.5ex]
 \hline\hline
 $D_{n,q=1}^{(m\,\text{odd})}$ & $\Z_m$ &$ \frac{k^2-4 k+3}{3 (2 k-1)}\quad m=2k-1$   & $\frac{1}{6(2k-1)}\quad m=2k-1$  &  $\frac{2 k^2-4 k+5}{6(2 k-1)}\quad m=2k-1$   \\[0.5ex]
 \hline

  $D_{n,q=1}^{(m\,\text{even})}$ & $\Z_{2m}$ & $\begin{array}{cl} \frac{2 k^2-6 k+1}{12 k} & m=2k \\ \frac{k^2-2 k}{3 (2 k+1)} & m=2k+1 \end{array} $  &  $\begin{array}{cl}\frac{1}{12k} & m=2k \\ 0 & m=2k+1 \end{array} $ &   $\begin{array}{cl} \frac{2k^2-6k}{12k} & m=2k \\ \frac{k^2-2k}{3(2k+1)} & m=2k+1 \end{array} $    \\[0.5ex]
 \hline
$T_m^{\:\!(1,5)}$  & $\Z_m$ & $\begin{array}{cl} \frac{k^2-5 k+2}{2 (6 k-1)} & m=6k-1 \\ \frac{k^2-7k}{2 (6 k+1)}  & m=6k+1 \end{array} $ &  $\begin{array}{cl} \frac{k}{6(6 k-1)}& m=6k-1 \\ \frac{5 k+1}{6 (6k+1)} & m=6k+1 \end{array} $  & $\begin{array}{cl} \frac{3 k^2-16k+6}{6 (6k-1)} & m=6k-1 \\ \frac{3 k^2-26 k-1}{6 (6 k+1)} & m=6k+1 \end{array} $   \\[0.5ex]
       \hline
       $T_m^{\:\!(3)}$ & $\Z_{3m}$ & $ \frac{9 k^2-63 k+32}{18 (6 k-3)} \quad m=6k-3$  & $ \frac{9 k-4}{18 (6 k-3)}\quad m=6k-3 $ &  $\frac{k^2-8 k+4}{2( 6k-3)} \quad m=6k-3$     \\[0.5ex]
       \hline
        $O_m$ & $\Z_m$ & $\begin{array}{cc} {\footnotesize -\frac{2}{5}, -\frac{2}{7},- \frac{2}{11},- \frac{7}{13}, {\footnotesize \text{$\dots$}} } \\ {\footnotesize \text{~~~for $m=5,7,11,13,\dots$ }} \end{array} $  & $ \begin{array}{cc} {\footnotesize + \frac{1}{10},+\frac{1}{14},+\frac{1}{66},+\frac{2}{13} , {\footnotesize \text{$\dots$}} } \\ {\footnotesize \text{~~~~~\,for $m=5,7,11,13,\dots$ }} \end{array} $ &  $ \begin{array}{cc} {\footnotesize -\frac{1}{2},-\frac{5}{14},-\frac{13}{66},-\frac{9}{13}, {\footnotesize \text{$\dots$}} } \\ {\footnotesize \text{~~~~for $m=5,7,11,13,\dots$ }} \end{array} $    \\[0.5ex]
       \hline
       $I_m$ &  $\Z_m$ & $\begin{array}{cc} -\frac{3}{7}, -\frac{6}{11},-\frac{6}{13}, -\frac{6}{17}, {\footnotesize \text{$\dots$}}  \\ {\footnotesize ~~\, \text{for $m=7,11,13,17, \dots$ }} \end{array} $ &  $\begin{array}{cc}  + \frac{2}{21},+\frac{7}{66},+\frac{5}{39},+\frac{2}{51}, {\footnotesize \text{$\dots$}}  \\ {\footnotesize ~~~~ \text{for $m=7,11,13,17, \dots$ }} \end{array} $  &  $\begin{array}{cc}  -\frac{11}{21},-\frac{43}{66},-\frac{23}{39},-\frac{20}{51}, {\footnotesize \text{$\dots$}}  \\ {\footnotesize ~~~~ \text{for $m=7,11,13,17, \dots$ }} \end{array} $     \\[0.5ex]
       \hline
\end{tabular}
}
\caption{ $\eta$\:\!-invariant computations and 1-form anomaly computations for small $\Gamma\subset SU(3)$. }
\label{tab:NonAbAnomalies}
\end{table}

We present our anomaly computations in table \ref{tab:NonAbAnomalies}. Here, we follow the steps outline in section \ref{sec:eta} and compute with respect to following representation $\rho$. Given the element $\gamma\in\Gamma$ (for $\Gamma\neq T_m^{\:\!(3)},D_{n,q}^{(m\,\text{even})}$) we represent it as a word
\be
\gamma=Y_{M}^{\:\!y} \dots B^b A^a \,,
\ee
with integers $y,\dots,b,a$ and appropriate $M$. Then, we set $\rho(\gamma)=\omega^{\:\!y}$  with
\be
\omega=\exp(2\pi i/|\text{Ab}(\Gamma/\Gamma_{\text{fix}})|)\,,
\ee
which is the representation induced by a representation of the Abelianization of $\Gamma/\Gamma_{\text{fix}}$ which sends the fixed point free generator $Y_{M}$ to $\omega$. For $\Gamma= T_m^{\:\!(3)}, D_{n,q}^{(m\,\text{even})}$ we proceed identically with powers of $Y_{6m}C, Y_{4m}A$ counted by $y$ respectively. Unfortunately, we do not find closed form expressions for the families $O_m,I_m$. The $\eta$\:\!-invariant relates to the $\beta$'s and $\gamma$'s as \eqref{eq:betagamma}.

Finally, we discuss Abelian non-cyclic quotients by groups $\Gamma\cong \Z_N\times \Z_M$ with $M>1$ dividing $N$ (every Abelian subgroup of $SU(3)$ must be of this type \cite{Ludl:2011gn}). Such actions admits a canonical parametrization with weights $\Z_N(m_1,m_2,m_3)$ and $\Z_M(0,1,M-1)$ and gcd$(N,m_1)$ and $M$ coprime.

In this parametrization, all elements of the $\Z_M$ subgroup generated by $(0,1)$ fix subsets of the boundary. As such, the defect group and higher symmetry groups of the 5D SCFT $\mathcal{T}_X$ engineered by $\mathbb{C}^3/\Gamma$ in M-theory are determined by the $\Z_N$ factor, placing us largely into the setup of section \ref{sec:NonIsolatedCyclic}. Our discussion there can then essentially be carried over, up to a minor technical detail in computing the 1-form self- and the mixed 1-form-gravitational anomaly.

\begin{table}[]
{\renewcommand*{\arraystretch}{1.35}

\hspace{-40pt}\begin{tabular}{||c ||   c| c | c || c|| c || }
 \hline
 $\begin{array}{cc} \Gamma \cong \Z_N(m_1,m_2,m_3) \\ ~~\;\!\times \Z_M(0,1,M-1)\end{array}$ &  $Q$ & $\raisebox{0.05ex}{$\frac{1}{2}\eta^{\text{\:\!$\slashed{D}$}}_{\raisebox{-0.3ex}{\scriptsize$L_Q$}}\!(S^5\!/\:\!\Z_N\times \Z_M)$}$   & $\raisebox{0.05ex}{$\frac{1}{2}\eta^{\text{\:\!$\slashed{D}$}}_{\raisebox{-0.3ex}{\scriptsize$L_{(1,0)}$}}\!(S^5\!/\:\!\Z_N\times \Z_M)$}$  \\ [0.5ex]
 \hline
$\begin{array}{cc} \Z_{2^{n+1}}(1,2^{n+1}-3,2) \\ \times\, \Z_2(0,1,1)\end{array}$ & $(4,0)$  & $\frac{1}{9}(-2^{n-1}+2^{-n+2}+(-1)^n)$    & $\frac{1}{9}(-2^{n-3}+5 \times 2^{-n-1}+(-1)^n)$ \\[0.5ex]
 \hline
  $\begin{array}{cc}  \Z_{3^{n+1}}(1,3^{n+1}-4,3)\\ \times\, \Z_3(0,1,2)\end{array}$ & $(9,0)$  & $ \frac{1}{16}(-3^n+3^{-n+3}+6(-1)^n)$   & $\frac{1}{16}(-3^{n-2}-5\times 3^{-n}-2(-1)^n)$  \\[0.5ex]
 \hline
$\begin{array}{cc}  \Z_{5^{n+1}}(1,5^{n+1}-6,5)\\ \times\, \Z_5(0,1,4)\end{array}$ & $(25,0)$  & $\frac{1}{72}(-5^{n+1}+5^{-n+4}+100(-1)^n)$  & $\frac{1}{72}(-5^{n-1}-2347 \times 5^{-n-1}-92(-1)^n)$   \\[0.5ex]
       \hline
\end{tabular}
}
\caption{Examples of $\eta$-invariants for groups $\Gamma\cong \Z_N\times \Z_M$ relevant to the pure 1-form anomaly and the mixed 1-form gravitational anomaly of 5D SCFTs engineered by M-theory on $\mathbb{C}^3/\Gamma$. The family $\Gamma=\Z_{2^{n+1}}\times \Z_2$ has line defects $\mathbb{D}^{(1)}_{\text{elec}}\cong\Z_{2^{n-1}}$ with $\Gamma_{\text{fix}}\cong \Z_{4} \times \Z_2$, $|\Gamma_{\text{fix}}|=8$, $|\text{FF}(\Gamma)|=2^{n+2}-6$. The family $\Gamma=\Z_{3^{n+1}}\times \Z_3$ has line defects $\mathbb{D}^{(1)}_{\text{elec}}\cong\Z_{3^{n-1}}$ with $\Gamma_{\text{fix}}\cong \Z_{9} \times \Z_3$, $|\Gamma_{\text{fix}}|=27$, $|\text{FF}(\Gamma)|=3^{n+2}-13$. The family $\Gamma=\Z_{5^{n+1}}\times \Z_5$ has line defects $\mathbb{D}^{(1)}_{\text{elec}}\cong\Z_{5^{n-1}}$ with $\Gamma_{\text{fix}}\cong \Z_{25} \times \Z_5$, $|\Gamma_{\text{fix}}|=125$, $|\text{FF}(\Gamma)|=5^{n+2}-33$. All of these expressions are quantized as expected, e.g., consider the bottom right entry $\frac{1}{2}\eta^{\text{\:\!$\protect\slashed{D}$}}_{\raisebox{-0.3ex}{\scriptsize$L_{(1,0)}$}}\!(S^5\!/\:\!\Z_{5^{n+1}}\times \Z_5)=-\frac{1}{25},-\frac{201}{125},+\frac{549}{625},-\frac{9451}{3125},\dots$ for $n=1,2,3,4,\dots$ with $\Gamma\cong \Z_{5^{n+1}}\times \Z_5$, i.e., denominators are $5^{n+1}$, which is the order of the line defect group $|\mathbb{D}^{(1)}_{\text{elec}}|=5^{n-1}$ refined by the first factor of $\Gamma_{\text{fix}}\cong \Z_{25}\times \Z_5$. A few amusing features of the entries of the last column are that $2347$ is prime, and $92 = 2^2 \times 23$.
}
\label{tab:TableForNonIsolated2}
\end{table}

This detail concerns the evaluation of the $\eta$-invariant \eqref{eq:Degeratu}. Instead of \eqref{eq:EtaFixed}, the sum runs now over $|\text{FF}(\Gamma)|$ elements which we now count. First, introduce $M'=N/M$. Then, the $k$-th circle $S^1_k\subset S^5$ cut out by $z_i,z_j=0$ with $\{i,j,k\}=\{1,2,3\}$ is fixed by the subgroup $\Z_{M\text{gcd}(M',m_1)}\subset \Gamma$ (i.e., there are always three flavor branes \cite{Cvetic:2022imb}). These groups only share the identity element of $\Gamma$. If any one group element were to fix two such circles, it would fix a full 3-sphere containing these, this however, would be in contradiction with the orbifold $\mathbb{C}^3/\Gamma$ being Calabi-Yau, which constrains singularities to be codimension-6 or codimension-4 and not codimension-2. With this we count
\be
|\text{FF}(\Gamma)|=NM-1-\sum_{k=1}^3 ( M\text{gcd}(M',m_k)-1)=M\lb N-\sum_{k=1}^3 \text{gcd}(M',m_k)  \rb+2\,.
\ee
From here, the $\eta$-invariant \eqref{eq:Degeratu} and anomalies $\beta,\gamma$ are straightforwardly computed. We give examples in table \ref{tab:TableForNonIsolated2} (with notation as in table \ref{tab:TableForNonIsolated}). In all cases, the flavor algebra is
\be\label{eq:Singularities}
\mathfrak{f}=\bigoplus_{k=1}^3\mathfrak{su}(M\text{gcd}(M',m_k))\,.
\ee
In the absence of Kawasaki's canonical generator, we compute here using the line bundle $L_Q$ associated to the charge $Q=N/|\Gamma/\Gamma_{\text{fix}}|$ representation of $\Z_N$ and refinements are given with respect to $Q=1$. In particular, as the $\Z_M$ factor does not affect 1-form symmetries or line defects, we set all representations of $\Z_M$ to the trivial representation.

\begin{table}[]
{\renewcommand*{\arraystretch}{1.35}

\hspace{-25pt}\begin{tabular}{||c ||   c| c | c || c|| c || }
\hline
       $\begin{array}{cc} \Gamma \cong \Z_N(m_1,m_2,m_3) \\ ~~\;\!\times \Z_M(0,1,M-1)\end{array}$ & $Q$&  ${\beta}_{Q}$   & ${\gamma}_{Q}$  \\ [0.5ex]
 \hline
$\begin{array}{cc} \Z_{2^{n+1}}(1,2^{n+1}-3,2) \\ \times\, \Z_2(0,1,1)\end{array}$ & $(4,0)$ &   $\frac{1}{9} \left(-2^{2-n}-\frac{(-1)^n}{2}+\frac{3}{2}\right)$    & $\frac{1}{9} \left(-2^{n-1}+2^{3-n}+ \frac{3 (-1)^n}{2}-\frac{3}{2}\right)$ \\[0.5ex]
 \hline
  $\begin{array}{cc}  \Z_{3^{n+1}}(1,3^{n+1}-4,3)\\ \times\, \Z_3(0,1,2)\end{array}$ & $(9,0)$  & $ \frac{1}{8} \left(-3^{2-n}-(-1)^n+2\right)$   & $\frac{1}{16} \left(8 (-1)^n+5\times 3^{2-n}-3^n-4\right)$  \\[0.5ex]
 \hline
$\begin{array}{cc}  \Z_{5^{n+1}}(1,5^{n+1}-6,5)\\ \times\, \Z_5(0,1,4)\end{array}$ & $(25,0)$  & $\frac{1}{36} \left(-5^{3-n}-10 (-1)^n+15\right)$  & $\frac{5}{72} \left(24 (-1)^n+7\times 5^{-n+2}-5^n-6\right)$   \\[0.5ex]
       \hline
\end{tabular}
}
\caption{Table \ref{tab:TableForNonIsolated2} continued. We list the 1-form anomaly data for some families $\Gamma\cong\Z_N\times \Z_M$.
}
\label{tab:TableForNonIsolated3}
\end{table}

Finally, we turn to generalize the refinement discussion given in subsection \ref{ssec:RefinedAnomaly} for the cyclic case $\Gamma$. We begin by discussing the flavor symmetry anomalies in \eqref{eq:WZterm} and \eqref{eq:WZterm2}. The term \eqref{eq:WZterm} makes sense for all ADE types as basic invariant polynomials of degree 2 exists for all simply laced Lie algebras. In contrast, the term \eqref{eq:WZterm2} relies on the existence of an invariant degree 3 polynomial, which only exists as $\text{Tr} F^3$ for A-type Lie algebras \cite{076a57d2-9415-3f82-b417-586ee96e50e3}, and therefore such terms, and consequently anomalies $\epsilon_k$, must be absent for flavor branes of type D and E.

Finally note that the natural generalization of \eqref{eq:Unrestricted} is
\be\label{eq:Unrestricted2}
\eta^{\text{\:\!$\slashed{D}$}}(S^5/\Gamma)\,:\quad  \text{Rep}(\Gamma)\rightarrow \Q/\Z\,, \qquad  R \mapsto \frac{1}{2}\eta^{\text{\:\!$\slashed{D}$}}_{\raisebox{-0.3ex}{\scriptsize$\mathscr{V}_R$}}\quad \text{mod}\,1\,,
\ee
where $\mathscr{V}_{R}=(R\times S^5)/\Gamma$ is an orbi-bundle. Instead of a single variable $Q$ as in the cyclic case (see, e.g., \eqref{eq:alphaDiracAp4}), we expect that the function \eqref{eq:Unrestricted2} may be expressed as quasi-polynomial with respect to as many variables as $\text{Rep}(\Gamma)$ has generators and coefficients in this polynomial carry anomaly data. We leave a detailed investigation of this structure to future work.

\section{Conclusions}
\label{sec:Conc}

In this paper we studied anomalies of 5D, 6D, 7D theories engineered by orbifolds $X=\mathbb{C}^n/\Gamma$ in string and M-theory. We determined these as functions of $\partial X=S^{2n-1}/\Gamma$ by expressing them using $\eta$\:\!-invariants.

Concretely, for all 5D SCFTs constructed by the Calabi-Yau orbifolds $\mathbb{C}^3/\Gamma$ in M-theory, we computed, for the discrete 1-form symmetry
\be
A\cong \text{Hom}( \text{Ab}(\Gamma/\Gamma_{\text{fix}}),U(1))\cong  \text{Ab}(\Gamma/\Gamma_{\text{fix}})^\vee\,,
\ee
the pure 1-form anomaly and mixed 1-form-gravitational anomaly as recorded by the SymTFT terms
 \be\label{eq:Anom2345234}\ba
\mathcal{S}_{\;\!\text{SymTFT}}^{(\text{anom})}&= 2\pi  \int_{\text{6D}}  \lb \beta_{L_Q} B_2^3+ \gamma_{L_Q} B_{2\:\!}  p_1/4  \rb+\dots \qquad \text{mod 1}\,.
\ea\ee
Here, $B_2$ is the 1-form symmetry background and $p_1$ the first Pontryagin class of spacetime and the anomalies were evaluated to
\be\label{eq:betagamma22}\ba
\beta_{L_Q}&=+\frac{1}{12}\eta^{\text{\:\!$\slashed{D}$}}_{\raisebox{-0.3ex}{\scriptsize$L_{2Q}$}}(\partial X) -\frac{1}{6} \eta^{\text{\:\!$\slashed{D}$}}_{\raisebox{-0.3ex}{\scriptsize$L_{Q}$}}(\partial X)   \qquad \,\;\!\text{mod}\,1 \,,\\[0.5em]
\gamma_{L_Q}&= -\frac{1}{12}\eta^{\text{\:\!$\slashed{D}$}}_{\raisebox{-0.3ex}{\scriptsize$L_{2Q}$}}(\partial X)  + \frac{2}{3} \eta^{\text{\:\!$\slashed{D}$}}_{\raisebox{-0.3ex}{\scriptsize$L_{Q}$}}(\partial X)  \qquad\; \text{mod}\,1 \,.
\ea \ee

\noindent The extra-dimensional line bundle $L_Q$ over $\partial X$ determines an internal basis for a dimensional reduction scheme with respect to which SymTFT terms we computed starting from topological terms of 11D supergravity. We matched these results with intersection computation which constitute the previous standard in the literature.

We provided evidence and conjectured that full set of twisted $\eta$\:\!-invariants of the Dirac operator mod 1, which may be collected into the function
\be\label{eq:NotQuiteK}
\eta^{\text{\:\!$\slashed{D}$}}(S^5/\Gamma)\,:\quad \text{Rep}(\Gamma)\rightarrow \Q/\Z\,, \qquad R \mapsto \frac{1}{2}\eta^{\text{\:\!$\slashed{D}$}}_{\raisebox{-0.3ex}{\scriptsize$\mathscr{V}_{R}$}}(S^5/\Gamma) \quad \text{mod}\,1\,,
\ee
on the set of orbi-bundles $\mathscr{V}_{R}=(R\times S^5)/\Gamma$, determines all anomalies between to the 1-form symmetry and non-Abelian flavor symmetries of the system. The obvious K-theoretic generalization of \eqref{eq:NotQuiteK}, when $X$ is not a global quotient, is
\be\label{eq:K}
\eta^{\text{\:\!$\slashed{D}$}}(\partial X)\,:\quad K^0(\partial X)\rightarrow \Q/\Z\,, \qquad \mathscr{V} \mapsto \frac{1}{2}\eta^{\text{\:\!$\slashed{D}$}}_{\raisebox{-0.3ex}{\scriptsize$\mathscr{V}$}}(\partial X) \quad \text{mod}\,1\,.
\ee
This gave rise to an interesting open question concerning how heavily a SCFT is constrained by its generalized symmetries is:

\noindent Among the set of 5D SCFTs with extra-dimensional construction in M-theory on a Calabi-Yau three-fold $X$, does the function $\eta^{\text{\:\!$\slashed{D}$}}(\partial X)$ uniquely determine a 5D SCFT?

The precise interpretation of $\eta^{\text{\:\!$\slashed{D}$}}(\partial X)$ away from the subring of vector bundles also requires more attention and technical clarification. It is interesting to note that $\eta$\:\!-invariants modulo 1 are bordism invariants and it would be of interest to connect the anomaly considerations in this note to the study of generalized charges as characterized by bordism groups \cite{McNamara:2019rup,  Montero:2020icj, Dierigl:2020lai, Buratti:2021yia, Debray:2021vob, Andriot:2022mri, Blumenhagen:2022bvh, Velazquez:2022eco, Debray:2023yrs, Braeger:2025kra, Chakrabhavi:2025bfi}.

Further, this paper demonstrated (perhaps unsurprisingly), that anomaly structures in geometric engineering can require refinements beyond singular cohomology, and are perhaps better formulated in K-theory. Streamlining this perspective in terms of a reduction procedure in the context of extra-dimensional approaches to symmetry theories is expected to be extremely useful, especially when studying theories in low dimensions, for which the engineering geometries become complicated. In such cases the distinctions between different cohomology theories will likely be more prominent. 
This will be addressed in future work \cite{WIP2}.

\newpage

\section*{Acknowledgements}

We thank C. C\'ordova, V. Chakrabhavi, M. Del Zotto, D. Freed, S. Meynet, V. Nevoa, S. Raman, Jiahua Tian, Yi-Nan Wang, and X. Yu for helpful discussions.
MH thanks the workshop ``Generalized Symmetries in HEP and CM" at Peking University, the 22nd Simons Summer Workshop at the Simons Center for Geometry and Physics, the Center for Mathematical Sciences and Applications at Harvard University and the High Energy Theory Group at Harvard for hospitality and support during stages of this work. The work of MC is supported by DOE (HEP)
Award DE-SC0013528, the Slovenian Research Agency
(ARRS No. P1-0306) and Fay R. and Eugene L. Langberg Endowed Chair funds. MC thanks the High Energy Theory Group at Harvard University for hospitality during final stages of this work.
The research of RD was partially supported by NSF grant DMS–2244978, FRG:
Collaborative Research: New birational invariants; by NSF grant DMS–2401422; and by
Simons Foundation Collaboration grant \#390287 ``Homological Mirror Symmetry.''
The work of JJH is supported by DOE (HEP) Award DE-SC0013528 as well as by BSF grant 2022100. RD thanks the Center for Mathematical Sciences and Applications at Harvard University for hospitality during final stages of this work. The
work of JJH is also supported in part by a University Research Foundation grant at the
University of Pennsylvania. The work of MH is supported by the Marie Skłodowska-Curie Actions under the European Union’s
Horizon 2020 research and innovation programme, grant agreement number \#101109804.
MH acknowledges support from the the VR Centre for Geometry and Physics (VR grant
No. 2022-06593).

\appendix

\section{Matching Bulk and Boundary Computations}
\label{sec:Check}

In this appendix we give examples of bulk intersection computations for 1-form anomalies of 5D SCFTs engineered by M-theory $X=\mathbb{C}^3/\Z_N$.  These computation functions as checks and are compared to the $\eta$\:\!-invariant computations with which they ultimately agree. We have evaluated $O(500)$ examples of the type $\mathbb{C}^3/\Z_N$ via \texttt{Mathematica}, both with isolated and non-isolated singularities, similar to the two representative examples below.

\subsection{Isolated Singularity Example: $\mathbb{C}^3/\Z_{2n+1}(1,1,2n-1)$}
\label{sec:Check1}

General examples quickly become unwieldy and therefore we spell out details only for the particularly favorable case of the 1-parameter family of orbifolds $\mathbb{C}^3/\Z_{2n+1}(1,1,2n-1)$. We will explicitly demonstrate that the bulk computation agrees with the boundary $\eta$\:\!-invariant computation and give the computations producing the first row in table \ref{tab:TableForIsolated}.

\begin{figure}
\centering
\scalebox{0.7}{
\begin{tikzpicture}
	\begin{pgfonlayer}{nodelayer}
		\node [style=none] (0) at (0, 1) {};
		\node [style=none] (1) at (-1, 0) {};
		\node [style=none] (2) at (1, -1) {};
		\node [style=none] (3) at (-2.5, 0.5) {$n=1$};
		\node [style=SmallCircle] (7) at (0, 0) {};
		\node [style=SmallCircleBrown] (8) at (0, 1) {};
		\node [style=SmallCircleBrown] (9) at (1, -1) {};
		\node [style=SmallCircleBrown] (10) at (-1, 0) {};
		\node [style=none] (11) at (0, -2) {};
		\node [style=none] (12) at (-1, -3) {};
		\node [style=none] (13) at (3, -4) {};
		\node [style=SmallCircle] (18) at (0, -3) {};
		\node [style=SmallCircleBrown] (19) at (0, -2) {};
		\node [style=SmallCircleBrown] (20) at (3, -4) {};
		\node [style=SmallCircleBrown] (21) at (-1, -3) {};
		\node [style=SmallCircle] (22) at (1, -3) {};
		\node [style=none] (23) at (0, -5) {};
		\node [style=none] (24) at (-1, -6) {};
		\node [style=none] (25) at (5, -7) {};
		\node [style=SmallCircle] (26) at (0, -6) {};
		\node [style=SmallCircleBrown] (27) at (0, -5) {};
		\node [style=SmallCircleBrown] (28) at (5, -7) {};
		\node [style=SmallCircleBrown] (29) at (-1, -6) {};
		\node [style=SmallCircle] (30) at (1, -6) {};
		\node [style=SmallCircle] (31) at (2, -6) {};
		\node [style=none] (32) at (0, -8) {};
		\node [style=none] (33) at (-1, -9) {};
		\node [style=none] (34) at (7, -10) {};
		\node [style=SmallCircle] (35) at (0, -9) {};
		\node [style=SmallCircleBrown] (36) at (0, -8) {};
		\node [style=SmallCircleBrown] (37) at (7, -10) {};
		\node [style=SmallCircleBrown] (38) at (-1, -9) {};
		\node [style=SmallCircle] (39) at (1, -9) {};
		\node [style=SmallCircle] (40) at (2, -9) {};
		\node [style=SmallCircle] (41) at (3, -9) {};
		\node [style=none] (42) at (-2.5, -2.5) {$n=2$};
		\node [style=none] (43) at (-2.5, -5.5) {$n=3$};
		\node [style=none] (44) at (-2.5, -8.5) {$n=4$};
		\node [style=none] (45) at (-1.5, -0.5) {$D_{z_3}$};
		\node [style=none] (46) at (-1.5, -3.5) {$D_{z_3}$};
		\node [style=none] (47) at (-1.5, -6.5) {$D_{z_3}$};
		\node [style=none] (48) at (-1.5, -9.5) {$D_{z_3}$};
		\node [style=none] (49) at (0, -4.5) {$D_{z_1}$};
		\node [style=none] (50) at (0, -1.5) {$D_{z_1}$};
		\node [style=none] (51) at (0, 1.5) {$D_{z_1}$};
		\node [style=none] (52) at (0, -7.5) {$D_{z_1}$};
		\node [style=none] (53) at (1.5, -1) {$D_{z_2}$};
		\node [style=none] (54) at (3.5, -4) {$D_{z_2}$};
		\node [style=none] (55) at (5.5, -7) {$D_{z_2}$};
		\node [style=none] (56) at (7.5, -10) {$D_{z_2}$};
		\node [style=none] (57) at (0, -10.75) {};
		\node [style=none] (58) at (8, -2) {};
		\node [style=none] (59) at (9, -2) {};
		\node [style=none] (60) at (8, -3) {};
		\node [style=none] (61) at (9, -1.5) {$y$};
		\node [style=none] (62) at (7.5, -3) {$x$};
	\end{pgfonlayer}
	\begin{pgfonlayer}{edgelayer}
		\draw [style=ThickLine] (0.center) to (2.center);
		\draw [style=ThickLine] (2.center) to (1.center);
		\draw [style=ThickLine] (1.center) to (0.center);
		\draw [style=ThickLine] (11.center) to (13.center);
		\draw [style=ThickLine] (13.center) to (12.center);
		\draw [style=ThickLine] (12.center) to (11.center);
		\draw [style=ThickLine] (23.center) to (25.center);
		\draw [style=ThickLine] (25.center) to (24.center);
		\draw [style=ThickLine] (24.center) to (23.center);
		\draw [style=ThickLine] (32.center) to (34.center);
		\draw [style=ThickLine] (34.center) to (33.center);
		\draw [style=ThickLine] (33.center) to (32.center);
		\draw (10) to (7);
		\draw (7) to (8);
		\draw (7) to (9);
		\draw (19) to (18);
		\draw (18) to (21);
		\draw (18) to (20);
		\draw (22) to (20);
		\draw (22) to (18);
		\draw (22) to (19);
		\draw (27) to (26);
		\draw (26) to (29);
		\draw (27) to (30);
		\draw (27) to (31);
		\draw (26) to (30);
		\draw (30) to (28);
		\draw (31) to (28);
		\draw (31) to (30);
		\draw (36) to (35);
		\draw (35) to (38);
		\draw (35) to (41);
		\draw (41) to (37);
		\draw (40) to (37);
		\draw (39) to (37);
		\draw (39) to (36);
		\draw (36) to (40);
		\draw (41) to (36);
		\draw [style=ArrowLineRight] (58.center) to (59.center);
		\draw [style=ArrowLineRight] (58.center) to (60.center);
	\end{pgfonlayer}
\end{tikzpicture}
}
\caption{Triangulated and refined toric diagrams for the orbifold $\mathbb{C}^3/\Z_{2n+1}$  with an action carrying weights $(1,1,2n-1)$. We give the triangulated diagrams for the first four cases $n=1,2,3,4$ associated with the crepant resolution. The black dots indicate $n$ compact toric divisors, located at $(x,y)=(0,0), \dots , (0,n-1)$. The 3 brown dots indicate the non-compact divisors $z_i=0$ denoted $D_{z_i}$ located at $(-1,0),(1,2n-1),(0,-1)$ for $i=1,2,3$ respectively. }
\label{fig:toricDs}
\end{figure}
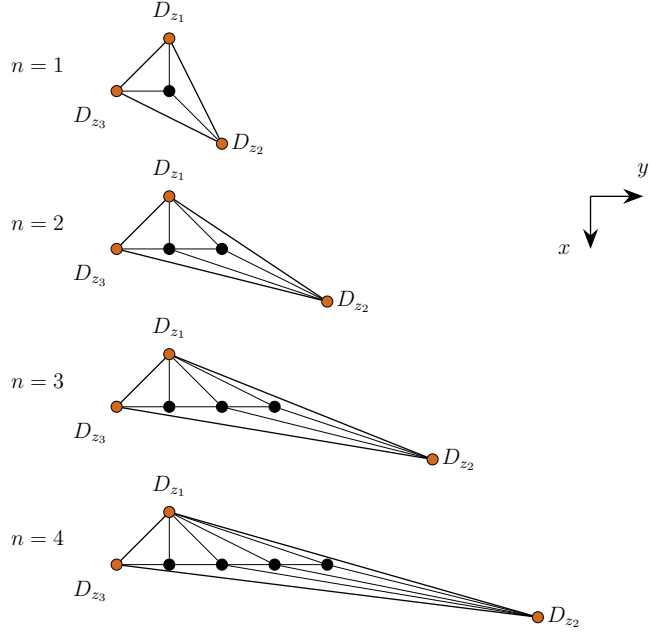

To proceed then, take $Z$ to be the crepant resolution of $\mathbb{C}^3/\Z_{2n+1}(1,1,2n-1)$. See figure \ref{fig:toricDs} for the triangulated toric diagrams corresponding to $n=1,2,3,4$. The bulk integral \eqref{eq:bulkint} for $\alpha$ will now be evaluated using the intersection ring of $Z$ which we therefore determine first via standard methods from toric geometry.

To begin, note that under toric resolution of $\mathbb{C}^3/\Z_{2n+1}$ with respect to the (in this case unique) triangulation depicted in  figure \ref{fig:toricDs}, one has that the isolated codimension-6 singularity is replaced by a collection of $n$ complex surfaces:
\be \label{eq:Everything}
 \begin{tikzpicture}
	\begin{pgfonlayer}{nodelayer}
		\node [style=none] (0) at (0.5, 0) {$\mathbb{F}_5$};
		\node [style=none] (1) at (-2.25, 0) {$\mathbb{F}_7$};
		\node [style=none] (2) at (-6.5, 0) {$\mathbb{F}_{2n-1}$};
		\node [style=none] (3) at (3.25, 0) {$\mathbb{F}_3$};
		\node [style=none] (4) at (6, 0) {$\P^2$};
		\node [style=none] (5) at (-1.75, 0) {};
		\node [style=none] (6) at (-2.75, 0) {};
		\node [style=none] (7) at (0, 0) {};
		\node [style=none] (8) at (1, 0) {};
		\node [style=none] (9) at (2.75, 0) {};
		\node [style=none] (10) at (3.75, 0) {};
		\node [style=none] (11) at (5.5, 0) {};
		\node [style=none] (12) at (-5.75, 0) {};
		\node [style=none] (13) at (5.3, 0.315) {$h$};
		\node [style=none] (14) at (4, 0.3) {$b_3$};
		\node [style=none] (15) at (2.5, 0.3) {$d_3$};
		\node [style=none] (16) at (1.25, 0.3) {$b_5$};
		\node [style=none] (17) at (-0.25, 0.3) {$d_5$};
		\node [style=none] (18) at (-1.5, 0.3) {$b_7$};
		\node [style=none] (19) at (-3, 0.3) {$d_7$};
		\node [style=none] (20) at (-5.25, 0.3) {$b_{2n-1}$};
		\node [style=none] (21) at (-4.25, 0) {};
		\node [style=none] (22) at (-3.75, 0) {};
	\end{pgfonlayer}
	\begin{pgfonlayer}{edgelayer}
		\draw [style=ThickLine] (5.center) to (7.center);
		\draw [style=ThickLine] (8.center) to (9.center);
		\draw [style=ThickLine] (10.center) to (11.center);
		\draw [style=ThickLine] (12.center) to (21.center);
		\draw [style=ThickLine] (22.center) to (6.center);
		\draw [style=DottedLine] (21.center) to (22.center);
	\end{pgfonlayer}
\end{tikzpicture}
\ee
Here $\mathbb{F}_n$ denotes the $n$-th Hirzebruch surface whose homology lattice of 2-cycles is generated by the curves $b_n,f_n$ with intersections $b_n\cdot b_n=-n$ and $f_n\cdot f_n=0$ and $f_n\cdot b_n=1$. Further, we have introduced the combination $d_n=b_n+nf_n$ and denote the hyperplane class of $\mathbb{P}^2$ by $h$ with intersection $h\cdot h=1$.

The notation in \eqref{eq:Everything} then indicates a gluing of surfaces over curves, following the notation in \cite{Bhardwaj:2018yhy, Bhardwaj:2018vuu}. The compact divisors organizing into a linear chain, i.e., line \eqref{eq:Everything}, is what makes this example particularly favorable.

The toric resolution is made explicit as follows: Starting from the fan with vertices
\be
u_1 = (-1,0,1)\,, \qquad  u_2 = (1,2n-1,1)\,, \qquad u_3 = (0,-1,1)\,,
\ee
and the fully triangulated refined toric diagram in figure \ref{fig:toricDs} we note that besides the compact toric divisors $\P^2,\mathbb{F}_3,\dots, \mathbb{F}_{2n-1}$ there are also three non-compact toric divisors. In the unresolved geometry $\mathbb{C}^3/\Z_{2n+1}$ these are simply obtained by setting each coordinate $z_i$ to zero in turn. We therefore denote these by $D_{z_i}$ indicated in the vanishing coordinate as a subscript. Modulo $2n+1$ we have $u_1+u_2+(2n-1)u_3=0$ and therefore the $u_i$ are in correspondence with the $D_{z_i}$.

To proceed, we denote the coordinates associated with the $n$ compact exceptional toric divisors by $e_0,e_1,\dots, e_{n-1}=0$ respectively. The divisors are $D_{e_0},D_{e_1},\dots D_{e_{n-1}}$ respectively, were $D_{e_k}$ corresponds to the vector $v_k=(0,k,1)$. Explicitly, $D_{e_0}=\P^2$ and $D_{e_k}=\mathbb{F}_{2k+1}$ for $k=1,\dots, n-1$. Overall, proceeding with standard techniques,\footnote{See e.g., \cite{Greene:1996cy,Hori:2003ic,Denef:2008wq} for reviews directed towards physicists.} these toric divisors are found to be subject to the linear equivalences
\be \ba
D_{z_1}- D_{z_2}&=0\,, \\
(2n-1)D_{z_1}-D_{z_3}+\sum_{r=1}^{n-1}rD_{e_r}&=0\,,\\
D_{z_1}+D_{z_2}+D_{z_3}+\sum_{k=0}D_{e_k}&=0\,,
\ea \ee
which we solve for the 3 non-compact divisors to express these as a rational linear combination of compact divisors
\be \ba
D_{z_1}&= - \frac{1}{2n+1}\sum_{k=0}^{n-1} (k+1)D_{e_{k}}\\
D_{z_2}&= - \frac{1}{2n+1}\sum_{k=0}^{n-1} (k+1)D_{e_{k}} \\
D_{z_3}&= -\frac{1}{2n+1}\sum_{\ell=0}^{n-1} (2n-1-2\ell)D_{e_\ell}\,.
\ea\ee
With this we can compute all triple intersections between toric divisors from those of the compact divisors which are
\be\label{eq:TripInt} \ba
\mathbb{P}^2\cdot \mathbb{P}^2\cdot \mathbb{P}^2&=9\,, \qquad \qquad &&\mathbb{P}^2\cdot \mathbb{P}^2\cdot \mathbb{F}_3 =-3\,, &&\quad \mathbb{F}_{k+2} \cdot \mathbb{F}_{k+2} \cdot \mathbb{F}_{k} =k\,,\\
 \mathbb{F}_k \cdot \mathbb{F}_k  \cdot \mathbb{F}_k&=8\,, \qquad\qquad && \mathbb{F}_3 \cdot \mathbb{F}_3 \cdot \mathbb{P}^2 = 1 \,,  && \quad \mathbb{F}_{k-2} \cdot \mathbb{F}_{k-2} \cdot \mathbb{F}_{k} =-k \,.
\ea \ee

Given this intersection ring, we now compute the $\alpha$ for the family of 5D SCFTs engineered by $\mathbb{C}^3/\Z_{2n+1}(1,1,2n-1)$ in M-theory. With respect to the completely smooth $Z$ the bulk integral \eqref{eq:bulkint} is expressed as an intersection as
\be
\frac{D_{z_3}\cdot D_{z_3} \cdot D_{z_3}}{6} - \frac{p_1(Z) \cdot D_{z_3}}{24}=\frac{n(n+1)}{6(2n+1)}\qquad  \text{mod}\,1\,.
\ee
where $p_1(Z) $ is given, following \cite{7d3c9f5cc0d546b3a3c103df4d4cb4f5}:
\be
p_1(Z)= \sum_{i=1}^3 D_{z_i}\cdot D_{z_i}+ \sum_{k=0}^{n-1} D_{e_k} \cdot D_{e_k} \,.
\ee
The above expression disagrees with the first entry of table \ref{tab:TableForIsolated} (even after the shift $n\rightarrow n+1$). In general we can not expect that $D_{z_3}$ is dual to the Kawasaki generator of $\Z_{2n+1}$ with respect to which the first entry of table \ref{tab:TableForIsolated} was computed. To make this explicit we compute
\be\label{eq:ApBulk}
D_{z_3}\cdot D_{z_3} \cdot D_{z_3}=\frac{n+1}{2n+1}\qquad \text{mod}\, 1\,,
\ee
which we compare to the Kawasaki linking
\be\label{eq:ApKawa}
\ell(v_2^2,v_2)=\frac{(2n-1)^3}{(2n-1)(2n+1)}=\frac{4}{2n+1}\qquad \text{mod}\, 1\,,
\ee
and note that \eqref{eq:ApBulk} and \eqref{eq:ApKawa} differ by a factor of $8=2^3$, i.e., $s_3=2$ (see the paragraph below \eqref{eq:HoloMap} for a discussion of $s_3$). We now normalize into the Kawasaki basis $D_{z_3}\rightarrow 2D_{z_3}$ and compute
\be \label{eq:IntIso}
\frac{2D_{z_3}\cdot 2D_{z_3} \cdot 2D_{z_3}}{6} - \frac{p_1(Z) \cdot 2D_{z_3}}{24}=\frac{n(n-2)}{3(2n+1)}
\ee
which agrees with the first entry in table \ref{tab:TableForIsolated} after the shift $n\rightarrow n+1$. Next, to determine the remaining entries in the first row of table \ref{tab:TableForIsolated}, we compute the charge, following \eqref{eq:LinkingCharge}, to
\be
\ell(v_2,w_4)=\frac{(2n-1)(2n-1)}{(2n-1)(2n+1)}=\frac{-2}{2n+1}\qquad \text{mod}\,1\,.
\ee
therefore $q=-2$. With this charge we have, by \eqref{eq:alphaDirac},
\be\ba\label{eq:alphaDiracAp}
\frac{1}{2}\eta^{\text{\:\!$\slashed{D}$}}_{\raisebox{-0.3ex}{\scriptsize$L_q$}}&=\frac{1}{2n+1}\sum_{k=1}^{2n}(-1)^k\omega^{-2k}\frac{1}{\omega^{-k/2}-\omega^{k/2}}\frac{1}{\omega^{-k/2}-\omega^{k/2}}\frac{1}{\omega^{-k(2n-1)/2}-\omega^{k(2n-1)/2}}\\[0.4em]
&=\frac{n(n-2)}{3(2n+1)}
\ea\ee
with $\omega=\exp(2\pi i/(2n+1))$ which is in perfect agreement with the bulk intersection computations \eqref{eq:IntIso}.

Next, we proceed to discuss the anomalies $\beta,\gamma$ in greater detail, in particular their homogeneity properties. The goal will be to understand how for example redefinitions $B_2\rightarrow m B_2$ in the SymTFT action \eqref{eq:Anom} are correctly captured by the anomalies given in \eqref{eq:betagamma} where, given the expansion $G_4=B_2T_2+\dots$, these ought to amount to $\beta_{L_{Q}},\gamma_{L_Q}\rightarrow \beta_{L_{mQ}},\gamma_{L_{mQ}}$.

For this, we begin writing out the linear system resulting in \eqref{eq:betagamma}, which is
\be\ba
\text{Index}(\slashed{D},L_Q',Z)&=\int_{Z} \text{ch}(L'_Q)\hat{\text{A}}_{Z} -\frac{1}{2}\etaDirac{Q}(\partial Z)\\[0.4em]
\text{Index}(\slashed{D},L_{2Q}',Z)&=\int_{Z} \text{ch}(L'_{2Q})\hat{\text{A}}_{Z} -\frac{1}{2}\etaDirac{2Q}(\partial Z)\,.
\ea \ee
Then, expanding the integral according to the two terms in \eqref{eq:IntIso}, we have with respect to Kawasaki's generator
\be\ba
\text{Index}(\slashed{D},L_q',Z)&=\beta_{L_q'}+\gamma_{L_q'}-\frac{1}{2}\etaDirac{q}(\partial Z)\\
\text{Index}(\slashed{D},L_{2q}',Z)&= 8\beta_{L_q'}+2\gamma_{L_q'}-\frac{1}{2}\etaDirac{2q}(\partial Z)\,.\\
\ea\ee
Let us denote the indices respectively by  $I_{L_q'}(Z),I_{L_{2q}'}(Z)$ then we have
\be\label{eq:Z6ambg}\ba
\beta_{L_q'}&=+\frac{1}{12}\etaDirac{2q}(\partial Z)-\frac{1}{6}\etaDirac{q}(\partial Z)+\frac{I_{L_{2q}'}(Z)-2I_{L_q'}(Z)}{6}\,, \\[0.4em]
\gamma_{L_q'}&=-\frac{1}{12}\etaDirac{2q}(\partial Z)+\frac{2}{3}\etaDirac{q}(\partial Z)+\frac{8I_{L_q'}(Z)-I_{L_{2q}'}(Z)}{6}\,.
\ea \ee
The point we make regarding \eqref{eq:Z6ambg} is that we have
\be\label{eq:scalingrules}\ba
\beta_{L_{2q}'}=8\beta_{L_q'}\,, \qquad \gamma_{L_{2q}'}=2\gamma_{L_q'}\,, \\
\beta_{L_{2q}}\neq 8\beta_{L_q}\,, \qquad \gamma_{L_{2q}}\neq 2\gamma_{L_q}\,,
\ea \ee
with the second line as in \eqref{eq:betagamma}. Note the primes and recall that $L_Q'|_{\partial Z}=L_Q$. The first line in \eqref{eq:scalingrules} is obvious given the bulk expressions
\be\ba
\beta_{L_{sq}'}&=\frac{2sD_{z_3}\cdot 2sD_{z_3} \cdot 2sD_{z_3}}{6} =-\frac{4ns^3}{3(2n+1)}\,, \\[0.5em]
\gamma_{L_{sq}'}&= - \frac{p_1(Z) \cdot 2sD_{z_3}}{24}=\frac{n(n+2)s}{3(2n+1)}\,.
\ea \ee
The second line in \eqref{eq:scalingrules} are boundary expressions. These compute, for example, to (matching table \ref{tab:TableForIsolated} after the shift $n\rightarrow n+1$)
\be\ba
\beta_{L_{q}}&=-\frac{2n-3}{6(2n+1)}\,, \qquad \gamma_{L_{q}}=\frac{2n^2-2n-3}{6(2n+1)}\,,
\ea \ee
and for $\beta_{L_{2q}},\gamma_{L_{2q}}$ we find no closed form expressions (nonetheless they are straightforwardly computable using \eqref{eq:betagamma}). Clearly, bulk and boundary computations differ. For the anomalies given in \eqref{eq:scalingrules} we compute the following differences
\be\label{eq:difference1}\ba
\beta_{L_{q}}-\beta_{L_{q}'}&=\frac{1}{2},\frac{1}{2} ,\frac{1}{2} ,\frac{1}{2} ,\frac{1}{2} ,\frac{1}{2} ,\frac{1}{2} ,\dots \,,\qquad \\[0.4em]
\gamma_{L_{q}}-\gamma_{L_{q}'}&=-\frac{1}{2},-\frac{1}{2},-\frac{1}{2},-\frac{1}{2},-\frac{1}{2},-\frac{1}{2},-\frac{1}{2},\dots \,, \\
\ea \ee
and
\be\label{eq:difference2}\ba
\beta_{L_{2q}}-\beta_{L_{2q}'}&=\frac{7}{2},\frac{13}{3},\frac{14}{3},\frac{14}{3},\frac{14}{3},\frac{14}{3},\frac{14}{3},\dots  \,, \\[0.4em]
\gamma_{L_{2q}}-\gamma_{L_{2q}'}&=-\frac{1}{2},-\frac{4}{3},-\frac{5}{3},-\frac{5}{3},-\frac{5}{3},-\frac{5}{3},-\frac{5}{3},\dots\,,
\ea \ee
where we list out $n=1,\dots,7$. There are two conclusions to draw from these computations. First, all differences in equations \eqref{eq:difference1} and \eqref{eq:difference2} are valued in $\Z_6$, and are equal and opposite modulo 1. In this sense $\beta_{L_Q}$ and $\gamma_{L_Q}$ have a single common $\Z_6$ ambiguity between them (which is unphysical by \eqref{eq:Dirac}). Second, the boundary expressions \eqref{eq:betagamma} in general satisfy the homogeneity relations $\beta_{L_{mq}}= m^3\beta_{L_q}$ and $\gamma_{L_{mq}}= m\gamma_{L_q}$ only up to this $\Z_6$ ambiguity.

\subsection{Non-Isolated Singularity Example: $\mathbb{C}^3/\Z_{2n+2}(1,1,2n)$}
\label{sec:Check2}

We now turn to the 1-parameter family of orbifolds $\mathbb{C}^3/\Z_{2n+2}(1,1,2n)$. This class of example is again particularly simple due to exceptional divisors organizing in a linear chain.  We will explicitly demonstrate that the bulk computation agrees with the boundary $\eta$\:\!-invariant computation and verify the first entry in table \ref{tab:TableForNonIsolated}.

This example has an $\mathfrak{su}(2)$ ADE singularity supported on a plane $\mathbb{C}/\Z_n\subset \mathbb{C}^3/\Z_{2n+2}$ which intersects the link $S^5/\Z_{2n+2}$ in a circle $S^1/\Z_{n+1}$. The bulk intersection will be computed with respect to the fully resolved geometry. The comparison will however be with the $\eta$-invariant associated to the singular model. These are found to agree, via the index theory we therefore find that the $\eta$-invariant is invariant under resolutions induced by crepant resolutions of the bulk as in \eqref{eq:EtaInvInv}.

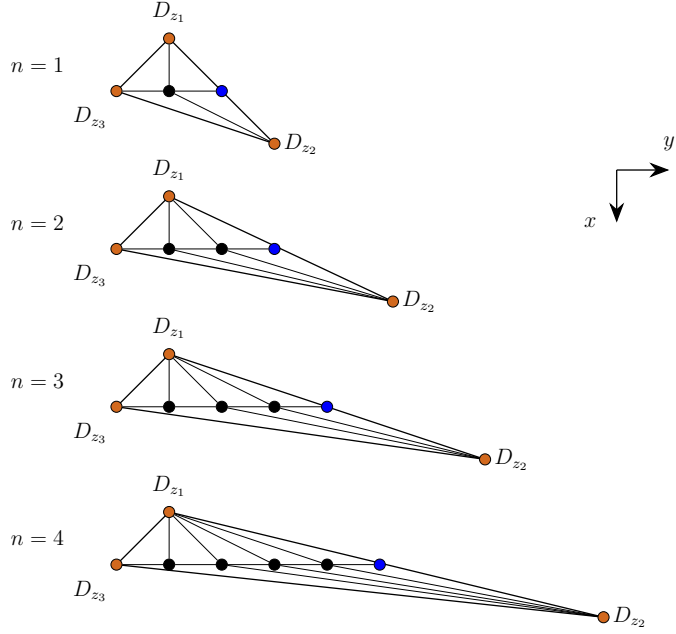
\begin{figure}[]
\centering
\scalebox{0.7}{
\begin{tikzpicture}
	\begin{pgfonlayer}{nodelayer}
		\node [style=none] (0) at (0, 4) {};
		\node [style=none] (1) at (-1, 3) {};
		\node [style=none] (2) at (2, 2) {};
		\node [style=none] (3) at (-2.5, 3.5) {$n=1$};
		\node [style=SmallCircle] (4) at (0, 3) {};
		\node [style=SmallCircleBrown] (5) at (0, 4) {};
		\node [style=SmallCircleBrown] (6) at (2, 2) {};
		\node [style=SmallCircleBrown] (7) at (-1, 3) {};
		\node [style=none] (8) at (0, 1) {};
		\node [style=none] (9) at (-1, 0) {};
		\node [style=none] (10) at (4.25, -1) {};
		\node [style=SmallCircle] (11) at (0, 0) {};
		\node [style=SmallCircleBrown] (12) at (0, 1) {};
		\node [style=SmallCircleBrown] (13) at (4.25, -1) {};
		\node [style=SmallCircleBrown] (14) at (-1, 0) {};
		\node [style=SmallCircle] (15) at (1, 0) {};
		\node [style=none] (16) at (0, -2) {};
		\node [style=none] (17) at (-1, -3) {};
		\node [style=none] (18) at (6, -4) {};
		\node [style=SmallCircle] (19) at (0, -3) {};
		\node [style=SmallCircleBrown] (20) at (0, -2) {};
		\node [style=SmallCircleBrown] (21) at (6, -4) {};
		\node [style=SmallCircleBrown] (22) at (-1, -3) {};
		\node [style=SmallCircle] (23) at (1, -3) {};
		\node [style=SmallCircle] (24) at (2, -3) {};
		\node [style=none] (25) at (0, -5) {};
		\node [style=none] (26) at (-1, -6) {};
		\node [style=none] (27) at (8.25, -7) {};
		\node [style=SmallCircle] (28) at (0, -6) {};
		\node [style=SmallCircleBrown] (29) at (0, -5) {};
		\node [style=SmallCircleBrown] (30) at (8.25, -7) {};
		\node [style=SmallCircleBrown] (31) at (-1, -6) {};
		\node [style=SmallCircle] (32) at (1, -6) {};
		\node [style=SmallCircle] (33) at (2, -6) {};
		\node [style=SmallCircle] (34) at (3, -6) {};
		\node [style=none] (35) at (-2.5, 0.5) {$n=2$};
		\node [style=none] (36) at (-2.5, -2.5) {$n=3$};
		\node [style=none] (37) at (-2.5, -5.5) {$n=4$};
		\node [style=none] (38) at (-1.5, 2.5) {$D_{z_3}$};
		\node [style=none] (39) at (-1.5, -0.5) {$D_{z_3}$};
		\node [style=none] (40) at (-1.5, -3.5) {$D_{z_3}$};
		\node [style=none] (41) at (-1.5, -6.5) {$D_{z_3}$};
		\node [style=none] (42) at (0, -1.5) {$D_{z_1}$};
		\node [style=none] (43) at (0, 1.5) {$D_{z_1}$};
		\node [style=none] (44) at (0, 4.5) {$D_{z_1}$};
		\node [style=none] (45) at (0, -4.5) {$D_{z_1}$};
		\node [style=none] (46) at (2.5, 2) {$D_{z_2}$};
		\node [style=none] (47) at (4.75, -1) {$D_{z_2}$};
		\node [style=none] (48) at (6.5, -4) {$D_{z_2}$};
		\node [style=none] (49) at (8.75, -7) {$D_{z_2}$};
		\node [style=none] (50) at (8.5, 1.5) {};
		\node [style=none] (51) at (9.5, 1.5) {};
		\node [style=none] (52) at (8.5, 0.5) {};
		\node [style=none] (53) at (9.5, 2) {$y$};
		\node [style=none] (54) at (8, 0.5) {$x$};
		\node [style=SmallCircleBlue] (55) at (1, 3) {};
		\node [style=SmallCircleBlue] (56) at (2, 0) {};
		\node [style=SmallCircleBlue] (57) at (3, -3) {};
		\node [style=SmallCircleBlue] (58) at (4, -6) {};
		\node [style=none] (59) at (2, -7.75) {};
	\end{pgfonlayer}
	\begin{pgfonlayer}{edgelayer}
		\draw [style=ThickLine] (0.center) to (2.center);
		\draw [style=ThickLine] (2.center) to (1.center);
		\draw [style=ThickLine] (1.center) to (0.center);
		\draw [style=ThickLine] (8.center) to (10.center);
		\draw [style=ThickLine] (10.center) to (9.center);
		\draw [style=ThickLine] (9.center) to (8.center);
		\draw [style=ThickLine] (16.center) to (18.center);
		\draw [style=ThickLine] (18.center) to (17.center);
		\draw [style=ThickLine] (17.center) to (16.center);
		\draw [style=ThickLine] (25.center) to (27.center);
		\draw [style=ThickLine] (27.center) to (26.center);
		\draw [style=ThickLine] (26.center) to (25.center);
		\draw (7) to (4);
		\draw (4) to (5);
		\draw (4) to (6);
		\draw (12) to (11);
		\draw (11) to (14);
		\draw (11) to (13);
		\draw (15) to (13);
		\draw (15) to (11);
		\draw (15) to (12);
		\draw (20) to (19);
		\draw (19) to (22);
		\draw (20) to (23);
		\draw (20) to (24);
		\draw (19) to (23);
		\draw (23) to (21);
		\draw (24) to (21);
		\draw (24) to (23);
		\draw (29) to (28);
		\draw (28) to (31);
		\draw (28) to (34);
		\draw (34) to (30);
		\draw (33) to (30);
		\draw (32) to (30);
		\draw (32) to (29);
		\draw (29) to (33);
		\draw (34) to (29);
		\draw [style=ArrowLineRight] (50.center) to (51.center);
		\draw [style=ArrowLineRight] (50.center) to (52.center);
		\draw (4) to (55);
		\draw (15) to (56);
		\draw (24) to (57);
		\draw (34) to (58);
	\end{pgfonlayer}
\end{tikzpicture}
}
\caption{Triangulated and refined toric diagrams for the orbifold $\mathbb{C}^3/\Z_{2n+2}$  with an action carrying weights $(1,1,2n)$. We give the triangulated diagrams for the first four cases $n=1,2,3,4$ associated with the crepant resolution. The black dots indicate $n$ compact toric divisors, located at $(x,y)=(0,0), \dots , (0,n-1)$. The 3 brown dots indicate the non-compact divisors $z_i=0$ denoted $D_{z_i}$ located at $(-1,0),(1,2n-1),(0,-1)$ for $i=1,2,3$ respectively. The single blue dot indicates the non-compact exceptional divisor at $(0,n)$  }
\label{fig:toricDs2}
\end{figure}

To proceed then, we take $Z$ to be the crepant resolution of $\mathbb{C}^3/\Z_{2n+2}(1,1,2n)$. See figure \ref{fig:toricDs2} for the triangulated toric diagrams corresponding to $n=1,2,3,4$. The bulk integral \eqref{eq:bulkint} for $\alpha$ is evaluated using the intersection ring of $Z$ which we therefore determine first via methods of toric geometry.

Under crepant resolution of $\mathbb{C}^3/\Z_{2n+2}$ the isolated codimension-6 singularity is replaced by a collection of $n$ compact complex surfaces:
\be \label{eq:Everything2}
 \begin{tikzpicture}
	\begin{pgfonlayer}{nodelayer}
		\node [style=none] (0) at (0.5, 0) {$\mathbb{F}_4$};
		\node [style=none] (1) at (-2.25, 0) {$\mathbb{F}_6$};
		\node [style=none] (2) at (-5.75, 0) {$\mathbb{F}_{2n}$};
		\node [style=none] (3) at (3.25, 0) {$\mathbb{F}_2$};
		\node [style=none] (4) at (7, 0) {$\frac{1}{2}\mathbb{F}_0=\mathbb{C}\times \P^1$};
		\node [style=none] (5) at (-1.75, 0) {};
		\node [style=none] (6) at (-2.75, 0) {};
		\node [style=none] (7) at (0, 0) {};
		\node [style=none] (8) at (1, 0) {};
		\node [style=none] (9) at (2.75, 0) {};
		\node [style=none] (10) at (3.75, 0) {};
		\node [style=none] (11) at (5.5, 0) {};
		\node [style=none] (12) at (-5.25, 0) {};
		\node [style=none] (13) at (5.3, 0.315) {$h$};
		\node [style=none] (14) at (4, 0.3) {$b_2$};
		\node [style=none] (15) at (2.5, 0.3) {$d_2$};
		\node [style=none] (16) at (1.25, 0.3) {$b_4$};
		\node [style=none] (17) at (-0.25, 0.3) {$d_4$};
		\node [style=none] (18) at (-1.5, 0.3) {$b_6$};
		\node [style=none] (19) at (-3, 0.3) {$d_6$};
		\node [style=none] (20) at (-4.875, 0.3) {$b_{2n}$};
		\node [style=none] (21) at (-4.25, 0) {};
		\node [style=none] (22) at (-3.75, 0) {};
	\end{pgfonlayer}
	\begin{pgfonlayer}{edgelayer}
		\draw [style=ThickLine] (5.center) to (7.center);
		\draw [style=ThickLine] (8.center) to (9.center);
		\draw [style=ThickLine] (10.center) to (11.center);
		\draw [style=ThickLine] (12.center) to (21.center);
		\draw [style=ThickLine] (22.center) to (6.center);
		\draw [style=DottedLine] (21.center) to (22.center);
	\end{pgfonlayer}
\end{tikzpicture}
\ee
Here we are using the same conventions and techniques as were used for \eqref{eq:Everything}. Here $h$ is the compact curve of $\P^1\times \C$ which is one half of $\mathbb{F}_0=\mathbb{P}^1\times \mathbb{P}^1$ in the sense that gluing two copies of  $\P^1\times \C$ by identifying the two copies of $\C$ via $z'=1/z$ gives $\mathbb{P}^1$.

This is made explicit in toric geometry. Starting from the fan with vertices
\be
u_1 = (-1,0,1)\,, \qquad  u_2 = (1,2n,1)\,, \qquad u_3 = (0,-1,1)\,,
\ee
we can apply standard techniques to construct $Z$. See figure \ref{fig:toricDs2} for the fully triangulated and refined toric diagram. Besides the compact toric divisors $\mathbb{F}_2,\dots, \mathbb{F}_{2n}$ there are also four non-compact toric divisors. In the unresolved geometry $\mathbb{C}^3/\Z_{2n+2}$ three of these these are simply obtained by setting each coordinate $z_i$ to zero in turn. We therefore denote these by $D_{z_i}$ indicated in the vanishing coordinate as a subscript. Modulo $2n+2$ we have $u_1+u_2+(2n)u_3=0$ and therefore the $u_i$ are in correspondence with the $D_{z_i}$. The final non-compact divisor is an exceptional toric divisor, topologically $\mathbb{C}\times \P^1$, which in the singular model has the $\mathbb{P}^1$ contracted to an $\mathfrak{su}(2)$ singularity.

We denote the coordinates associated with the $n$ compact exceptional toric divisors by $e_1,\dots, e_{n}=0$ respectively. The divisors are denoted respectively by $D_{e_1},\dots D_{e_{n}}$. Here $D_{e_k}$ corresponds to the vector
\be
v_k=(0,k-1,1)\,.
\ee
We also introduce the non-compact exceptional divisor $D_{f}$ with vertex $v_f=(0,n,1)$. Explicitly, $D_{f}=\C\times \P^1$ and $D_{e_k}=\mathbb{F}_{2k}$ for $k=1,\dots, n$. Overall, proceeding with standard techniques, these toric divisors are found to be subject to three linear equivalences which when solved for the 3 non-compact divisors $D_{z_i}$ expresses these as a rational linear combination of compact divisors
\be \ba
D_{z_1}&= -\frac{1}{2}D_f - \frac{1}{2n+2}\sum_{k=1}^{n} kD_{e_{k}}\\
D_{z_2}&=  -\frac{1}{2}D_f   - \frac{1}{2n+2}\sum_{k=1}^{n} kD_{e_{k}} \\
D_{z_3}&= -\frac{1}{n+1}\sum_{\ell=1}^{n} (n+1-\ell)D_{e_\ell}\,.
\ea\ee
With this we can compute all triple intersections between toric divisors from those of the exceptional divisors, which are
\be\label{eq:TripInt2} \ba
(\mathbb{P}^1\times \mathbb{C} )\cdot(\mathbb{P}^1\times \mathbb{C} )\cdot (\mathbb{P}^1\times \mathbb{C} )&=1\,, \\
 \mathbb{F}_k \cdot \mathbb{F}_k  \cdot \mathbb{F}_k&=8\,,
\ea \ee
where we used $\mathbb{P}^1\times \mathbb{C} =\mathbb{F}_0/2$, and
\be \ba
&(\mathbb{P}^1\times \mathbb{C} )\cdot(\mathbb{P}^1\times \mathbb{C} )\cdot \mathbb{F}_2 =-2\,, &&\quad \mathbb{F}_{k+2} \cdot \mathbb{F}_{k+2} \cdot \mathbb{F}_{k} =k\,,\\
& \mathbb{F}_2 \cdot \mathbb{F}_2 \cdot (\mathbb{P}^1\times \mathbb{C} ) = 0 \,,  && \quad \mathbb{F}_{k-2} \cdot \mathbb{F}_{k-2} \cdot \mathbb{F}_{k} =-k \,.
\ea \ee
Next, we discuss the choice of divisor representing $T_2$ in \eqref{eq:bulkint}. We have
\be
H^2(S^5/\Z_{2n})=\Z_n\,,
\ee
which is the subgroup $\Z_n\subset \Z_{2n}\cong \Gamma$. Considering coefficients modulo 1 (which realizes the bulk to boundary screening) we see that $D_{z_3}$ is a good non-compact divisor which on the boundary corresponds to a generator of $\Z_n$.
Then, given the above intersection ring, we now compute the pure 1-form anomaly $\alpha$ of the 5D SCFTs engineered by M-theory on $\mathbb{C}^3/\Z_{2n+2}(1,1,2n)$. With respect to the completely smooth $Z$ the bulk integral \eqref{eq:bulkint} is expressed as an intersection as
\be
\frac{D_{z_3}\cdot D_{z_3} \cdot D_{z_3}}{6} - \frac{p_1(Z) \cdot D_{z_3}}{24}=-\frac{n(n-1)}{6(n+1)}\qquad  \text{mod}\,1\,.
\ee
The above expression disagrees with the first entry of table \ref{tab:TableForNonIsolated} by a sign. Therefore, $-D_{z_3}$ is dual to the Kawasaki generator, i.e., $s_3=-1$. Next, to determine the remaining entries in the table, we compute the charge, following \eqref{eq:LinkingCharge}, to
\be
\ell(v_2,w_4)=\frac{(2n)^2}{(2n)(2n+2)}=\frac{n}{n+1}=\frac{-2}{2n+2}\qquad \text{mod}\,1\,.
\ee
therefore $q=-2$. With this charge we have with \eqref{eq:EtaFixed}
\be\ba\label{eq:alphaDiracAp2}
\frac{1}{2}\eta^{\text{\:\!$\slashed{D}$}}_{\raisebox{-0.3ex}{\scriptsize$L_q$}}&=\frac{1}{2n+2}\lb \sum_{k=1}^{n}+\sum_{k=n+2}^{2n+1}\rb (-1)^k\omega^{-2k}\frac{1}{\omega^{-k/2}-\omega^{k/2}}\frac{1}{\omega^{-k/2}-\omega^{k/2}}\frac{1}{\omega^{-k(2n-2)/2}-\omega^{k(2n-2)/2}}\\[0.5em]
&=\frac{n(n-1)}{3(2n+2)}=0,\frac{1}{9},\frac{1}{4},\frac{2}{5},\frac{5}{9},\frac{5}{7},\frac{7}{8},\frac{28}{27},\dots\,,
\ea\ee
with $\omega=\exp(2\pi i/(2n+2))$. Following \eqref{eq:Refined} this refines to
\be\ba\label{eq:alphaDiracAp3}
\frac{1}{2}\eta^{\text{\:\!$\slashed{D}$}}_{\raisebox{-0.3ex}{\scriptsize$L_{\frac{q}{2}=-1}$}}&=\frac{1}{2n+2}\lb \sum_{k=1}^{n}+\sum_{k=n+2}^{2n+1}\rb (-1)^k\omega^{-k}\frac{1}{\omega^{-k/2}-\omega^{k/2}}\frac{1}{\omega^{-k/2}-\omega^{k/2}}\frac{1}{\omega^{-k(2n-2)/2}-\omega^{k(2n-2)/2}}\\[0.5em]
&=\frac{n(n+2)}{6(2n+2)}=\frac{1}{8},\frac{2}{9},\frac{5}{16},\frac{2}{5},\frac{35}{72},\frac{4}{7},\frac{21}{32},\frac{20}{27},\dots\,,
\ea\ee
We have reproduce the first entry in table \ref{tab:TableForNonIsolated} and find a match between bulk and boundary.

\bibliographystyle{utphys}
\bibliography{KvsH}

\end{document}